Fully Countering Trusting Trust through Diverse Double-Compiling

A dissertation submitted in partial fulfillment of the requirements for the degree of Doctor of Philosophy at George Mason University

By

David A. Wheeler
Master of Science
George Mason University, 1994
Bachelor of Science
George Mason University, 1988

Co-Directors: Dr. Daniel A. Menascé and Dr. Ravi Sandhu, Professors
The Volgenau School of Information Technology & Engineering

Fall Semester 2009
George Mason University
Fairfax, VA

# Dedication

This is dedicated to my wife and children, who sacrificed many days so I could perform this work, to my extended family, and to the memory of my former mentors Dennis W. Fife and Donald Macleay, who always believed in me.

Soli Deo gloria—Glory to God alone.

# Acknowledgments

I would like to thank my PhD committee members and former members Dr. Daniel A. Menascé, Dr. Ravi Sandhu, Dr. Paul Ammann, Dr. Jeff Offutt, Dr. Yutao Zhong, and Dr. David Rine, for their helpful comments.

The Institute for Defense Analyses (IDA) provided a great deal of help. Dr. Roger Mason and the Honorable Priscilla Guthrie, former directors of IDA's Information Technology and Systems Division (ITSD), partly supported this work through IDA's Central Research Program. Dr. Margaret E. Myers, current IDA ITSD director, approved its final release. I am very grateful to my IDA co-workers (alphabetically by last name) Dr. Brian Cohen, Aaron Hatcher, Dr. Dale Lichtblau, Dr. Reg Meeson, Dr. Clyde Moseberry, Dr. Clyde Roby, Dr. Ed Schneider, Dr. Marty Stytz, and Dr. Andy Trice, who had many helpful comments on this dissertation and/or the previous ACSAC paper. Reg Meeson in particular spent many hours carefully reviewing the proofs and related materials, and Clyde Roby carefully reviewed the whole dissertation; I thank them both. Aaron Hatcher was immensely helpful in working to scale the Diverse Double-Compiling (DDC) technique up to a real-world application using GCC. In particular, Aaron helped implement many applications of DDC that we thought should have worked with GCC, but didn't, and then helped to determine *why* they didn't work. These "Edison successes" (which successfully found out what did *not* work) were important in helping to lead to a working application of DDC to GCC.

Many others also helped create this work. The work of Dr. Paul A. Karger, Dr. Roger R. Schell, and Ken Thompson made the world aware of a problem that needed solving; without knowing there was a problem, there would have been no work to solve it. Henry Spencer posted the first version of this idea that eventually led to this dissertation (though this dissertation expands on it far beyond the few sentences that he wrote). Henry Spencer, Eric S. Raymond, and the anonymous ACSAC reviewers provided helpful comments on the ACSAC paper. I received many helpful comments and other information after publication of the ACSAC paper, including comments from (alphabetically by last name) Bill Alexander, Dr. Steven M. Bellovin, Terry Bollinger, Ulf Dittmer, Jakub Jelinek, Dr. Paul A. Karger, Ben Laurie, Mike Lisanke, Thomas Lord, Bruce Schneier, Brian Snow, Ken Thompson, Dr. Larry Wagoner, and James Walden. Tawnia Wheeler proofread both the ACSAC paper and this document; thank you! My thanks to the many developers of the OpenDocument specification and the OpenOffice.org implementation, who made developing this document a joy.

# Table of Contents

# List of Tables

# List of Figures

# List of Abbreviations and Symbols

| | |
|---|---|
| -A | not A. Equivalent to $\neg A$ |
| A & B | A and B (logical and). Equivalent to $A \wedge B$ |
| A \| B | A or B (logical or). Equivalent to $A \vee B$ |
| A -> B | A implies B. Equivalent to $A \rightarrow B$ and $(\neg A) \vee B$ |
| ACL2 | A Computational Logic for Applicative Common Lisp |
| ACSAC | Annual Computer Security Applications Conference |
| aka | also known as |
| all X A | for all X, A. Equivalent to $\forall X. A$ |
| ANSI | American National Standards Institute |
| API | Application Programmer Interface |
| ASCII | American Standard Code for Information Interchange |
| BIOS | Basic input/output system |
| BSD | Berkeley Software Distribution |
| cA or $c_A$ | Compiler $c_A$, the compiler-under-test executable (see $s_A$) |
| cGP or $c_{GP}$ | Compiler $c_{GP}$, the putative grandparent of $c_A$ and putative parent of $c_P$ |
| CNSS | Committee on National Security Systems |
| cP or $c_P$ | Compiler P, the putative parent of $c_A$ |
| CP/M | Control Program for Microcomputers |
| CPU | Central Processing Unit |
| cT or $c_T$ | Compiler $c_T$, a "trusted" compiler (see section 4.3) |
| DDC | Diverse Double-Compiling |
| DoD | Department of Defense (U.S.) |
| DOS | Disk Operating System |
| DRAM | Dynamic Random Access Memory |
| e1 | Environment that produces stage1 |
| e2 | Environment that produces stage2 |
| eA | Environment that putatively produced $c_A$ |
| eArun | Environment that $c_A$ and stage2 are intended to run in |
| EBCDIC | Extended Binary Coded Decimal Interchange Code |
| ECC | Error Correcting Code(s) |
| eP | Environment that putatively produced $c_P$ |
| FOL | First-Order Logic (with equality), aka first-order predicate logic |
| FS | Free Software |
| FLOSS | Free-Libre/Open Source Software |
| FOSS | Free/Open Source Software |
| FSF | Free Software Foundation |
| GAO | General Accounting Office (U.S.) |
| GCC | GNU Compiler Collection (formerly the GNU C compiler) |
| GNU | GNU's not Unix |
| GPL | General Public License |
| HOL | Higher Order Logic |
| IC | Integrated Circuit |
| IDA | Institute for Defense Analyses |
| iff | if and only if |
| I/O | input/output |
| IP | Intellectual Property |
| ISO | International Organization for Standardization (sic) |
| ITSD | Information Technology and Systems Division |
| MDA | Missile Defense Agency (U.S.); formerly named SDIO |
| MS-DOS | Microsoft Disk Operating System (MS-DOS) |
| NEL | Newline (#x85), used in OS/360 |
| NIST | National Institute of Science and Technology (U.S.) |
| OpenBSD | Open Berkeley Software Distribution |
| OS/360 | IBM System/390 operating-system |
| OSI | Open Source Initiative |
| OSS | Open Source Software |
| OSS/FS | Open Source Software/Free Software |
| PITAC | President's Information Technology Advisory Committee |
| ProDOS | Professional Disk Operating System |
| PVS | Prototype Verification System |
| QED | Quod erat demonstrandum ("which was to be demonstrated") |
| RepRap | Replicating Rapid-prototyper |
| S-expression | Symbolic expression |
| sA or $s_A$ | putative source code of $c_A$ |
| SAMATE | Software Assurance Metrics And Tool Evaluation (NIST project) |
| SDIO | Strategic Defense Initiative Organization (U.S.); later renamed to the Missile Defense Agency (MDA) |
| SHA | Secure Hash Algorithm |
| sic | spelling is correct |
| sP or $s_P$ | putative source code of $c_P$ |
| SQL | Structured Query Language |
| STEM | Scanning Transmission Electron Microscope |
| tcc or TinyCC | Tiny C Compiler |
| UCS | Universal Character Set |
| URL | Uniform Resource Locator |
| U.S. | United States |
| UTF-8 | 8-bit UCS/Unicode Transformation Format |
| UTF-16 | 16-bit UCS/Unicode Transformation Format |
| VHDL | VHSIC hardware description language |
| VHSIC | Very High Speed Integrated Circuit |
| $\Phi, \Psi, \Lambda$ | Arbitrary FOL formula |
| $\tau_x$ | Arbitrary FOL term number x |

See appendix E for key definitions.

# Abstract

FULLY COUNTERING TRUSTING TRUST THROUGH DIVERSE DOUBLE-COMPILING

David A. Wheeler, PhD

George Mason University, 2009

Dissertation Directors: Dr. Daniel A. Menascé and Dr. Ravi Sandhu

An Air Force evaluation of Multics, and Ken Thompson's Turing award lecture ("Reflections on Trusting Trust"), showed that compilers can be subverted to insert malicious Trojan horses into critical software, including themselves. If this "trusting trust" attack goes undetected, even complete analysis of a system's source code will not find the malicious code that is running. Previously-known countermeasures have been grossly inadequate. If this attack cannot be countered, attackers can quietly subvert entire classes of computer systems, gaining complete control over financial, infrastructure, military, and/or business systems worldwide. This dissertation's thesis is that the trusting trust attack can be detected and effectively countered using the "Diverse Double-Compiling" (DDC) technique, as demonstrated by (1) a formal proof that DDC can determine if source code and generated executable code correspond, (2) a demonstration of DDC with four compilers (a small C compiler, a small Lisp compiler, a small maliciously corrupted Lisp compiler, and a large industrial-strength C compiler, GCC), and (3) a description of approaches for applying DDC in various real-world scenarios. In the DDC technique, source code is compiled twice: the source code of the compiler's parent is compiled using a trusted compiler, and then the putative compiler source code is compiled using the result of the first compilation. If the DDC result is bit-for-bit identical with the original compiler-under-test's executable, and certain other assumptions hold, then the compiler-under-test's executable corresponds with its putative source code.

# 1 Introduction

Many software security evaluations examine source code, under the assumption that a program's source code accurately represents the executable actually run by the computer[1]. Naïve developers presume that this can be assured simply by recompiling the source code to see if the same executable is produced. Unfortunately, the "trusting trust" attack can falsify this presumption.

For purposes of this dissertation, an executable that does not correspond to its putative source code is *corrupted*[2]. If a corrupted executable was intentionally created, we can call it a *maliciously corrupted* executable. The *trusting trust attack* occurs when an attacker attempts to disseminate a compiler executable that produces corrupted executables, at least one of those produced corrupted executables is a corrupted compiler, and the attacker attempts to make this situation self-perpetuating. The attacker may use this attack to insert other Trojan horse(s) (software that appears to the user to perform a desirable function but facilitates unauthorized access into the user's computer system).

---

[1]An *executable* is data that can be directly executed by a computing environment. An executable may be code for an actual machine or for a simulated machine (e.g., a "byte code"). A common alternative term for executable is "binary" (e.g., [Sabin2004]), but this term is misleading; in modern computers, *all* data is represented using binary codes. For purposes of this dissertation, "object code" is a synonym for "executable". *Source code* is a representation of a program that can be translated into an executable, and is typically human-readable. A *compiler* is an executable that when executed translates source code into an executable (it may also perform other actions). An *assembler* is a compiler for a language whose instructions are primarily a close approximation of the executing environment's instructions. The process of using a compiler to translate source code into an executable is termed *compiling*.

[2]An executable e corresponds to source code s if and only if execution of e always behaves as specified by s when the execution environment of e behaves correctly.

Information about the trusting trust attack was first published in [Karger1974]; it became widely known through [Thompson1984]. Unfortunately, there has been no practical way to fully detect or counter the trusting trust attack, because repeated in-depth review of industrial compilers' executable code is impractical.

For source code evaluations to be strongly credible, there must be a way to justify that the source code being examined accurately represents what is being executed—yet the trusting trust attack subverts that very claim. Internet Security System's David Maynor argues that the risk of attacks on compilation processes is increasing [Maynor2004] [Maynor2005]. Karger and Schell noted that the trusting trust attack was still a problem in 2000 [Karger2000], and some technologists doubt that computer-based systems can ever be secure because of the existence of this attack [Gauis2000]. Anderson et al. argue that the general risk of subversion is increasing [Anderson2004].

Recently, in several mailing lists and blogs, a technique to detect such attacks has been briefly described, which uses a second (diverse) "trusted" compiler (as will be defined in section 4.3) and two compilation stages. This dissertation terms the technique "diverse double-compiling" (DDC). In the DDC technique, the source code of the compiler's parent is compiled using a trusted compiler, and then the putative compiler source code is compiled using the result of the first compilation (chapter 4 further explains this). If the DDC result is bit-for-bit identical with the original compiler-under-test's executable, and certain other assumptions hold, then the compiler-under-test's executable corresponds with its putative source code (chapter 5 justifies this claim). Before this work began, there had been no examination of DDC in detail which identified its assumptions, proved its correctness or effectiveness, or discussed practical issues in applying it. There had also not been any public demonstration of DDC.

This dissertation's thesis is that the trusting trust attack can be detected and effectively countered using the "Diverse Double-Compiling" (DDC) technique, as demonstrated by (1) a formal proof that DDC can determine if source code and generated executable code correspond, (2) a demonstration of DDC with four compilers (a small C compiler, a small Lisp compiler, a small maliciously corrupted Lisp compiler, and a large industrial-strength C compiler, GCC), and (3) a description of approaches for applying DDC in various real-world scenarios.

This dissertation provides background and a description of the threat, followed by an informal description of DDC. This is followed by a formal proof of DDC, information on how diversity (a key requirement of DDC) can be increased, demonstrations of DDC, and information on how to overcome practical challenges in applying DDC. The dissertation closes with conclusions and ramifications. Appendices have some additional detail. Further details, including materials sufficient to reproduce the experiments, are available at:

http://www.dwheeler.com/trusting-trust/

This dissertation follows the guidelines of [Bailey1996] to enhance readability. In addition, this dissertation uses logical (British) quoting conventions; quotes do not enclose punctuation unless they are part of the quote [Ritter2002]. Including extraneous characters in a quotation can be grossly misleading, especially in computer-related material [Raymond2003, chapter 5].

# 2 Background and related work

This chapter provides background and related work. It begins with a discussion of the initial revelation of the trusting trust attack by Karger, Schell, and Thompson, including a brief description of "obvious" yet inadequate solutions. The next sections discuss work on corrupted or subverted compilers, the compiler bootstrap test, general work on analyzing software, and general approaches for using diversity to improve security. This is followed by evidence that software subversion is a real problem, not just a theoretical concern. This chapter concludes by discussing the DDC paper published by the Annual Computer Security Applications Conference (ACSAC) [Wheeler2005] and the improvements to DDC that have been made since that time.

## 2.1 Initial revelation: Karger, Schell, and Thompson

Karger and Schell provided the first public description of the problem that compiler executables can insert malicious code into themselves. They noted in their examination of Multics vulnerabilities that a "penetrator could insert a trap door into the... compiler... [and] since the PL/I compiler is itself written in PL/I, the trap door can maintain itself, even when the compiler is recompiled. Compiler trap doors are significantly more complex than the other trap doors... However, they are quite practical to implement" [Karger1974].

Ken Thompson widely publicized this problem in his 1984 Turing Award presentation ("Reflections on Trusting Trust"), clearly explaining it and demonstrating that this was both a practical and dangerous attack. He described how to modify the Unix C compiler to inject a Trojan horse, in this case to modify the operating system login program to surreptitiously give him root access. He also added code so that the compiler would inject a Trojan Horse when compiling itself, so the compiler became a "self-reproducing program that inserts both Trojan horses into the compiler". Once this is done, the attacks could be removed from the source code. At that point no source code examination—even of the compiler—would reveal the existence of the Trojan horses, yet the attacks could persist through recompilations and cross-compilations of the compiler. He then stated that "No amount of source-level verification or scrutiny will protect you from using untrusted code... I could have picked on any program-handling program such as an assembler, a loader, or even hardware microcode. As the level of program gets lower, these defects will be harder and harder to detect" [Thompson1984]. Thompson's demonstration also subverted the disassembler, hiding the attack from disassembly. Thompson implemented this attack in the C compiler and (as a demonstration) successfully subverted another Bell Labs group, the attack was never detected.

Thompson later gave more details about his demonstration, including assurances that the maliciously corrupted compiler was never released outside Bell Labs [Thornburg2000].

Obviously, this attack invalidates security evaluations based on source code review, and recompilation of source code using a potentially-corrupted compiler does not eliminate the risk. Some simple approaches appear to solve the problem at first glance, yet fail to do so or have significant weaknesses:

- Compiler executables could be manually compared with their source code. This is impractical given compilers' large sizes, complexity, and rate of change.

- Such comparison could be automated, but optimizing compilers make such comparisons extremely difficult, compiler changes make keeping such tools up-to-date difficult, and the tool's complexity would be similar to a compiler's.

- A second compiler could compile the source code, and then the executables could be compared automatically to argue semantic equivalence. There is some work in determining the semantic equivalence of two different executables [Sabin2004], but this is very difficult to do in practice.

- Receivers could require that they only receive source code and then recompile everything themselves. This fails if the receiver's compiler is already maliciously corrupted; thus, it simply moves the attack location. An attacker could also insert the attack into the compiler's source; if the receiver accepts it (due to lack of diligence or conspiracy), the attacker could remove the evidence in a later version of the compiler (as further discussed in section 8.4).

- Programs can be written in interpreted languages. But eventually an interpreter must be implemented by machine code, so this simply moves the attack location.

## 2.2 Other work on corrupted compilers

Some previous papers outline approaches for countering corrupted compilers, though their approaches have significant weaknesses. Draper [Draper1984] recommends screening out maliciously corrupted compilers by writing a "paraphrase" compiler (possibly with a few dummy statements) or a different compiler executable, compiling once to remove the Trojan horse, and then compiling a second time to produce a Trojan horse-free compiler. This idea is expanded upon by McDermott [McDermott1988], who notes that the alternative compiler could be a reduced-function compiler or one with large amounts of code unrelated to compilation. Lee's

"approach #2" describes most of the basic process of diverse double-compiling, but implies that the results might not be bit-for-bit identical [Lee2000]. Luzar makes a similar point as Lee, describing how to rebuild a system from scratch using a different trusted compiler but not noting that the final result should be bit-for-bit identical if other factors are carefully controlled [Luzar2003].

None of these papers note that it is possible to produce a result that is bit-for-bit identical to the original compiler executable. This is a significant advantage of diverse double-compiling (DDC), because determining if two different executables are "functionally equivalent" is extremely difficult[3], while determining if two executables are bit-for-bit identical is extremely easy. These previous approaches require each defender to recompile their compiler themselves before using it; in contrast, DDC can be used as an after-the-fact vetting process by multiple third parties, without requiring a significant change in compiler delivery or installation processes, and without requiring that all compiler users receive the compiler source code. All of these previous approaches simply move the potential vulnerability somewhere else (e.g., to the process using the "paraphrase" compiler). In contrast, an attacker who wishes to avoid detection by DDC must corrupt *both* the original compiler and *every* application of DDC to that executable, so each application of DDC can further build confidence that a specific executable corresponds with its putative source code. Also, none of these papers demonstrate their technique.

Magdsick discusses using different versions of a compiler, and different compiler platforms such as central processing unit (CPU) and operating system, to check executables. However, Magdsick presumes that the compiler itself will be the same base compiler (though possibly a different version). He does note the value of recompiling "everything" to check it [Magdsick2003]. Anderson notes that cross-compilation does not help if the attack is in the compiler [Anderson2003]. Mohring argues for the use of recompilation by GCC to check other components, presuming that the GCC executables themselves in some environments would be pristine [Mohring2004]. He makes no notice that all GCC executables used might be maliciously corrupted, or of the importance of diversity in compiler implementation. In his approach different compiler versions may be used, so outputs would be "similar" but not identical; this leaves the difficult problem of comparing executables for "exact equivalence" unresolved.

---

[3]Determining if two executables are equivalent is undecidable in general; see section 5.6.1.

A great deal of effort has been spent to develop proofs of correctness for compilers, either of the compiler itself and/or its generated results [Dave2003] [Stringer-Calvert1998] [Bellovin1982]. This is quite difficult even for simple languages, though there has been progress. [Leinenbach2005] discusses progress in verifying a subset C compiler using Isabelle/Higher Order Logic (HOL). "Compcert" is a compiler that generates PowerPC assembly code from Clight (a large subset of the C programming language); this compiler is primarily written using the specification language of the Coq proof assistant, and its correctness (that the generated assembly code is semantically equivalent to its source program) has been entirely proved within the Coq proof assistant [Leroy2006] [Blazy2006] [Leroy2008] [Leroy2009]. [Goerigk1997] requires formal specifications and correspondence proofs, along with double-checking of resulting transformations with the formal specifications. It does briefly note that "if an independent (whatever that is) implementation of the specification will generate an equal bootstrapping result, this fact might perhaps increase confidence. Note however, that, in particular in the area of security... We want to guarantee the correctness of the generated code, e.g., preventing criminal attacks" [Goerigk1997, 17]. However, it does not explain what independence would mean, nor what kind of confidence this equality would provide. [Goerigk1999] specifically focuses on countering Trojan horses in compilers, through formal verification techniques, but again this requires having formal specifications and performing formal correspondence proofs. Goerigk recommends "a posteriori code inspection based on syntactic code comparison" to counter the trusting trust attack, but such inspection is very labor-intensive on industrial-scale compilers that implement significant optimizations. DDC can be dramatically strengthened by having formal specifications and proofs of compilers (which can then be used as the trusted compiler), but DDC does not require them. Indeed, DDC and formal proofs of compilers can be used in a complementary way: A formally-proved compiler may omit many useful optimizations (as they can be difficult or time-consuming to prove), but it can still be used as the DDC "trusted compiler" to gain confidence in another (production-ready) compiler.

Spinellis argues that "Thompson showed us that one cannot trust an application's security policy by examining its source code... The recent Xbox attack demonstrated that one cannot trust a platform's security policy if the applications running on it cannot be trusted" [Spinellis2003]. It is worth noting that the literature for change detection (such as [Kim1994] and [Forrest1994]) and intrusion detection do not easily address this problem, because a compiler is *expected* to accept source code and generate object code.

Faigon's "Constrained Random Testing" process detects compiler defects by creating many random test programs, compiling them with a compiler-under-test and a reference compiler, and detecting if running them produces different results [Faigon]. Faigon's approach may be useful for finding some compiler errors, but it is extremely unlikely to find maliciously corrupted compilers.

## 2.3 Compiler bootstrap test

A common test for errors used by many compilers (including GCC) is the so-called "compiler bootstrap test". Goerigk formally describes this test, crediting Niklaus Wirth's 1986 book

*Compilerbau* as proposing this test for detecting errors in compilers [Goerigk1999]. In this test, if c(s,b) is the result of compiling source s using compiler executable b, and $\overline{m}$ is some other compiler (the "bootstrap" compiler), then[4]:

> *If m0 and s are both correct and deterministic, $\overline{m}$ is correct, m0=c(s,$\overline{m}$), m1=c(s,m0), m2=c(s,m1), all compilations terminate, and if the underlying hardware works correctly, then m1=m2.*

The compiler bootstrap test goes through steps to determine if m1=m2; if not, there is a compiler error of some kind. This test finds many unintentional errors, which is why it is popular. But [Goerigk1999] points out that this test is insufficient to make strong claims, in particular, m1 may equal m2 even if $\overline{m}$, m0, or s are *not* correct. For example, it is trivial to create compiler source code that passes this test, yet is incorrect, since this test only tests features used in the compiler itself. More importantly (for purposes of this dissertation), if $\overline{m}$ is a maliciously corrupted compiler, a compilation process can pass this test yet produce a maliciously corrupted compiler m2. Note that the compiler bootstrap test does *not* consider the possibility of using two different bootstrap compilers ($\overline{m}$ and $\overline{m}'$) and later comparing their different compiler results (m2 and m2′) to see if they produce the same (bit-for-bit) result. Therefore, the DDC technique is *not* the same as the compiler bootstrap test. However, DDC *does* have many of the same preconditions as the compiler bootstrap test. Since the compiler bootstrap test is popular, many DDC preconditions are already met by typical industrial compilers, making DDC easier to apply to typical industrial compilers.

## 2.4 Analyzing software

All programs can be analyzed to find intentionally-inserted or unintentional security issues (aka vulnerabilities). These techniques can be broadly divided into static analysis (which examines a static representation of the program, such as source code or executable, without executing it) and dynamic analysis (which examines what the program does while it is executing). Formal methods, which are techniques that use mathematics to prove programs or program models are correct, can be considered a specific kind of static analysis technique.

---

[4]This is theorem 2 (the bootstrap test theorem) of [Goerigk1999]. For clarity, the text has been modified so that its notation is the same as the notation used in this dissertation.

Since compilers are programs, these general analysis techniques (both static and dynamic) that are not specific to compilers can be used on compilers as well.

## 2.4.1 Static analysis

Static analysis techniques examine programs (their source code, executable, or both) without executing them. Both programs and humans can perform static analysis.

There are many static analysis programs (aka tools) available; many are focused on identifying security vulnerabilities in software. The National Institute of Science and Technology (NIST) Software Assurance Metrics And Tool Evaluation (SAMATE) project (http://samate.nist.gov) is "developing methods to enable software tool evaluations, measuring the effectiveness of tools and techniques, and identifying gaps in tools and methods". SAMATE has collected a long list of static analysis programs for finding security vulnerabilities by examining source code or executable code. There are also a number of published reports comparing various static analysis tools, such as [Zitser2004], [Forristal2005], [Kratkiewicz2005], and [Michaud2006]. A draft functional specification for source code analysis tools has been developed [Kass2006], proposing a set of defects that such tools would be required to find and the code complexity that they must be able to handle while detecting them.

Although [Kass2006] briefly notes that source code analysis tools might happen to find malicious trap doors, many documents on static analysis focus on finding *unintentional* errors, not maliciously-implanted vulnerabilities. [Kass2006] specifies a specific set of security-relevant errors that have been made many times in real programs, and limits the required depth of the analysis (to make analysis time and reporting manageable). [Chou2006] also notes that in practice, static analyzers give up on error classes that are too hard to diagnose. For unintentional vulnerabilities, this is sensible; unintentional errors that have commonly occurred in the past are likely to recur (so searching for them can be very helpful). Unfortunately, these approaches are less helpful against an adversary who is *intentionally* inserting malicious code into a program. An adversary could intentionally insert one of these common errors, perhaps because they have high deniability, but ensure that it is so complex that a tool is unlikely to find it. Alternatively, an adversary could insert code that is an attack but not in the list of patterns the tools search for. Indeed, an adversary can repeatedly use static analysis tools until he or she has verified that the malicious code will *not* be detected later by those tools.

Static analysis tools also exist for analyzing executable files, instead of source code files. Indeed, [Balakrishnan2005] argues that program analysis should begin with executables instead of source code, because only the executables are actually run and source code analysis can be misled. To address this, there are efforts to compute better higher-level constructs from executable code, but in the general case this is still a difficult research area [Linger2006].

[Wysopal] presents a number of heuristics that can be used to statically detect some application backdoors in executable files. This includes identifying static variables that "look like" usernames, passwords, or cryptographic keys, searching for network application programmer interface (API) calls in applications where they are unexpected, searching for standard date/time API calls (which may lead to a time bomb), and so on. Unfortunately, many malicious programs

will not be detected by such heuristics, and as noted above, attackers can develop malicious software in ways that specifically avoid detection by the heuristics of such tools.

Many static analysis tools for executables use the same approach as many static analysis tools for source code: they search for specific programs or program fragments known to be problematic. The most obvious case are virus-checkers; though it is possible to examine behavior, and some anti-virus programs are increasingly doing so, historically "anti-virus" programs have a set of patterns of known viruses, which is constantly updated and used to search various executables (e.g., in a file or boot record) to see if these patterns are present [Singh2002] [Lapell2006]. However, as noted in Fred Cohen's initial work on computer viruses [Cohen1985], viruses can mutate as they propagate, and it is not possible to create a pattern listing all-and-only malicious programs. [Christodorescu2003] attempts to partially counter this; this paper regards malicious code detection as an obfuscation-deobfuscation game between malicious code writers and researchers, and presents an architecture for detecting known malicious patterns in executables that are hidden by common obfuscation techniques. Even this more robust architecture does not work against different malicious patterns, nor against different obfuscation techniques.

Of course, even if tools cannot find malicious code, detailed human review can be used at the source or executable level if the software is critical enough to warrant it. For example, the Open Berkeley Software Distribution (OpenBSD) operating system source code is regularly and purposefully examined by a team of people with the explicit intention of finding and fixing security holes, and as a result has an excellent security record [Payne2002]. The Strategic Defense Initiative Organization (SDIO), now named the Missile Defense Agency (MDA), even developed a set of process requirements to counter malicious and unintentional vulnerabilities, emphasizing multi-person knowledge and review along with configuration management and other safeguards [SDIO1993].

Unfortunately, the trusting trust attack can render human reviews moot if there is no technique to counter the attack. The trusting trust attack immediately renders examination of the source code inadequate, because the executable code need not correspond to the source code. Thompson's attack subverted the symbolic debugger, so in that case, even human review of the executable could fail to detect the attack. Thus, human reviews are less convincing unless the trusting trust attack is itself countered.

Human review also presumes that other humans examining source code or executables will be able to detect malicious code. In large code bases, this can be a challenge simply due to their size and complexity. In addition, it is possible for an adversary to create source code that *appears* to work correctly, yet actually performs a malevolent action instead. This dissertation uses the term *maliciously misleading code* for any source code that is intentionally designed to look benign, yet creates a vulnerability (including an attack). The topic of maliciously misleading code is further discussed in section 8.11.

### 2.4.2 Dynamic analysis

It is also possible to use dynamic techniques in an attempt to detect and/or counter vulnerabilities by examining the activities of a system, and then halting or examining the system when those activities are suspicious. A trivial example is execution testing, where a small set of inputs are provided and the inputs are checked to see if they are correct. However, dynamic analysis is completely inadequate for countering the trusting trust attack.

Traditional execution testing is unlikely to counter the trusting trust attack. Such attacks will only "trigger" on very specific inputs, as discussed in section 3.2, so even if the executable is examined in detail, it is extremely unlikely that traditional execution testing will detect this problem.

Detecting at run-time arbitrary corrupted code in a compiler or the executable code it generates is very difficult. The fundamental behavior of a corrupted compiler – that it accepts source code and generates an executable – is no different from a uncorrupted one. Similarly, any malicious code a compiler inserts into other programs can often be made to behave normally in most cases. For example, a login program with a trap door (a hidden username and/or password) has the same general behavior: It decides if a user may log in and what privileges to apply. Indeed, it may act completely correctly as long as the hidden username and/or password are not used.

In theory, continuous comparison of an executable's behavior at run-time to its source code could detect differences between the executable and source code. Unfortunately, this would need to be done all the time, draining performance. Even worse, tools to do this comparison, given modern compilers producing highly optimized code, would be far more complex than a compiler, and would themselves be vulnerable to attack.

Given an extremely broad definition of "system", the use of software configuration management tools and change detection tools like Tripwire [Kim1994] could be considered dynamic techniques for countering malicious software. Both enable detection of changes in the behavior of a larger system. Certainly a configuration management system could be used to record changes made to compiler source, and then used to enable reviewers to examine just the differences. But again, such review presupposes that any vulnerability in an executable could be revealed by analyzing its source code, a presupposition the trusting trust attack subverts.

A broader problem is that once code is running, *some* programs must be trusted, and at least some of that code will almost certainly have been generated by a compiler. Any program that attempts to monitor execution might itself be subverted, just as Thompson subverted the symbolic debugger, unless there is a technique to prevent it. In any case, it would be better to detect and counter malicious code *before* it executed, instead of trying to detect malicious code's execution while or after it occurs.

## 2.5 Diversity for security

There are a number of papers and articles about employing diversity to aid computer security, though they generally do not discuss or examine how to use diversity to counter Trojan horses inside compilers themselves or the compilation environment.

Geer et al. strongly argue that a monoculture (an absence of diversity) in computing platforms is a serious security problem [Geer2003] [Bridis2003], but do not discuss employing compiler diversity to counter this particular attack.

Forrest et al argue that run-time diversity in general is beneficial for computer security. In particular, their paper discusses techniques to vary final executables by "randomized" transformations affecting compilation, loading, and/or execution. Their goal was to automatically change the executable (as seen at run-time) in some random ways sufficient to make it more difficult to attack. The paper provides a set of examples, including adding/deleting nonfunctional code, reordering code, and varying memory layout. They demonstrated the concept through a compiler that randomized the amount of memory allocated on a stack frame, and showed that the approach foiled a simple buffer overflow attack [Forrest1997]. Again, they do not attempt to counter corrupted compilers.

John Knight and Nancy Leveson performed an experiment with "N-version programming" and showed that, in their experiment, "the assumption of independence of errors that is fundamental to some analyses of N-version programming does not hold" [Knight1986] [Knight1990]. As will be explained in section 4.7, this result does not invalidate DDC.

## 2.6 Subversion of software is a real problem

Subversion of software is not just a theoretical possibility; it is a current problem. One book on computer crime lists various kinds of software subversion as attack methods (e.g., trap doors, Trojan horses, viruses, worms, salamis, and logic bombs) [Icove1995, 57-58]. CERT[5] has published a set of case studies of "persons who used programming techniques to commit malicious acts against their organizations" [Cappelli2008]. Examples of specific software subversion or subversion attempts include:

- Michael Lauffenburger inserted a logic bomb into a program at defense contractor General Dynamics, his employer. The bomb would have deleted vital rocket project data in 1991, including much that was unrecoverable, but another employee stumbled onto it before it was triggered [AP1991] [Hoffman1991].
- Timothy Lloyd planted a 6-line logic bomb into the systems of Omega Engineering, his employer, that went off on July 31, 1996. This erased all of the company's contracts and proprietary software used by their manufacturing tools, resulting in an estimated $12 million in damages, 80 people permanently losing their jobs, and the loss of their competitive edge in the electronics market space. Plant manager Jim Ferguson stated flatly, "We will never recover". On February 26, 2002, a judge sentenced Lloyd to 41 months in prison, three years of probation, and ordered him to pay more than $2 million in damages to Omega [Ulsh2000] [Gardian].

---
[5]CERT is not an acronym.

- Roger Duronio worked at UBS PaineWebber's offices in Weehawken, N.J., and was with the company for two years as a system administrator. Apparently dissatisfied with his pay, he installed a logic bomb to detonate on March 4, 2002, and resigned from the company. When the logic bomb went off, it caused over 1,000 of their 1,500 networked computers to begin deleting files. This cost UBS PaineWebber more than $3 million to assess and repair the damage, plus an undetermined amount from lost business. Duronio was sentenced to 97 months in federal prison (the maximum per the U.S. sentencing guidelines), and ordered to make $3.1 million in restitution [DoJ2006] [Gaudin2006b]. The attack was only a few lines of C code, which examined the time to see if it was the detonation time, and then (if so) executed a shell command to erase everything [Gaudin2006a].

- An unnamed developer inside Borland inserted a back door into the Borland/Inprise Interbase Structured Query Language (SQL) database server around 1994. This was a "superuser" account ("politically") with a known password ("correct"), which could not be "changed using normal operational commands, nor [deleted] from existing vulnerable servers". Versions released to the public from 1994 through 2001 included this back door. Originally Interbase was a proprietary program sold by Borland/Inprise. However, it was released as open source software[6] in July 2000, and less than six months later the open source software developers discovered the vulnerability [Havrilla2001a] [Havrilla2001b]. The Firebird project, an alternate open source software package based on the same Interbase code, was also affected. Jim Starkey, who launched InterBase but left in 1991 before the back door was added to the software in 1994, stated that he believed that this back door was not malicious, but simply added to enable one part of the database software to communicate with another part [Shankland2001]. However, this code had the hallmarks of many malicious back doors: It added a special account that was (1) undocumented, (2) cannot be changed, and (3) gave complete control to the requester.

---

[6]Open source software is, briefly, software where users have the right to use the software for any purpose, review it, modify it, and redistribute it (modified or not) without requiring royalty payments [Wheeler2007]. The Open Source Definition [OSI2006] and the Free Software Definition [FSF2009] have more formal definitions for this term or the related term "Free software". There is quantitative data showing that, in many cases, using open source software/Free software (abbreviated as OSS/FS, FLOSS, or FOSS) is a reasonable or even superior approach to using their proprietary competition according to various measures [Wheeler2007]. In almost all cases, it is commercial software [Wheeler2009f].

- An unknown attacker attempted to insert a malicious back door in the Linux kernel in 2003. The two new lines were crafted to *appear* legitimate, by using an "=" where a "==" would be expected. The configuration management tools immediately identified a discrepancy, and examination of the changes by the Linux developers quickly determined that it was an attempted attack [Miller2003] [Andrews2003].

More recently, in 2009 the Win32.Induc virus was discovered in the wild. This virus attacks Delphi compiler installations, modifying the compiler itself. Once the compiler is infected, all programs compiled by that compiler will be infected [Mills2009] [Feng2009]. Thus, countering subverted compilers is no longer an academic exercise; attacks on compilers have already occurred.

Many have noted insertion of malicious code into software as an important risk:

- Many have noted subversion of software as an issue in electronic voting machines [Saltman1988] [Kohno2004] [Feldman2006] [Barr2007].
- The U.S. Department of Defense (DoD) established a "software assurance initiative" in 2003 to examine software assurance issues in defense software, including how to counter intentionally inserted malicious code [Komaroff2005]. In 2004, the U.S. General Accounting Office (GAO) criticized the DoD, claiming that the DoD "policies do not fully address the risk of using foreign suppliers to develop weapon system software...

policies [fail to focus] on insider threats, such as the insertion of malicious code by software developers..." [GAO2004]. The U.S. Committee on National Security Systems (CNSS) defines Software Assurance (SwA) as "the level of confidence that software is free from vulnerabilities, either intentionally designed into the software or accidentally inserted at anytime during its lifecycle, and that the software functions in the intended manner" [CNSS2006]. Note that intentionally-created vulnerabilities inserting during software development are specifically included in this definition.

- The President's Information Technology Advisory Committee (PITAC) found that "Vulnerabilities in software that are introduced by mistake or poor practices are a serious problem today. In the future, the Nation may face an even more challenging problem as adversaries – both foreign and domestic – become increasingly sophisticated in their ability to insert malicious code into critical software" [PITAC2005, 9]. The U.S. National Strategy to Secure Cyberspace reported that a "spectrum of malicious actors can and do conduct attacks against our critical information infrastructures. Of primary concern is the threat of organized cyber attacks capable of causing debilitating disruption to our Nation's critical infrastructures, economy, or national security.... [and could subvert] our infrastructure with back doors and other means of access." [PCIB2003,6]
- In 2003, China's State Council announced a plan requiring all government ministries to buy only locally produced software when upgrading, and to increase use of open source software, in part due to concerns over "data spyholes installed by foreign powers" in software they procured for government use [CNETAsia2003].

In short, as software becomes more pervasive, subversion of it becomes ever more tempting to powerful individuals and institutions. Attackers can even buy legitimate software companies, or build them up, to widely disseminate quality products at a low price... but with "a ticking time bomb inside" [Schwartau1994, 304-305].

Not all articles about subversion specifically note the trusting trust attack as an issue, but as noted earlier, for source code evaluations to be strongly credible, there must be a way to justify that the source code being examined accurately represents what is being executed—yet the trusting trust attack subverts that very claim. Internet Security System's David Maynor argues that the risk of attacks on compilation processes is increasing [Maynor2004] [Maynor2005]; Karger and Schell noted that the trusting trust attack was still a problem in 2000 [Karger2000], and some technologists doubt that computer-based systems can ever be secure because of the existence of this attack [Gauis2000]. Anderson et al. argue that the general risk of subversion is increasing [Anderson2004]. Williams argues that the risk from malicious developers should be taken seriously, and describes a variety of techniques that malicious programmers can use to insert and hide attacks in an enterprise Java application [Williams2009].

## 2.7 Previous DDC paper

Initial results from DDC research were published by the Annual Computer Security Applications Conference (ACSAC) in [Wheeler2005]. This paper was well-received, for example, Bruce Schneier wrote a glowing review and summary of the paper [Schneier2006], and the Spring 2006 class "Secure Software Engineering Seminar" of Dr. James Walden (Northern Kentucky University) included it in its required reading list.

This dissertation includes the results of [Wheeler2005] and refines it further:
- The definition of DDC is generalized to cover the case where the compiler is not self-regenerating. Instead, a compiler-under-test may have been generated using a different

"parent" compiler. Self-regeneration (where the putative source code of the parent and compiler-under-test are the same) is now a special case.

- A formal proof of DDC is provided, including a formalization of DDC assumptions. The earlier paper includes only an informal justification. The proof covers cases where the environments are different, including the effect of different text representation systems.

- A demonstration of DDC with a known maliciously corrupted compiler is shown. As expected, DDC detects this case.

- A demonstration of DDC with an industrial-strength compiler (GCC) is shown.

- The discussion on the application of DDC is extended to cover additional challenges, including its potential application to hardware.

# 3 Description of threat

Thompson describes how to perform the trusting trust attack, but there are some important characteristics of the attack that are not immediately obvious from his presentation. This chapter examines the threat in more detail and introduces terminology to describe the threat. This terminology will be used later to explain how the threat is countered. For a more detailed model of this threat, see [Goerigk2000] and [Goerigk2002] which provide a formal model of the trusting trust attack.

The following sections describe what might motivate an attacker to actually perform such an attack, and the mechanisms an attacker uses that make this attack work (triggers, payloads, and non-discovery).

## 3.1 Attacker motivation

Understanding any potential threat involves determining the benefits to an attacker of an attack, and comparing them to the attacker's risks, costs, and difficulties. Although this trusting trust attack may seem exotic, its large benefits may outweigh its costs to some attackers.

The potential benefits are immense to a malicious attacker. A successful attacker can completely control all systems that are compiled by that executable and that executable's descendants, e.g., they can have a known login (e.g., a "backdoor password") to gain unlimited privileges on entire classes of systems. Since detailed source code reviews will not find the attack, even defenders who have highly valuable resources and check all source code are vulnerable to this attack.

For a widely-used compiler, or one used to compile a widely-used program or operating system, this attack could result in global control. Control over banking systems, financial markets, militaries, or governments could be gained with a single attack. An attacker could possibly acquire enormous funds (by manipulating the entire financial system), acquire or change extremely sensitive information, or disable a nation's critical infrastructure on command.

An attacker can perform the attack against multiple compilers as well. Once control is gained over all systems that use one compiler, trust relationships and network interconnections could be exploited to ease attacks against other compiler executables. This would be especially true of a patient and careful attacker; once a compiler is subverted, it is likely to stay subverted for a long time, giving an attacker time to use it to launch further attacks.

An attacker (either an individual or an organization) who subverted a few of the most widely used compilers of the most widely-used operating systems could effectively control, directly or indirectly, almost every computer in existence.

The attack requires knowledge about compilers, effort to create the attack, and access (gained somehow) to the compiler executable, but all are achievable. Compiler construction techniques are standard Computer Science course material. The attack requires the insertion of relatively small amounts of code, so the attack can be developed by a single knowledgeable person. Access rights to change the relevant compiler executables are usually harder to acquire, but there are clearly some who have such privileges already, and a determined attacker may be able to acquire such privileges through a variety of means (including network attack, social engineering, physical attack, bribery, and betrayal).

The amount of power this attack offers is great, so it is easy to imagine a single person deciding to perform this attack for their own ends. Individuals entrusted with compiler development might succumb to the temptation if they believed they could not be caught. Today there are many virus writers, showing that many people are willing to write malicious code even without gaining the control this attack can provide.

It is true that there are *other* devastating attacks that an attacker could perform in the current environment. Many users routinely download and install massive executables, including large patches and updates, that could include malicious code, and few users routinely examine executable machine code or byte code. Few users examine source code even when they *can* receive it, and in many cases users are not legally allowed to examine the source code. As a result, here are some other potentially-devastating attacks that could be performed besides the trusting trust attack:

- An attacker can find unintentional vulnerabilities in existing executables, and then write code to exploit them.
- An attacker could modify or replace a widely-used/important executable during or after its compilation, but before its release by its supplier. For example, an attacker might be able to do this by bribing or extorting a key person in the supplying organization, by becoming a key person, or by subverting the supplier's infrastructure.
- Even when users only accept source code and compile the source code themselves, an attacker could insert an intentional attack in the source code of a widely-used/important program in the hope that no one will find it later.

- An attacker with a long-range plan could develop a useful program specifically so that they can embed or eventually embed an attack (using the two attacks previously noted). In such cases the attacker might become a trusted (but not trustworthy) supplier.

However, there is a *fundamental difference* with the attacks listed above and the trusting trust attack: there are *known detection techniques* for these attacks:

- Static and dynamic analysis can detect many unintentional vulnerabilities, because they tend to be caused by common implementation mistakes. In addition, software designs can reduce the damage from such mistakes, and some implementation languages can completely eliminate certain kinds of mistakes. Many documents discuss how to develop secure software for those trying to do so, including [Wheeler2003s] and [NDIA2008].
- If an attacker swaps the expected executable with a malicious executable, without using a trusting trust attack, the attack can be discovered by recompiling the source code to see if it produces the same results (presuming a deterministic compiler is used). Even if it is not discovered, recompilation of the next version of the executable will often eliminate the attack if it is not a "trusting trust" attack.
- If an attacker inserts an intentional attack or vulnerability in the source code, this can be revealed by examining the source code (see section 8.11 for a discussion on attacks which are intentionally difficult to find in source code).
- If the user does not fully trust the supplier to perform such tests, then these tests could be performed by the user (if the user has the necessary information), or by a third party who is trusted by the user and supplier (if the supplier is unwilling to give necessary information to the user, but are willing to give it to such a third party). If the supplier is unwilling to provide the necessary information to either the user or a third party, the user could reasonably conclude that using such suppliers is a higher risk than using suppliers who *are* willing to provide this information, and then take steps based on that conclusion.

In contrast, there has been *no* known effective detection technique for the trusting trust attack. Thus, even if all of these well-known detection techniques were used, users would *still* be vulnerable to the trusting trust attack. What is more, the subversion can persist indefinitely; the longer it remains undetected, the more difficult it will be to reliably identify the perpetrator even if it *is* detected.

Given such extraordinarily large benefits to an attacker, and the lack of an effective detection mechanism, a highly resourced organization (such as a government) might decide to undertake it. Such an organization could supply hundreds of experts, working together full-time to deploy attacks over a period of decades. Defending against this scale of attack is far beyond the defensive abilities of most companies and non-profit organizations who develop and maintain popular compilers.

In short, this is an attack that can yield complete control over a vast number of systems, even those systems whose defenders perform independent source code analysis (e.g., those who have especially high-value assets), so it is worth defending against.

## 3.2 Triggers, payloads, and non-discovery

The trusting trust attack depends on three things: triggers, payloads, and non-discovery. For purposes of this dissertation, a "trigger" is a condition determined by an attacker in which a malicious event is to occur (e.g., when malicious code is to be inserted into a program). A "payload" is the code that actually performs the malicious event (e.g., the inserted malicious code and the code that causes its insertion). The attack also depends on non-discovery by its victims, that is, it depends on victims not detecting the attack (before, during, or after it has been triggered)[7].

For this attack to be valuable, there must be at least two triggers that can occur during compilation: at least one to cause a malicious attack directly of value to the attacker (e.g., detecting compilation of a "login" program so that a Trojan horse can be inserted into it), and one to propagate attacks into future versions of the compiler executable.

If a trigger is activated when the attacker does not intend the trigger to be activated, the probability of detection increases. However, if a trigger is not activated when the attacker intends it to be activated, then that particular attack will be disabled. If all the attacks by the compiler against itself are disabled, then the attack will no longer propagate; once the compiler is recompiled, the attacks will disappear. Similarly, if a payload requires a situation that (through the process of change) disappears, then the payload will no longer be effective (and its failure may reveal the attack).

In this dissertation, "fragility" is the susceptibility of the trusting trust attack to failure, i.e., that a trigger will activate when the attacker did not wish it to (risking a revelation of the attack), fail to trigger when the attacker would wish it to, or that the payload will fail to work as intended by the attacker. Fragility is unfortunately less helpful to the defender than it might first appear. An attacker can counter fragility by simply incorporating many narrowly-defined triggers and payloads. Even if a change causes one trigger to fail, another trigger may still fire. By using multiple triggers and payloads, an attacker can attack multiple points in the compiler and attack different subsystems as final targets (e.g., the login system, the networking interface, and so on). Thus, even if some attacks fail over time, there may be enough vulnerabilities in the resulting system to allow attackers to re-enter and re-insert new triggers and payloads into a malicious compiler. Even if a compiler misbehaves from malfunctioning malware, the results could appear to be a mysterious compiler defect; if programmers "code around" the problem, the attack will stay undetected.

---

[7]Even if the attack is eventually detected, if the attacker can be assured that the attack will not be detected for a very long time, the attacker may still find it valuable. The attacker could, for example, use this lengthy time to successfully perform other attacks and subvert an infrastructure in many other ways. Also, if the original attack is not detected for a long time, it is often increasingly difficult to determine the identity of the attacker or at least an important intermediary. For a summary of techniques that can resolve this "attribution" problem, see [Wheeler2003t].

Since attackers do not want their malicious code to be discovered, they may limit the number of triggers/payloads they insert and the number of attacked compilers. In particular, attackers may tend to attack only "important" compilers (e.g., compilers that are widely-used or used for high-asset projects), since each compiler they attack (initially or to add new triggers and payloads) increases the risk of discovery. However, since these attacks can allow an attacker to deeply penetrate systems generated with the compiler, maliciously corrupted compilers make it easier for an attacker to re-enter a previously penetrated development environment to refresh an executable with new triggers and payloads. Thus, once a compiler has been subverted, it may be difficult to undo the damage without a process for ensuring that there are no attacks left.

The text above might give the impression that only the compiler itself, as usually interpreted, can influence results (or how they are run), yet this is obviously not true. Assemblers and loaders are excellent places to place a trigger (the popular GCC C compiler actually generates assembly language as text and then invokes an assembler). An attacker could place the trigger mechanism in the compiler's supporting infrastructure such as the operating system kernel, libraries, or privileged programs.

# 4 Informal description of Diverse Double-Compiling (DDC)

The idea of diverse double-compiling (DDC) was first created and posted by Henry Spencer in 1998 [Spencer1998] in a very short posting. It was inspired by McKeeman et al's exercise for detecting compiler defects [McKeeman1970] [Spencer2005]. Since this time, this idea has been posted in several places, typically with very short descriptions [Mohring2004] [Libra2004] [Buck2004]. This chapter describes the graphical notation for describing DDC that is used in this dissertation. This is followed by a brief informal description of DDC, an informal discussion of its assumptions, a clarification that DDC does *not* require that arbitrary *different* compilers produce the same executable output given the same input, and a discussion of a common special case: Self-parenting compilers. This chapter closes by answering some questions, including: Why not *always* use the trusted compiler, and why is this different from N-version programming?

## 4.1 Terminology and notation

This dissertation focuses on compilers. For purposes of this dissertation, compilers execute in some environment, receiving as input *source code* as well as other input from the environment, and producing a result termed an *executable*. A compiler is, itself, an executable.

Figure 1 illustrates the notation used in this dissertation. A shaded box shows a compilation step, which executes a compiler (input from the top), processing source code (input from the left), and uses other input (input from the right), all to produce an executable (output exiting down). To distinguish the different steps, each compilation step will be given a unique name (shown here as "n"). Source code that is purported to be the source code for the executable Y is notated as $s_Y$. The result of a compilation step using compiler X, source code $s_Y$, other input I (e.g., run-time libraries, random number results, and thread schedule), and environment E is an executable, notated here as compile($s_Y$, $c_X$, I, E). Where the environment can be determined from context (e.g., it is all the same) that parameter is omitted; where that is true and any other input (if relevant) can be inferred, both are omitted yielding the notation compile($s_Y$, $c_X$). In some cases, this will be further abbreviated as c($s_Y$, $c_X$).

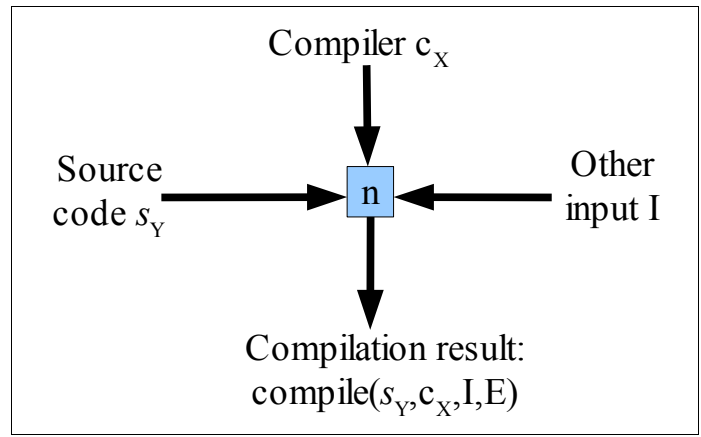

*Figure 1: Illustration of graphical notation*

The widely-used "T-diagram" (aka "Bratman") notation is not used in this dissertation. T-diagrams were originally created by Bratman [Bratman1961], and later greatly extended and formalized by Earley and Sturgis [Earley1970]. T-diagrams can be very helpful when discussing certain kinds of bootstrapping approaches. However, they are not a universally perfect notation, and this dissertation intentionally uses a different notation because the weaknesses of T-diagrams make DDC unnecessarily difficult to describe:

- T-diagrams combining multiple compilation steps can be very confusing [Mogensen2007, 219]. This is a serious problem when representing DDC, since DDC is fundamentally about multiple compilation steps.
- T-diagrams quickly grow in width when multiple steps are involved; since paper is usually taller than it is wide, this can make complex situations more difficult to represent on the printed page. Again, applying DDC involves multiple steps.
- T-diagrams do not handle multiple sub-components well (e.g., a library embedded in a compiler). The notation can be "fudged" to do this (see [Early1970, 609]) but the resulting graphic is excessively complex. Again, compilation of real compilers using DDC often involves handling multiple sub-components, making this weakness more important.
- T-diagrams create unnecessary clutter when applied to DDC. In a T-diagram, every compiler source code and compiler executable, as well as their executions, are represented by a T. This creates unnecessary visual clutter, making it difficult to see what is executed and what is not.

Niklaus Wirth abandoned T-diagrams in his 1996 book on compilers, without even mentioning them [Wirth1996], so clearly T-diagrams are not absolutely required when discussing compiler bootstrapping. The notation of this dissertation uses a single, simple box for each execution of a compiler, instead of a trio of T-shaped figures. As DDC application becomes complex, this simplification matters.

## 4.2 Informal description of DDC

In brief, to perform DDC, source code must be compiled twice. First, use a separate "trusted" compiler to compile the source code of the "parent" of the compiler-under-test. Then, run that resulting executable to compile the purported source code of the compiler-under-test. Then, check if the final result is *exactly* identical to the original compiler executable (e.g., bit-for-bit equality) using some trusted means. If it is, then the purported source code and executable of the compiler-under-test correspond, given some assumptions to be discussed later.

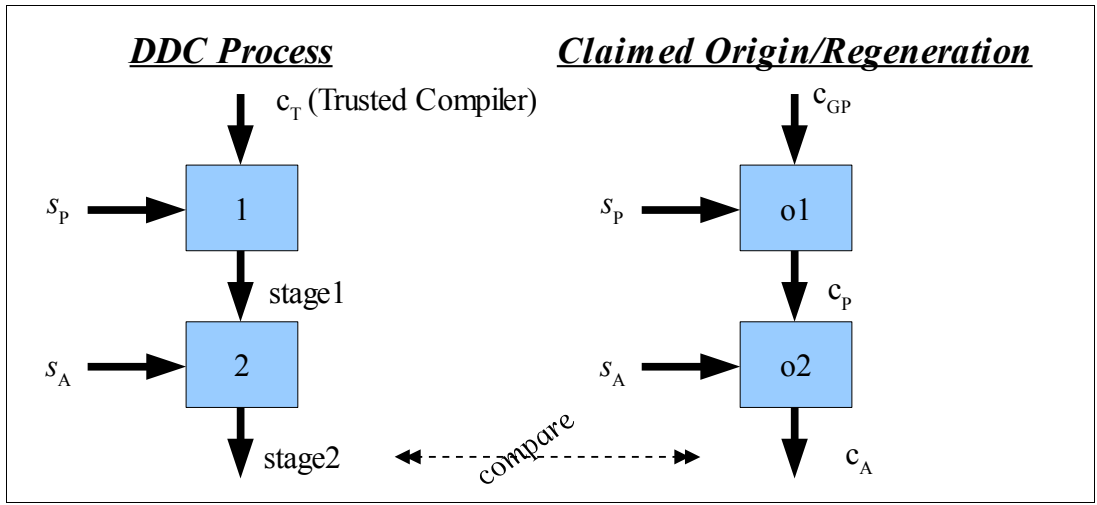

*Figure 2: Informal graphical representation of DDC*

Figure 2 presents an informal, simplified graphical representation of DDC, along with the claimed origin of the compiler-under-test (this claimed original process can be re-executed as a check for self-regeneration). The dashed line labeled "compare" is a comparison for exact equality. This figure uses the following symbols:

- $c_A$: Executable of the compiler-under-test, which may be corrupt (maliciously corrupted compilers are by definition corrupt).
- $s_A$: Purported source code of compiler $c_A$. Our goal is determine if $c_A$ and $s_A$ correspond.
- $c_P$: Executable of the compiler that is purported to have generated $c_A$ (it is the purported "parent" of $c_A$).

- $s_P$: Purported source code of parent $c_P$. Often a variant/older version of $s_A$.

- $c_T$: Executable of a "trusted" compiler, which must be able to compile $s_P$. The exact meaning of "trusted" will be explained later.

- 1, 2, o1, o2: Stage identifiers. Each stage executes a compiler.

- stage1, stage2: The outputs of the DDC stages. Stage1 is a function of $c_T$ and $s_P$, and can be represented as $c(s_P, c_T)$ where "c" means "compile". Similarly, stage2 can be represented as $c(s_A, \text{stage1})$ or $c(s_A, c(s_P, c_T))$.

The right-hand-side shows the process that purportedly generated the compiler-under-test executable $c_A$ in the first place. The right-hand-side shows the DDC process. The process graphs are very similar, so it should not be surprising that the results should be identical. This dissertation formally proves this (given certain conditions) and demonstrates that this actually occurs with real-world compilers.

Before performing DDC itself, it is wise to perform a regeneration check, which checks to see if we can regenerate $c_A$ using exactly the same process that was supposedly used to create it originally[8]. Since $c_A$ was supposed to have been created this way in the first place, regeneration should produce the same result. In practice, the author has found that this is often not the case. For example, many organizations' configuration control systems do not record all the information necessary to accurately regenerate a compiled executable, and the ability to perform regeneration is necessary for the DDC process. In such cases, regeneration acts like the control of an experiment; it detects when we do not have proper control over all the relevant inputs or environment. Corrupted compilers can also pass the regeneration test, so by itself the regeneration test is not sufficient to reliably detect corrupted compilers.

---

[8]DDC will not create an identical executable unless the regeneration check would succeed, and so from that perspective the regeneration check is mandatory. *Performing* the regeneration check has not been made mandatory, because there may be other evidence that it would succeed, but in most cases it is strongly recommended.

We then perform DDC by compiling twice. These two compilation steps are the origin of this technique's name: we compile twice, the first time using a different (diverse) trusted compiler. All compilation stages (stage 1 and stage 2, as well as the regeneration test) could be performed on the same or on different environments. Libraries can be handled in DDC by considering them as part of the compiler (if they are executed in that stage) or part of the source code (if they are used as input data but not executed in that stage).

Note that the DDC technique uses a separate trusted compiler as a check on the compiler-under-test. The trusting trust attack assumes that all later generations of the compiler will be descendants of a corrupted compiler; using a completely different second compiler can invalidate this assumption. The trusted compiler and its environment may be malicious, as long as that does not impact their result during DDC, and they may be very slow.

The formalized DDC model, along with formalized assumptions and its proof, are presented in chapter 5.

## 4.3 Informal assumptions

All approaches have assumptions. These will be formally and completely stated later, but a brief statement of some key assumptions should help in understanding the approach:

- DDC must be performed only by trusted programs and processes, including a trusted compiler $c_T$, trusted environment(s) to run DDC, a trusted comparer, and trusted processes and tools to acquire the compiler-under-test $c_A$ and the source code $s_P$ and $s_A$.

In this dissertation, something is "trusted" if we have *justified confidence* that it does not have triggers and payloads that would affect the results of DDC. A trusted program or process may have triggers and payloads, as long as they do not affect the result. A trusted program or process may have defects, though as shown later, any defects that affect its result in DDC are likely to be detected. Methods to increase the level of confidence are discussed in chapter 6.

- Compiler $c_T$ must have the same semantics for the same constructs as required by $s_P$. For example, a Java$^{(TM)}$ compiler cannot be used directly as $c_T$ if $s_P$ is written in the C language. If $s_P$ uses any nonstandard language extensions, or depends on a construct not defined by a published language specification, then $c_T$ must implement them in the way required by $s_P$. Any defect in $c_T$ can also cause problems if it affects compiling $s_P$ (otherwise it is irrelevant for DDC).

- The compiler defined by $s_P$ should be deterministic given its inputs. That is, once compiled, and then executed multiple times given the same inputs, it should produce exactly the same outputs each time. If the compiler described by $s_P$ is non-deterministic, in some cases it could be handled by running the process multiple times, but it is often easier to control enough inputs to make the compiler deterministic. Note that the regeneration process is helpful in detecting undesired non-determinism.

DDC does not determine if the source code is free of malicious code; DDC can only show if source code corresponds to a given executable. If the goal is to show that the compiler $c_A$ is not malicious, then the source code ($s_A$ and $s_P$) must also be reviewed to determine that the source code is not malicious. This is still an important change—it is typically far easier to review source code than to review executables. In some cases $s_A$ and $s_P$ are extremely similar; in such cases they can be simultaneously reviewed by reviewing one and then reviewing their differences. There is also an important special case—when $s_P = s_A$—that is described in section 4.5.

But first, we must clarify that DDC does *not* require something that is unlikely.

## 4.4 DDC does not require that different compilers produce identical executables

DDC does *not* require that arbitrary *different* compilers produce the same executable output, even given the same input. Indeed, this would be extremely unlikely for source code the size of typical compilers. Compiler executables $c_A$, $c_P$, and $c_T$ might even run on or generate code for different CPU architectures, making identical results extremely unlikely.

Instead, DDC runs a different executable; under certain conditions, this must produce the "same" result. This is perhaps best explained by example. Imagine two properly-working C compilers, both of which are given this source code to print the result of calculating 2+2:

```
#include <stdio.h>
main() {
   printf("%d\n", 2+2);
}
```

The executables produced by the two compilers are almost certainly different, but *running* these two programs on their respective environments must produce the same result for this line (once converted into the same text encoding format). Obviously, this depends on them implementing the same language (for the purposes of the given Source).

The conditions where this occurs are defined more formally in chapter 5. In particular, see section 5.7.9, where this is examined in more detail.

## 4.5 Special case: Self-parenting compiler

An important special case is when $s_P=s_A$, that is, when the putative source code of the parent compiler is the same as the putative source code of the compiler-under-test. There are often good reasons for releasing executables generated this way. For example, a compiler typically includes many optimization operations; each new version of a compiler may add new or improved optimization operations. By releasing a self-parented compiler (a compiler generated by setting $s_P=s_A$ and compiling twice), the supplier would release a compiler executable that uses the latest versions of those optimizations, giving the compiler itself maximum performance. Many existing compilers (including as GCC) use the compiler bootstrap test (essentially the self-regeneration check) to test themselves, so a compiler's build and test process may already include an automated way to create a self-parenting compiler. Figure 3 shows how figure 2 simplifies in this case.

Because this is a common case, the older paper [Wheeler2005] only considered this case. In contrast, this dissertation considers the more general case, subsuming self-parenting as a special case.

Having a self-parenting compiler can simplify the application of DDC. As discussed in more detail below, DDC only shows that source code and executable correspond, so review of compiler source code is still required if the goal is to show that there is no malicious code in an executable. In the general case, both $s_A$ and $s_P$ must be reviewed. Since $s_A=s_P$ in a self-parented compiler, reviewing both $s_A$ and $s_P$ can be done by reviewing just $s_A$, simplifying the use of DDC. Also, when a compiler is its own parent, a simplified regeneration check may be used to detect many problems without performing the complete regeneration test. This test, which can be termed

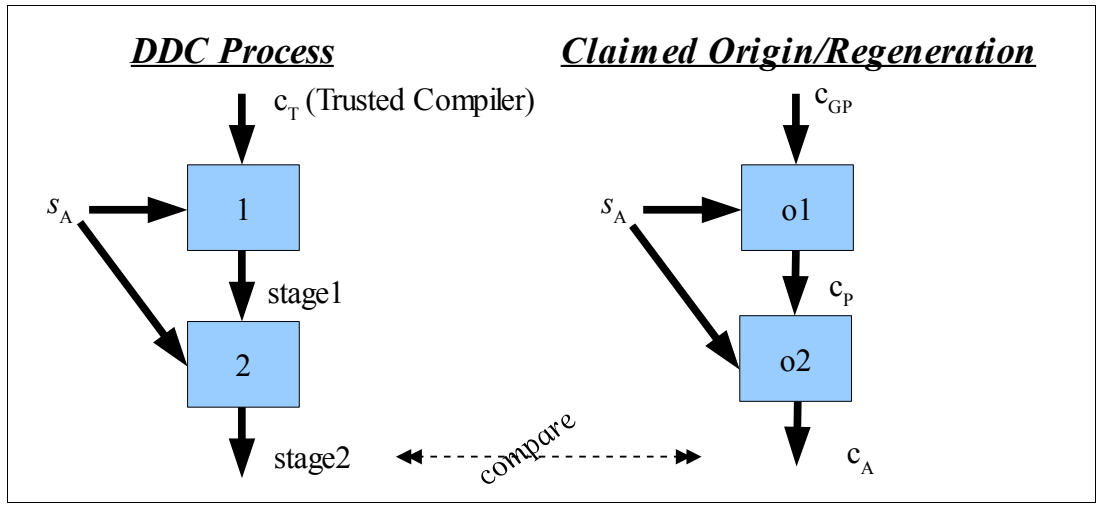

*Figure 3: Informal graphical representation of DDC for self-regeneration case*

"self-regeneration", simply uses $c_A$ to compile its putative source code $s_A$; the regeneration is successful if the generated executable is the same as the original $c_A$.

It is still useful to be able to handle the general case. Compiler $c_P$ need not be a radically different compiler; it might simply be an older version of $c_A$, differ only in its use of different compilation flags, or differ only in that it embeds a different version of a library executable. Nevertheless, if $c_P$ and $c_A$ are different, the general form of DDC must be used. Also, it is possible to have a "loop" of compilers that mutually depend on each other for self-regeneration (e.g., a Java compiler written in C and a C compiler written in Java might be generated using each other). In this case, the more general form of DDC is needed to break the loop.

## 4.6 Why not always use the trusted compiler?

DDC uses a second "trusted" compiler $c_T$, which is trusted in the sense that we have a justified confidence that $c_T$ does not have triggers or payloads that affect recompiling $s_P$ and $s_A$ (see section 4.3). We can now answer an obvious question: Why not *always* use the trusted compiler $c_T$?

First, there are many reasons compiler $c_T$ might not be suitable for general use. For example, compiler $c_T$ may be slow, produce slow code, generate code for a different CPU architecture than desired, be costly, or have undesirable software license restrictions. It may lack many useful functions necessary for general-purpose use (in DDC, trusted compiler $c_T$ only needs to be able to compile $s_P$). It is possible that the only purpose of the trusted compiler is to operate as a trusted checker for the more widely-used compiler, in fact, there are good reasons to do so. It is much easier to verify (and possibly formally prove) a simple compiler that has limited functionality and few optimizations; such compilers might not be suitable for general production use, but would be ideal as trusted compilers used to check production compilers. The trusted compiler could even be a "secret" compiler that is never publicly released (as source, executable, or a service); an attacker would find it extremely difficult to avoid detection by DDC if such a trusted compiler were used.

Second, using a different trusted compiler $c_T$ greatly increases the confidence that the compiler executable $c_A$ corresponds with source code $s_A$. When a second compiler $c_T$ is used as part of DDC, an attacker must subvert *multiple* executables and executable-generation processes to perform the "trusting trust" attack without detection. It is true that the trusted compiler $c_T$ could be used as a "trusted bootstrap" compiler that would always be used to generate each new version of $c_A$. This could be done even if $c_T$ is not suitable for general use. However, if we always generate updated versions of $c_A$ this way, and never use DDC, we have merely moved the trusting trust attack to a different location: We must now perfectly protect $c_T$ and the bootstrap process used to create each new version of $c_A$. Should the protection of $c_T$ ever fail, an attacker might change $c_T$ into a maliciously corrupted compiler $c_T'$, resulting in the potential corruption of future versions of $c_A$. By using DDC with a different trusted compiler $c_T$, $c_T$ is used as a separate check, requiring an attacker to subvert *two* different compilers and compiler-generation processes to avoid detection. Indeed, DDC could be performed multiple times using different compilers as $c_T$ and/or different environments, requiring an attacker to subvert *all* of the DDC processes to avoid detection. Using DDC with a different compiler $c_T$ greatly increases the confidence that $c_A$ exactly corresponds with $s_A$; using DDC multiple times can increase that confidence still further.

## 4.7 Why is DDC different from N-version programming?

N-version programming "has been proposed as a method of incorporating fault tolerance into software. Multiple versions of a program (i.e., 'N') are prepared and executed in parallel. Their outputs are collected and examined by a voter, and, if they are not identical, it is assumed that the majority is correct. This method [assumes] that programs that have been developed independently will fail independently" [Knight1986].

John Knight and Nancy Leveson performed an experiment with N-version programming and showed that, in their experiment, "the assumption of independence of errors that is fundamental to some analyses of N-version programming does not hold" [Knight1986] [Knight1990]. Instead, they found that if one program has a failure when processing a particular input, there was an increased likelihood of failure (compared to random failure) for another program with the same input, given that both programs were written to the same specification. This is an important result. It is not hard to see why this might be true; for example, if certain areas of the specification are unusually complex, two different programmers might both fail to meet it.

However, this result does not invalidate DDC, because the circumstances in DDC are very different from this and similar experiments.

In the Knight and Leveson work, N different programs were developed by different developers attempting to implement the *same* specification. In contrast, the purpose of applying DDC is to detect when two different compiler executables have been developed to implement *different* specifications, that is, when one program is written to attempt to compile source code accurately, while another program is written to produce corrupted results in certain cases. However:

- These changes are *extremely unlikely* to happen *unintentionally* (and in the same way) in both the trusted compiler and the original process used to create the compiler-under-test. Creating a corrupting compiler that is self-perpetuating and selectively corrupts other programs requires clever programming [Thompson1984] and significantly changes the compiler executable (for an example, see the differences shown in section A.5).
- These changes are *extremely unlikely* to happen *intentionally* in the trusted compiler and DDC process in general. This is by definition of the term "trusted"—we have justified confidence that the DDC process (including the trusted compiler) is unlikely to have triggers or payloads that affect DDC results.
- Since the kind of differences that motivate DDC are extremely unlikely to occur unintentionally *or* intentionally, the entire scenario is extremely unlikely.

Also, in the Knight and Leveson experiment, the issue was to determine if the different programs would produce identical results across all permitted inputs to the different programs. Their experiment simulated use of the N programs using one million test inputs, corresponding to about twenty years of operational use "if the program is executed once per second and unusual events occur every ten minutes". In contrast, in DDC, there is only *one* relevant input: the source code pair $s_P$ and $s_A$. Granted, these inputs will have a complex internal structure, but these are the *only* inputs that matter, as compared to the wide range of possible inputs a compiler might accept. Thus, in DDC we do *not* have the situation where there is a wide variety of potential test inputs; we have only one pair of inputs, and they are the only ones that matter.

There is a special case where the Knight and Leveson results *do* directly apply to DDC. This is when the original compiler and trusted compiler *both* fail to correctly compile the source code ($s_P$ and $s_A$), *and* this failure happens to produce the same results. DDC will not detect that both compilers are performing incorrectly in the same way. The Knight and Leveson paper shows that such program failures are not completely statistically independent, and thus this kind of failure is somewhat more likely than an independence model would predict. However, there are several reasons to believe that this special case is rare for mature compilers. First, mature compilers typically pass a large test suite, reducing the risk of such defects. Second, compilers are usually part of their own test suite, reducing the likelihood that a compiler will fail to correctly compile itself. Third, section 7.1.3 demonstrates that even when a compiler fails to correctly compile itself, DDC may still detect it. But all of this is beside the point. Since the purpose of applying DDC is to detect intentional self-perpetuating attacks, and not to prove total correctness, this special case does not invalidate the use of DDC to detect and counter the "trusting trust" attack.

Thus, the Knight and Leveson results do not invalidate DDC for the purpose of detecting and countering the "trusting trust" attack.

## 4.8 DDC works with randomly-corrupting compilers

DDC works even if an ancestor of $c_A$ randomly corrupts its results. If the compiler-under-test was not corrupted, DDC will correctly report this; otherwise, DDC will expose the corruption.

# 5 Formal proof

This chapter presents a formal proof of DDC. The first section presents a more complete graphical model of both the DDC process and how the compiler-under-test is claimed to have been created. This is followed by a description of the formal notation used (first-order logic (FOL) with equality), the rationales used in proof steps (aka the derivation rules or rules of inference), the tools used, and various proof conventions. After this, the three key proofs are presented. Each proof presents a set of predicates, functions, and assumptions about DDC in the formal notation, and shows how they lead to the concluding proof goal. The three proofs are:

- Proof #1, goal source_corresponds_to_executable: This is the key proof for DDC. It shows that given certain assumptions, if stage2 (the result of the DDC process) and $c_A$ (the original compiler-under-test) are equal, then the executable $c_A$ and the source code $s_A$ exactly correspond.

- Proof #2, goal always_equal: This proves that, under "normal conditions" (such as when compiler executables have not been rigged and thus *do* correspond to their respective source code), $c_A$ and stage2 are in fact always equal. Thus, the first proof is actually useful, because its assumptions will often hold. This also implies that if $c_A$ and stage2 are *not* equal, then at least one of its assumptions is *not* true.

- Proof #3, goal cP_corresponds_to_sP: The previous "always_equal" proof does not require that a "grandparent" compiler exist, but having one is a common circumstance. This third proof shows that if there *is* a grandparent compiler, one of the assumptions of proof #2 can be proved given other assumptions that may be easier to verify (potentially making DDC even easier to apply in this common case).

## 5.1 Graphical model for formal proof

Figure 4 graphically represents the DDC stages and how the compiler-under-test $c_A$ was putatively created. This is a more rigorous version of figure 2; the formal model includes more detail to accurately model potentially-different compilation environments and the effects these environments have on the compilation processes.

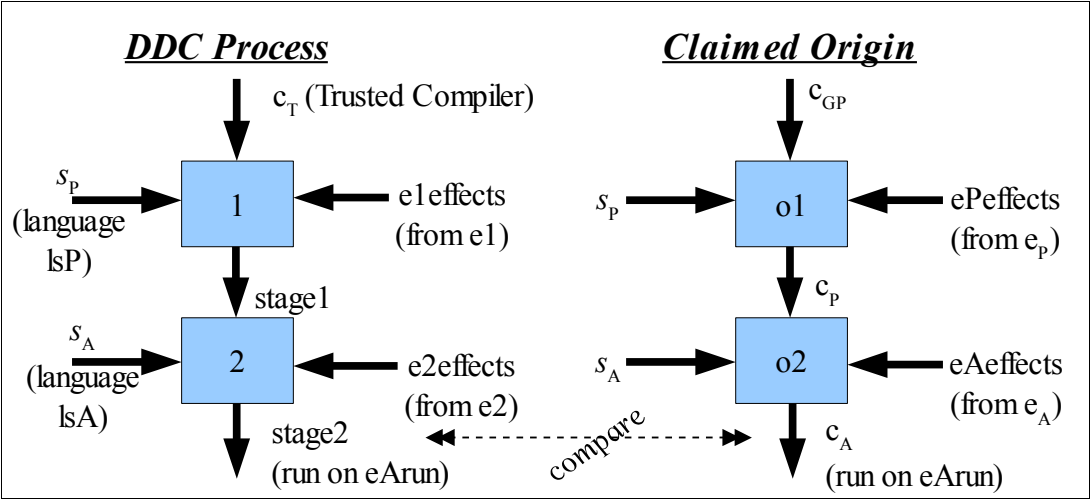

Figure 4: Graphical representation of DDC formal model

This dissertation argues that if the DDC process produces a "stage2" that is identical to the $c_A$, and certain other assumptions are true, then the executable stage2 corresponds to the source code $s_A$. The similarity of the DDC process and claimed origin figures suggest that this might be reasonable, but the challenge is to formalize exactly what those assumptions are, and then prove that this is true from those assumptions.

## 5.1.1 Types

Although types (sorts) are not directly used in the proof, it is easier to explain the graph and proofs by assigning types to the various constants used. There are four basic types:

- *Data:* For our purposes, data is information that is used as source code (input) and/or is the resulting executable (output) of a compilation. Some of the data could be both source and executable (e.g., a library object file could be executed during compilation and also copied into the final executable). Thus, as implied by its definition, data can be either (or both):

    - *Executable:* Data that can be executed by a computing environment. Compilers produce executables, and compilers themselves are executables.

    - *Source*: Data that can be compiled by a compiler to produce an executable. Any source (aka source code) is written in some language.

- *Environment:* A platform that can run executables. This would include the computer hardware (including the central processing unit) and any software that supports or could influence the compiler's result (e.g., the operating system). It could include a byte code interpreter or machine simulator.

- *Language*: The language, used by some source, that defines the meaning of the source.

- *Effects:* All information or execution timing arising from the environment that can affect the results of a compilation, but is not part of the input source code. This is used to model random number generators, thread execution ordering, differences between platforms allowed by the language, and so on. Note that this is not simply data in the usual sense, since other issues such as thread execution ordering are included as effects.

## 5.1.2 DDC components

The DDC process, as shown in figure 4, includes the following components, with the following types and meanings:

- $c_T$: Executable. The trusted compiler. It is trusted in the sense that it is trusted to not have triggers or payloads that will activate when compiling source $s_P$.
- $s_P$: Source. The (putative) source code of the "parent" compiler.
- $s_A$: Source. The (putative) source code of the compiler-under-test ($c_A$).
- e1: Environment. The environment that executes compilation step 1, which uses $c_T$ to compile $s_P$ and produce stage1.
- e2: Environment: The environment that executes compilation step 2, which uses stage1 to compile $s_A$ and produce stage2.
- eArun: Environment: The environment that stage2 is intended to run on.
- lsP, lsA: Language. The languages used by source $s_P$ and $s_A$, respectively.
- e1effects: Effects. The effects sent from environment e1 to compilation step 1.
- e2effects: Effects. The effects sent from environment e2 to compilation step 2.
- stage1: Executable. The result of DDC compilation step 1. This will be defined, using the functional notation below, as compile($s_P$, $c_T$, e1effects, e1, e2).
- stage2: Executable. The result of DDC compilation step 2. This will be defined as compile($s_A$, stage1, e2effects, e2, eArun).

Note that $s_A$ may be equal to $s_P$, e1 may be equal to e2 or eArun, e2 may be equal to eArun, and lsA may be equal to lsP. These identities are permitted but not required by DDC. All processes (including the compilations and their underlying environments, the process for acquiring $c_A$, and the process for comparing $c_A$ and stage2) must be trusted (i.e., they must not have triggers or payloads that affect their operation during DDC).

## 5.1.3 Claimed origin

The compiler-under-test $c_A$ was putatively developed by a similar process. This "claimed origin" process can also be modeled, with the following components not already described in the DDC process:

- $c_{GP}$: Executable. The grandparent compiler, if there is one.
- eP: Environment. The environment that executes compilation step o1, which uses $c_{GP}$ to compile source $s_P$ and produce executable $c_P$.
- eA: Environment: The environment that executes compilation step o2, which uses $c_P$ to compile $s_A$ and produce $c_A$.
- ePeffects: Effects. The effects sent from $e_P$ to compilation step o1.
- eAeffects: Effects. The effects sent from $e_A$ to compilation step o2.
- $c_P$: Executable. Putative parent compiler.
- $c_A$: Executable. The compiler-under-test, which putatively was developed by the process above.

Note that compiler-under-test $c_A$ may, in fact, be different than if it were really generated through this process. But if $c_A$ *was* generated through this process, we can prove that certain outcomes will result, given certain assumptions, as described below.

## 5.2 Formal notation: First-Order Logic (FOL)

The formal logic used in this dissertation is classical first-order logic (FOL) with equality, aka first-order predicate logic. FOL was selected because it is a widely understood and accepted formal logic system[9]. This dissertation uses the FOL notation and conventions defined in [Huth2004, 93-139]. In FOL, every expression is a *term* or a *formula*.

A *term* (which denotes an object) is defined as: a variable, a constant, or a function application of form $f(\tau_1, \tau_2, \ldots, \tau_n)$ where each of the zero or more comma-separated parameters is a term. In this dissertation, variables begin with an uppercase letter, while constants begin with a lowercase letter (this is the same convention used by Prolog).

A *formula* (which denotes a truth value) is defined as: $\neg \Phi$, $\Phi \wedge \Psi$, $\Phi \vee \Psi$, $\Phi \rightarrow \Psi$, $\forall \chi \Phi$, $\tau_1 = \tau_2$, $\tau_1 \neq \tau_2$, or a predicate of form $p(\tau_1, \tau_2, \ldots, \tau_n)$ where each of the one or more comma-separated parameters is a term. This definition requires that $\Phi$ and $\Psi$ are formulas, $\chi$ is an unbound variable, and anything beginning with $\tau$ is a term.

In some sense, a formula is a boolean expression that represents true or false, while a term represents any non-boolean type. Functions and predicates have the same syntax if they have any parameters. Table 1 shows the traditional FOL notation for FOL expressions (terms and formulas), an equivalent American Standard Code for Information Interchange (ASCII) representation, and a summary of its meaning[10]:

---

[9] For an "analysis and interpretation of the process that led to First-Order Logic and its consolidation as a core system of modern logic" see [Ferreirós2001]. An alternative to classical logic is intuitionist logic, which does not accept the equivalence of $\neg \neg \Phi$ and $\Phi$ as being universally true; [Hesseling2003] describes in detail the early history of intuitionist logic.

[10] As a notation, FOL does have weaknesses. For example, predicates and functions cannot have formulas (booleans) as parameters, so traditional FOL cannot express a function *if_then_else(formula1, term1, term2)* that returns term1 if formula1 is true, else it returns term2. FOL also does not include built-in support for types (sorts). There are extensions and alternatives which remove these weaknesses.

*Table 1: FOL notation*

| Traditional Notation | ASCII Representation | Meaning |
|---|---|---|
| $\neg \Phi$ | - PHI | *not* $\Phi$, aka negation. If $\Phi$ is true, $\neg \Phi$ is false; if $\Phi$ is false, $\neg \Phi$ is true. $\neg \neg \Phi$ is equivalent to $\Phi$. |
| $\Phi \wedge \Psi$ | PHI & PSI | $\Phi$ *and* $\Psi$, aka conjunction, aka "logical and". Both $\Phi$ and $\Psi$ must be true for the expression to be true. |
| $\Phi \vee \Psi$ | PHI \| PSI | $\Phi$ *or* $\Psi$, aka disjunction, aka "logical inclusive or". $\Phi$, $\Psi$, or both must be true for the expression to be true. |
| $\Phi \rightarrow \Psi$ | PHI -> PSI | $\Phi$ *implies* $\Psi$, aka implication, entailment, or "if $\Phi$, then $\Psi$". Equivalent to $(\neg \Phi) \vee \Psi$. |
| $\forall \chi \Phi$ | all Chi PHI | *For-all*, aka universal quantification. For all values of variable $\chi$, $\Phi$ is true. In this dissertation, this is optional; all unbound variables are universally quantified. |
| $\tau_1 = \tau_2$ | tau_1 = tau_2 | $\tau_1$ *equals* $\tau_2$. If true, $\tau_2$ can substitute for $\tau_1$. |
| $\tau_1 \neq \tau_2$ | tau_1 != tau_2 | $\tau_1$ *is not equal to* $\tau_2$. Equivalent to $\neg(\Phi = \Psi)$. |
| $x(\tau_1, \tau_2, ..., \tau_n)$ | x(tau_1, tau_2, ..., tau_n) | *Function or predicate x* with terms $\tau_1, \tau_2, ..., \tau_n$. A predicate is like a function that returns a boolean. |

Parentheses are used to indicate precedence. FOL also has a "there exists" notation (using $\exists$) which is not directly used in this dissertation. A formula is either true or false (this is the principle of the excluded middle); thus, $\Phi \vee \neg \Phi$ is true for any formula $\Phi$. In this dissertation, a top-level FOL formula is terminated by a terminating period (".").

For example, the following FOL formula could represent "all men are mortal":

```
man(X) -> mortal(X).
```

This formula can be read as "for all values of X, if X is a man, then X is mortal". Note that "X" is a variable, not a constant, because it begins with a capital letter. Also note that since X is not bound, an implied "all X ..." surrounds the entire formula.

---

However, since these FOL weaknesses do not interfere in the proof of DDC, and since traditional FOL is both widely-understood and widely-implemented, FOL is used in this dissertation.

In addition, the following formula could be used to represent "Socrates is a man":

```
man(socrates).
```

From these two formulas, it can be determined that "Socrates is mortal":

```
mortal(socrates).
```

FOL is a widely-used general notation, and not designed for proofs about specific fields (such as compilation). Thus, as with most uses of FOL, additional "non-logical" symbols must be added before particular problems can be analyzed. In this dissertation, these additions are the various constant terms in the graphical model described in 5.1 (above), as well as various predicates and functions that will be defined below. The proofs below will introduce these predicates and functions, as well as various assumptions, and then show that certain important conclusions (termed "goals") can be formally proved from them. Some assumptions define a term, predicate, or function; these assumptions are also called "definitions" in this dissertation.

All formal models, including the one in this dissertation, must include lowest-level items (such as predicates, functions, and constants) that are not defined in the formal model itself. Therefore, it is unreasonable to protest that these lowest-level items are not defined in this model, since that is necessarily true. The key is that the lowest-level items should accurately model the real world, thus forming a rational basis for proving something about the real world.

## 5.3 Proof step rationales (derivation rules or rules of inference)

Every step in each formal proof must have a rationale (aka a derivation rule or rule of inference). In this dissertation, only the following rationales are permitted in the formal proofs (for clarity, the terminating "." in top-level formulas is omitted in this list):

- *Assumption*: Given assumption. All definitions are assumptions.
- *Goal*: The given goal to be proved.

- ***Clausify***: Transform a previous step (formula) into a normalized clausal form. In particular, all expressions of the form $\Phi \rightarrow \Psi$ are transformed into $(\neg \Phi) \vee \Psi$. For example, using the example in section 5.2, "man(X) -> mortal(X)" can be transformed into "-man(X) | mortal(X)". See [McCune2008] and [Duffy1991] for a detailed description.

- ***Copy...flip***: Copy a previous result but reverse the order of an equality statement. Thus, given $\Phi = \Psi$, this rationale can produce $\Psi = \Phi$.

- ***Deny***: Negate a previous step; this processes the goal statement. All formal proofs in this chapter are proofs by contradiction; the goal is negated by the "Deny" rule, and the rest of the proof shows that this leads to a contradiction.

- ***Resolve***: Resolution (aka general resolution), that is, produce a resolvant from two clauses. Resolution is a generalized version of ground (propositional) resolution, so to explain resolution, we will first explain ground resolution.

    Ground resolution is a derivation rule that applies to clauses in propositional logic (a simpler logic than FOL that lacks terms, predicates, functions, quantification (for-all and there-exists), and equality; variables are true or false). Ground resolution requires two ground clauses (formulas) which can be reordered into the forms $\Lambda \vee \Phi$ and $\Lambda' \vee \Psi$, where $\Lambda'$ is a complement (negation) of formula $\Lambda$, and where $\Phi$, $\Psi$, or both may be empty. From that, ground resolution can derive $\Phi \vee \Psi$ removing any duplicates (this can be informally viewed as combining the two clauses with $\Lambda$ and $\Lambda'$ "canceling" each other). If both $\Phi$ and $\Psi$ are empty, the empty clause (false) is derived. For example, given both $P \vee Q$ and $\neg P \vee R$, ground resolution can derive $Q \vee R$. Ground resolution is a sound rule for reasoning because any formula $\Lambda$ must be either true or false: If $\Lambda$ is

false, and $\Lambda \vee \Phi$ is true, then $\Phi$ must be true. If $\Lambda$ is true, then $\Lambda'$ is false, and since $\Lambda' \vee \Psi$ is true, then $\Psi$ must be true. Since *either $\Phi$ or $\Psi$ must* be true, it follows that $\Phi \vee \Psi$ is *always* true. The traditional logic rule *modus ponens (*given $\Phi$ and $\Phi \rightarrow \Psi$, then $\Psi$) is a special case of ground resolution; $\Phi \rightarrow \Psi$ can be rewritten (using clausify) as $\neg \Phi \vee \Psi$, and ground resolution can combine $\Phi$ with $\neg \Phi \vee \Psi$ to derive $\Psi$.

The full resolution rule extends ground resolution so that it can handle quantifiers and predicates. It does this by using unification, the process of replacing the variables in the expressions with terms to make the modified expressions identical to each other. For details, see section 3.3 of [Duffy1991] or [Robinson2001].

For example, given "-man(X) | mortal(X)", we can substitute "X=socrates" yielding "-man(socrates) | mortal(socrates)"; this can then be combined with "man(socrates)" to prove "mortal(socrates)".

- *Para*: Paramodulation, a rule that adds support for the equality relation. This replaces an expression with another expression it is equal to, including any parameter substitutions. For example, given "f(d, e, X)" and "f(A, B, C)=g(C, B, A)", paramodulation can derive "g(X, e, d)". The precise definition of this rule is complex (e.g., it handles cases where the equality holds only under certain conditions); for details, see section 3.3.7 of [Duffy1991] or [Robinson2001].

These proof step rationales (aka derivation rules or rules of inference) were used because they are the rationales supported by the selected proof tools.

## 5.4 Tools and rationale for confidence in the proofs

### 5.4.1 Early DDC proof efforts

Early versions of these proofs were developed by hand. Unfortunately, it was very difficult to rigorously check or amend those hand-created proofs[11].

The tool named Prototype Verification System (PVS) was then used for some time, in part because it has a powerful notation that supports type-checking (which can eliminate some errors) and higher-order logic [Owre2001]. At the time, it was thought that higher-order logic would be especially helpful, since a compiler can be viewed as a computational function that produces a computational function. However, while PVS is very good at what it does, and several proofs were created using PVS, PVS required a large amount of manual effort to produce the proofs. These early proofs showed that higher-order logic was not necessary or especially helpful in modeling this particular problem, and that other logic systems and provers could be used instead. Many other tools have less powerful notations (e.g., first-order logic without types) but can better automate proof development.

### 5.4.2 Prover9, mace4, and ivy

The final proofs, as presented in this dissertation, were developed and checked with the assistance of several related tools: prover9, mace4, and ivy:

- Prover9 is an automated theorem prover for first-order and equational (classical) logic, which uses an ASCII representation of FOL. All of the proofs given in this chapter were developed by prover9 version Aug-2007.

---

[11] For example, the original hand-created proofs did not account for the possibility of different environments. When attempting to modify the proofs to account for the different environments, the painful "bookkeeping" required to keep the proof accurate soon led the author to look for an automated tool.

- Mace4 is a tool paired with prover9 that searches for finite structures satisfying first-order and equational statements (the same kind of statement that Prover9 accepts). From a logic point of view, mace4 produces interpretations which are models of the input formulas; from a mathematical point of view, mace4 produces structures satisfying the input formulas. Put simply, mace4 tries to find an assignment of integers $0..n-1$ (the "domain") to each constant term, to each function (given their possible inputs in the domain), and true/false values for each predicate that will satisfy the given set of statements. By default, mace4 starts searching for a structure of domain size 2, and then it increments the size until it succeeds or reaches some limit.
- Ivy is a separate proof checker that can accept and verify the proof as output by prover9. Ivy is written using A (sic) Computational Logic for Applicative Common Lisp (ACL2) and has itself been proven sound using ACL2 [McCune2000]. All of the prover9 proofs were verified by ivy. Indeed, one reason prover9 was chosen over some other tools was the availability of ivy.

Far more detail about prover9 is provided in [McCune2008]; its general approach (in particular, information on resolution and paramodulation) is discussed in detail in texts such as [Duffy1991] and [Robinson2001]. For purposes of this dissertation, prover9 is given a set of assumptions and a goal statement, using first-order logic (FOL) with equality. Prover9 negates the goal, transforms all assumptions and the goal into simpler clauses, and then attempts to find a proof by contradiction. Should prover9's search algorithm find a proof, it can print the sequence of steps and the rationale for each step that leads to the proof.

## 5.4.3 Tool limitations

Unlike PVS, traditional FOL and the prover9 tool (which implements FOL) do not directly support types (sorts). It is possible to implement types (sorts) using FOL: types of constants can be declared as assertions (e.g., "executable(cA)" could represent "$c_A$ is an executable"), assertions about compilers could be modified to state the types of compiler inputs and outputs, and the goal could be extended to include type requirements. However, because prover9 does not directly support type declaration, implementing types in prover9 makes the proofs far more complicated. These complications do not add value, because the types of compiler input and output are not in doubt (and thus do not need proof). In this dissertation types are only used as part of the comments to clarify the proof results, and are not directly expressed in the proof notation.

It should be noted that these tools did not make creating the proofs trivial. In particular, prover9 can only find a proof given a correct goal and assumptions. When prover9 cannot prove a goal, it either halts with a declaration that it cannot prove the result or it times out. In either case it is often difficult to determine *why* the proof cannot be found. The companion tool mace4 may be able to find a counter-example, but even then it is often not obvious what is wrong. In practice, the proofs were developed by first creating very simplified models of the world, and then expanding them stepwise to model additional complexities of the real world.

Prover9 will sometimes use information it does not need, leading to overly-complicated proofs. To counteract this, each proof was developed separately and includes only the statements necessary for the proof.

## 5.4.4 Proofs' conclusions follow from their assumptions

There are many reasons to have very high confidence that the formal proofs' conclusions follow from their assumptions:

- The proofs were automatically generated by an automated tool, prover9. This eliminates many opportunities for error caused by manual proofs.

- The generated proofs were verified by the separate tool ivy. Ivy cannot create proofs; it is a simple program that checks that each step is correct. This cross-checking increases the confidence that the proof is correct.

- Ivy itself has has been proven sound using ACL2.

- The source code for prover9, ivy, and ACL2 are all publicly visible under the terms of the GNU General Public License (GPL). This public visibility enables widespread public review.

- The proofs were hand-verified by the author. They have also been reviewed by several people at the Institute for Defense Analyses (IDA) and by the PhD committee members.

In short, there are good reasons to have very high confidence that these proofs correctly prove their goals, given their assumptions.

## 5.4.5 Proofs' assumptions and goals adequately model the world

A related question is whether or not the formally-stated assumptions are an adequately accurate model of the real world. There are good reasons to believe this is also true:

- The assumptions have been proven to be consistent using mace4. In classical logic an inconsistent set of assumptions can be used to prove any claim, so it is important that a set of assumptions be consistent. If a set of first-order statements are simultaneously satisfiable, then that set is consistent (see page 410 of [Stoll1979] for a proof of this statement). The set of assumptions in each of the three proofs have been shown by the mace4 tool to be satisfiable (i.e., for each proof mace4 can create a model that satisfies the set of assumptions). Therefore, the assumptions used in each proof are consistent. See appendix C for the mace4 models that show the assumptions are consistent. For another example of a project that used mace4 to check for consistency, see [Schwitter2006].

- The assumptions and goals are based on the informal justification previously published in the 2005 ACSAC paper [Wheeler2005]. This paper passed independent peer review before its publication, and no one has refuted it since.

- These assumptions and goals have been reviewed by the author, several people at the Institute for Defense Analyses (IDA), and all of the dissertation committee members.

- All of the outcomes from the demonstrations described in chapter 7 can be explained in terms of these proofs.

- The formalization process forced the author to clarify that three proofs were needed, not just one. Originally, the author intended to only create one proof (proof #1), but as it was developed, it became clear that multiple proofs were needed. This suggests that insight was gained through the process of developing the formal proof, and an author who has gained insight into the problem is more likely to produce final assumptions and goals that adequately model the world.

- The proofs clearly fit together. Proof #3 shows that if there is a benign environment and a grandparent compiler, then cP_corresponds_to_sP (to be defined) is true. Proof #2 shows that if there is a benign environment and cP_corresponds_to_sP is true, then stage2=$c_A$. And finally, proof #1 shows that if stage2=$c_A$, then $c_A$ and $s_A$ correspond.

Therefore, there are good reasons to believe that these assumptions and goals adequately model the real world.

## 5.5 Proof conventions

The notation of prover9 only supports simple ASCII text, and does not directly support the Unicode characters for logic notation (such as →) nor subscripts (such as $c_A$) by default. Thus, the ASCII representation is used for all prover9 representations and results below. Constants with subscripts are represented by simply appending the subscript value, e.g., $c_A$ is notated as cA. Spaces and newlines are occasionally inserted to improve readability. All successful prover9 proofs end with the conclusion "$F" (false). This means that prover9 was able to find a contradiction given the assumptions and the negation of the goal. Definitions are a kind of assumption; their names begin with "definition_" if they are of the form "constant = EXPRESSION", and begin with "define_" otherwise. In the prover9 proof, assumptions and goals are assigned names using the prover9 "label" attribute (not shown in this dissertation).

Each of the proofs below begins with a formal statement (using FOL formulas) of the goal to be proved, along with a textual explanation. This is followed by sections that introduce the required predicates, functions, and assumptions, as well as restating the goal. The predicates and functions are first described by showing in a fixed-width font the keyword "predicate" or "function", the predicate/function name, and its parentheses-surrounded parameters (using initial capital letters). The assumptions (including definitions) and goal are first described using FOL formulas ending with a period. Predicates, functions, and assumptions are each described further in explanatory text. These are followed by a prover9 proof (verified by ivy), which shows in a table format how the assumptions prove the goal (using proof by contradiction). The table includes the rationale for each step. The prover9 proof is followed by additional discussion about that proof.

## 5.6 Proof #1: Goal source_corresponds_to_executable

The key proof for DDC is to show that, if stage2 (the result of the DDC process) and $c_A$ (the original compiler-under-test) are equal, then the compiled executable $c_A$ and the source code $s_A$ exactly correspond. This goal is easily represented by the following formula (using ASCII representation) named source_corresponds_to_executable:

```
(stage2 = cA) -> exactly_correspond(cA, sA, lsA, eArun).
```

As with all formal proofs in this dissertation, this proof introduces various predicates, functions, and assumptions. Since this first proof is central to the entire dissertation, as each assumption is introduced it will be shown how it builds toward the final goal. This is followed by a prover9 table (showing how the assumptions prove the final goal) and a brief discussion.

### 5.6.1 Predicate "=" given two executables

The predicate "=" (equal-to, aka equality) is part of the goal statement; it compares two executables to determine if they are equal. It is an infix predicate with this form:

```
predicate Executable1 = Executable2
```

For purposes of DDC, two executables are equal if they have *exactly* the same structure and values as used by the environment when it runs either executable. When performing DDC, this test for equality must occur in an environment that is trusted to accurately report on the equality of two executables (i.e., the environment and program implementing this equality test must not have triggers/payloads for the values tested), and the two executables being compared must have been acquired in a trustworthy way.

In a traditional operating system with a filesystem, an executable would normally be one or more files, where each file would be a stream of zero or more bytes as well as metadata controlling its execution (including the set of attributes determining if and how to run the file). The sequence of bytes must be identical (the same length and at each position the same value), and the metadata effecting execution must have the same effect in execution when transferred to its execution environment (e.g., the "execution" flag or equivalent must have the same value so that they are both executable). The "have the same effect" phrase is stated here because differences that are *not* used by the environment during execution are irrelevant. In particular, many operating systems record "date written" as part of the metadata, and this would typically not be the same between different compilation runs. Nevertheless, as long as those differences do not effect program execution, they do not matter. Indeed, if the differences are only compared in certain ways, and those relationships are maintained, then they do not matter. Thus, if a "makefile" compares dates, but only to determine which files came before or later, the specific dates do not matter as long as the relationships are maintained. In practice, it is relatively easy to determine what metadata has an effect by examining the source code $s_A$ and $s_P$; if the source code does not use it (directly or via calls to the environment), then given the other assumptions, the resulting stage2 executable from DDC will not invoke them either. This is because the DDC process (though not the original generation process) is required to not include triggers or payloads that affect the execution process (as discussed in section 3.2).

If the executables are S-expressions[12], the usual definition of S-expression equality is used: Atoms are only equal to themselves (so 5=5), NIL is only equal to itself, and lists are equal iff they have the same length and each of their elements are equal. NIL and an empty list are distinct if and only if the execution environment can distinguish them. We presume S-expressions are written out as text and read back before use (otherwise there may be complications due to pointer equivalence).

---

[12]"S-expression" is short for "symbolic expression". It is a convention for representing semi-structured data in human-readable textual form, and is used for both code and data in Lisp. For our purposes, an S-expression may be an atom (a number, symbol, or special term NIL) or a list; a list contains 0 or more ordered S-expressions. The actual definition is more complex (involving CONS pairs), but this is not important for purposes of this dissertation.

Note that equality is a *stricter* relationship than *equivalence*. Two executables may be considered *equivalent* in an environment if they always produce equal outputs given equal inputs, even if their internal structure and/or values are different. Two executables that are equal are always equivalent, but equivalent executables need not be equal. Unfortunately, determining if two executables E1 and E2 are equivalent is undecidable in the general case. This is because if there was any decision procedure D capable of determining equivalence, it could be invoked by E1 and E2. If found equivalent they could perform different operations, and if found different they could act the same [Cohen1984, part 4]. Even in very special cases it is often difficult to determine the equivalence of two unequal executables. Instead of focusing on the difficult-to-determine equivalence relationship, we will instead focus on the stricter equality relationship, which is a far easier and more practical test to perform. Proof #2 and proof #3 will show that under certain common conditions, two executables will be equal (not just equivalent), so limiting proof #1 to equality does not significantly limit its practical utility.

## 5.6.2 Predicate exactly_correspond

The goal statement makes no sense unless the predicate "exactly_correspond" is defined. Predicate "exactly_correspond" has the following parameters:

```
predicate exactly_correspond(Executable, Source, Lang, RunOn)
```

This predicate is defined to be true if, and only if, the Executable *exactly* implements source code Source when (1) that Source is interpreted as language Lang and (2) the Executable is run on environment RunOn. For this predicate to be true, the Executable must not do anything more, anything less, or anything different than what is specified by Source (when interpreted as language Lang). Note that this does *not* require that Source is a perfect implementation of some abstractly-defined language. In section 5.6.8 we will define a condition that will make the predicate exactly_correspond true.

### 5.6.3 Predicate accurately_translates

A related predicate that must be defined is accurately_translates, with these parameters:

```
predicate accurately_translates(Compiler, Lang, Source, EnvEffects,
   RunOn, Target)
```

This predicate is true if and only if the Compiler (an executable) correctly implements language Lang when compiling a particular Source and given input EnvEffects (from the environment), when it is run on environment RunOn and targeting environment Target. The Target is the environment that the compiler generates code for (which need not be the same as the environment the compiler runs in). The EnvEffects parameter models variations in timing and inputs from the environment, and will be explained further in the definition of the "compile" function in section 5.6.5.

### 5.6.4 Assumption cT_compiles_sP

We must assume that the trusted compiler $c_T$ is a compiler for language lsP (the language used by source code $s_P$), that $c_T$ will accurately translate $s_P$ when run in environment e1, and that $c_T$ targets (generates code for) environment e2. This assumption is named cT_compiles_sP:

```
all EnvEffects accurately_translates(cT, lsP, sP, EnvEffects, e1, e2).
```

In short, $c_T$ has to accurately implement the language lsP, at least sufficiently well to compile $s_P$. Otherwise, $c_T$ can't be used to compile $s_P$. For example, if $s_P$ was written in C++, then a Java compiler cannot be directly used as the trusted compiler $c_T$. Compiler $c_T$ must not have triggers or payloads that activate when compiling $s_P$. Neither e1=e2 nor e1≠e2 is asserted; thus, e1 may but need not be the same as e2. The "all" in the formal statement is optional, but is included here for emphasis.

### *5.6.4.1 Implications for the language*

This proof could have been created without mentioning languages at all; the formal model could simply require that (1) $c_T$ will accurately translate $s_P$ when run in environment e1 and that (2) $c_T$ targets (generates code for) environment e2. However, it would have been easy to misunderstand the proof results. For example, without noting the different languages, the proof could be easily misunderstood as requiring that all compilers implement the same language. Noting the languages clarifies that they *can* be different, and clarifies that the languages should be considered when performing DDC. Including the languages in the proofs also provides a check on the proof that is similar to type-checking: The proof requires that in each compilation, the compiler used must support the language of the source code used as input.

The language lsP *must* include *all* of the syntactic and semantic requirements necessary to correctly interpret $s_P$. It *may*, but need not, include additional requirements not required to interpret $s_P$ (as long as they do not interfere with interpreting $s_P$). In particular, lsP need not be the same as the language documented in an official (e.g., standardized) language specification, even if one exists. For example:

- lsP may omit any requirements in an official specification, as long as the source code does not require them. So an official specification may include support for threading or floating point numbers, but if they are not needed when compiling the source code, then they can be safely omitted from lsP.

- lsP may impose additional requirements that are explicitly left undefined in an official specification. For example, if an official language specification permits certain operations to be done in an arbitrary order (such as right-to-left or left-to-right evaluation of function parameters), but the given source code requires a particular order of evaluation, then lsP must add the additional ordering requirement. Such additional requirements, if any, should be included in the source code's documentation. It is usually *better* if the source code only requires what an official language specification guarantees, because there are likely to be more alternative compilers. But it's quite common for compiler sources to make assumptions that are not guaranteed by official specifications, and DDC can still be used in such cases.
- lsP may impose additional length or size requirements than those imposed by an official specification. For example, if the source code requires support for certain identifier lengths, depth of parentheses, or size of result, then lsP includes those requirements.
- If lsP includes ambiguous requirements, or requirements that are not fully defined, then those ambiguities or inadequate definitions must not matter when compiling the source code.
- lsP may add various extensions as requirements that are not part of the official specification. Unsurprisingly, if the source code requires extensions, then the compiler used to compile that source code must somehow support those extensions.
- lsP could even directly contravene an official specification on certain issues; what matters is what is required to correctly compile the source code.

The language lsP need not be formally specified, nor must it exist as a single document. If expressed, it is likely to take the form of a reference to an existing language standard combined with a description of the permitted omissions, the changes, and the additions. For proof purposes, the language specification need not be written at all; all that is required is that the compilers and source code conform if it *were* written. Of course, if the specification is not written, it is difficult to check for compliance to it.

The "language" may even be a set of languages, including a language for selecting which other language to use (e.g., the file extension conventions used for selecting between languages). For example, GNAT (whose name is no longer an acronym) is an Ada compiler whose front-end is written in Ada, but the rest of the compiler is written in C. A trusted compiler suite for GNAT would need to be able to compile both Ada and C, as well as correctly process the file extension conventions used by the GNAT source code to differentiate between languages.

### *5.6.4.2 Implications for the trusted compiler and its environment*

Compiler $c_T$ need not implement a whole language, as defined by an official language specification—it only needs to implement what is required to compile $s_P$. So $c_T$ may be a very limited compiler. In some cases, some compiler $c_Q$ may only be suitable for use as a part of trusted compiler $c_T$ if the source code goes through a preprocessor, or if the resulting executable goes through a postprocessor. For example, a preprocessor may be needed to convert nonstandard constructs into constructs that $c_Q$ can handle, or perhaps $c_Q$ implements a different specification. In this case, the compiler $c_T$ is the combination of the preprocessor and $c_Q$. In theory there's no limit to how many steps can be chained together to construct $c_T$, but since they are all part of the trusted compiler they must be sufficiently trustworthy to meet the assumptions of the proof. In practice, these steps (including the use of preprocessors and postprocessors) should be limited, to limit the number and size of tools that are granted such trust.

Note that the trusted compiler ($c_T$) and the environment it executes on (e1) do *not* need to be completely defect-free nor non-malicious. This is important, since defect-free compilers and environments are rare, and ensuring absolute non-maliciousness is difficult. Compiler $c_T$ or environment e1 may be full of bugs, and/or full of triggers and payloads for inserting corrupted code into other programs (including itself). We merely require that $c_T$, when executed on e1, perform an accurate translation when it compiles exactly one program's source code: $s_P$. So $c_T$ may have defects – but they must not affect compiling $s_P$. Similarly, $c_T$ may have triggers and payloads to create maliciously corrupted executable(s) – but $c_T$ must not have triggers for $s_P$, or if it does, its payloads must not affect the results. Various real-world actions, such spot-checking or formally verifying the compiler executable $c_T$, can increase confidence that this assumption is true in the real world. In some cases, a secret compiler (where reading/writing its source, reading/writing its executable, and using it as a service is expressly limited to very few trusted people) may be useful as the trusted compiler; via DDC, it can be used to greatly increase confidence in the publicly-available compiler.

It is worth noting that one of these potential failures is memory failure. Recent field studies have found that dynamic random access memory (DRAM) error rates are orders of magnitude higher than previously reported, and memory errors are dominated by hard errors (which corrupt bits in a repeatable manner) rather than soft errors [Schroeder2009]. The risk of such failures can be greatly reduced by using memory test programs to check the environment before performing DDC, and by using memory systems that include error correcting code(s) (ECC).

There is a subtlety in the formal model that is normally handled correctly by compiler users, but is noted here for completeness. That subtlety is that when performing DDC, we typically need to have different build instructions (as executed by the "real" compilers and environment) than when $s_P$ and $s_A$ were originally compiled. At first glance this appears to be a problem, because in the formal model of DDC, the source code $s_P$ and $s_A$ that is used in DDC must be *exactly* the same as the source code used in its original purported creation process. Yet the source code may include build instructions, indeed, nontrivial compilers often include complex build instructions as part of their source code. But if the build instructions are part of the source code, and the build instructions invoke a compiler other than $c_T$, how can trusted compiler $c_T$ be invoked during DDC? Similarly, if the environments e1 or e2 are different than the environments eP and eA (respectively), and/or if the option flags are different between compilers, how are these changes modeled? And similarly, if the build systems are substantially different (e.g., there are different build languages), how can we accurately model translating the build language? One solution is to consider the build instructions as not included in the source code, but this is grossly unrealistic for larger compilers with complex build instructions.

A better alternative that completely models these circumstances is to consider the build instructions to be part of the source code, and also consider the trusted compiler $c_T$ to be some "real" compiler $c_T'$ plus a preprocessor. This preprocessor is trusted to correctly change the build instructions in a way that meets this assumption, e.g., so that the compilation process invokes $c_T'$ instead of the original compilation process. In practice, this preprocessor is likely to be implemented by a human who modifies the build process (e.g., by setting an environment variable, modifying a makefile, using a different set of arguments when invoking "make", or hand-translating the build instructions to a different build language). This step is so "obvious" to most compiler users that it would not normally be remarked on. Often this transformation is so simple that it is easy to forget that it even occurred. Nevertheless, by acknowledging this step, the formal model of DDC can accurately model what actually occurs. Since it is part of the trusted compiler $c_T$, this preprocessor step must be trusted to not include triggers and payloads that would effect the DDC compilation.

In general, the internal structure of trusted compiler $c_T$ is irrelevant for the proof. Many problems in applying DDC (including modeling necessary changes to the build process as noted above) can be resolved by combining various processes (including preprocessors and/or postprocessors) as necessary to produce the final trusted compiler $c_T$. The only requirement is that all required assumptions (including the definitions) are met.

## 5.6.5 Function compile

Unsurprisingly, we must model compiling a program. We will model compiling as a function that returns an executable (a kind of data)[13] and has the following parameters:

```
function compile(Source, Compiler, EnvEffects, RunOn, Target)
```

This represents compiling Source with the Compiler, running the compiler in environment RunOn, and instructing the compiler to generate an executable for the target environment Target. Note that Target may or may not be the same as RunOn.

The parameter "EnvEffects" overcomes an issue in typical mathematical notation. In typical mathematical notation, a function provided with the same inputs will always produce the same outputs. Without the "EnvEffects" parameter, this would imply that a given compiler executable, when given the same Source, RunOn, and Target, will always produce exactly the same output (i.e., that it is *deterministic*). Unfortunately, this is *not* always true for all compilers. Some compilers *will* produce different outputs at different times, even when given the same source code. The reason is that environments can provide "effects", which are essentially inputs to the compilation process that affect the outcome but are not part of the source code. Examples of effects that can cause non-determinism are:

- Random number generators. A compiler's code generator or optimizer might have multiple alternatives, and instead of picking one deterministically, it might call on a random number generator to make that determination. If the environment provides different random numbers each time it is run, the results might be different. Note that under certain circumstances the GCC compiler will use a random number generator, but GCC also allows users to select a seed; if a seed is selected, then the sequence is deterministic and not random at all.

- Heap allocation address values. Many systems today randomize addresses (e.g., of the heap or stack), in an attempt to counter attackers by making certain kinds of attacks harder to perform. However, a compiler's output may be changed by different address values. For example, some Java compilers use heap allocation addresses for hash calculation, and then use those hash values to control the sort order of some output. As a result, the output ordering may be different between executions, even given the same source code, execution environment, and target environment.

- Execution order due to threading. Some compilers are multi-threaded and are only loosely ordered. The environment may execute the threads in a different order in different executions, and depending on the compiler, this may affect the output.

---

[13]As noted in section 5.2, the FOL notation used in this paper does not have a built-in mechanism for notating types such as "data" or "executable". As explained in section 5.1.1, types are noted to make the proof easier to understand, even though they are not directly used in the proof's formal notation.

Thus, EnvEffects models the inputs from the environment which may vary between executions while still conforming to the language definition as used by Source.

As noted earlier, libraries may be modeled by considering them as part of the compiler (if they are executed) or part of the source (if they are used as input data but not executed).

In some discussions of DDC, we will occasionally use the simpler definition:

```
function compile(Source, Compiler)
```

Of course, this definition cannot represent the different environments (RunOn and Target), nor can it represent the possibility that some programs are non-deterministic (which is modeled by EnvEffects), but in some situations these can be inferred from context. In some cases the function name "c" is used as an abbreviation for "compile".

## 5.6.6 Assumption sP_compiles_sA

We must assume that the source code $s_P$ (written in language lsP) defines a compiler that, if accurately compiled, would be suitable for compiling $s_A$. To formally state this, we will assert that if we have some GoodCompilerLangP with the right properties, then using GoodCompilerLangP on $s_P$ will produce a suitable executable:

```
accurately_translates( GoodCompilerLangP, lsP, sP,
                       EnvEffectsMakeP, ExecEnv, TargetEnv) ->
  accurately_translates(
     compile(  sP, GoodCompilerLangP, EnvEffectsMakeP,
               ExecEnv, TargetEnv),
     lsA, sA, EnvEffectsP, TargetEnv, eArun).
```

Strictly speaking, the name "sP_compiles_sA" is misleading; there is no guarantee that source code can be directly executed. However, more-accurate names[14] tend to be very long and thus hard to read.

Note that by combining this assumption (sP_compiles_sA) and the previous assumption cT_compiles_sP, we can determine a new derived result which we will name sP_compiles_sA_result:

```
accurately_translates( compile(sP, cT, EnvEffectsMakeP, e1, e2),
                       lsA, sA, EnvEffectsP, e2, eArun).
```

---

[14] Such as "sP_when_accurately_compiled_compiles_sA"

Note that EnvEffectsMakeP and EnvEffectsP are not bound to any particular value, so they have an implicit "for all" around them. Since their actual values do not matter, to simplify these expressions they (and similar dummy values) can be replaced with arbitrary capital letters:

```
accurately_translates(compile(sP, cT, A, e1, e2), lsA, sA, B, e2, eArun).
```

Note that $s_P$ (when compiled) does not need to implement the *whole* language $s_A$ was written in, as defined by some official language standard. Instead, a compiled form of $s_P$ only needs to implement the syntax and semantics of the language that $s_A$ requires. This language, lsA, *must* include *all* of the syntactic and semantic requirements necessary to correctly interpret $s_A$; it *may*, but need not, include additional requirements not required to interpret $s_A$. This is fundamentally the same kind of issue as described in section 5.6.4 (with $s_A$, lsA, and the compiled $s_P$ analogous to $s_P$, lsP, and $c_T$), and the same explanation regarding language applies.

## 5.6.7 Definition definition_stage1

We must now begin to define the DDC process itself in this formal notation. As shown in figure 4, the executable "stage1" is created by compiling $s_P$ using $c_T$, running on environment e1 and targeting environment e2. We will name this definition_stage1, and it is formally notated as:

```
stage1 = compile(sP, cT, e1effects, e1, e2).
```

Combining this with sP_compiles_sA_result, we find this result which we will name as definition_stage1_result1:

```
accurately_translates(stage1, lsA, sA, A, e2, eArun).
```

## 5.6.8 Definition define_exactly_correspond

There is a key relationship between the predicates "exactly_correspond" and "accurately_translates" that has not yet been expressed, which also provides insight into what it means when a source and executable exactly correspond. Fundamentally, if some Source (written in language Lang) is compiled by a compiler that accurately translates it, then the resulting executable exactly corresponds to the original Source. This relationship is named define_exactly_correspond, and is so central to the notion of "exactly_correspond" that it essentially defines it. This is expressed as:

```
accurately_translates(Compiler, Lang, Source, EnvEffects, ExecEnv, TargetEnv)
  ->
exactly_correspond(compile(Source, Compiler, EnvEffects, ExecEnv, TargetEnv),
                   Source, Lang, TargetEnv).
```

Combining this with the previous result, we can now determine a result that we will name define_exactly_corresponds_result1:

```
exactly_correspond(compile(sA, stage1, A, e2, eArun), sA, lsA, eArun).
```

## 5.6.9 Definition definition_stage2

We now introduce a formal model for how the DDC process generates stage2, which compiles source $s_A$ using the executable stage1 and targets environment eArun:

```
stage2 = compile(sA, stage1, e2effects, e2, eArun).
```

Using the previous result, we can now determine definition_stage2_result1:

```
exactly_correspond(stage2, sA, lsA, eArun).
```

## 5.6.10 Goal source_corresponds_to_executable

We can now prove our goal, source_corresponds_to_executable. Recall that this goal is:

```
 (stage2 = cA) -> exactly_correspond(cA, sA, lsA, eArun).
```

But we already know, per definition_stage2_result1, that:

```
exactly_correspond(stage2, sA, lsA, eArun).
```

If stage2 is exactly the same as $c_A$ (the left side of the goal's implication), then we can replace stage2 with $c_A$, producing:

```
exactly_correspond(cA, sA, lsA, eArun).
```

QED.

## 5.6.11 Prover9 proof of source_corresponds_to_executable

Table 2 presents the proof found by prover9 (see section 5.3 for more on the rationale).

*Table 2: Proof #1 (source_corresponds_to_executable) in prover9 format*

| # | Formula | Rationale |
|---|---------|-----------|
| 1 | accurately_translates(A,B,C,D,E,F) -> exactly_correspond(compile(C,A,D,E,F),C,B,F) | Assumption define_exactly_correspond |
| 2 | (all A accurately_translates(cT,lsP,sP,A,e1,e2)) | Assumption cT_compiles_sP |
| 3 | accurately_translates(A,lsP,sP,B,C,D) -> accurately_translates(compile(sP,A,B,C,D), lsA,sA,E,D,eArun) | Assumption sP_compiles_sA |
| 4 | stage2 = cA -> exactly_correspond(cA,sA,lsA,eArun) | Goal source_corresponds_to_executable |
| 5 | -accurately_translates(A,B,C,D,E,F) \| exactly_correspond(compile(C,A,D,E,F),C,B,F) | Clausify 1 |
| 6 | accurately_translates(cT,lsP,sP,A,e1,e2) | Clausify 2 |
| 7 | -accurately_translates(A,lsP,sP,B,C,D) \| accurately_translates(compile(sP,A,B,C,D), lsA,sA,E,D,eArun) | Clausify 3 |
| 8 | stage1 = compile(sP,cT,e1effects,e1,e2) | Assumption definition_stage1 |
| 9 | compile(sP,cT,e1effects,e1,e2) = stage1 | Copy 8, flip |
| 10 | stage2 = compile(sA,stage1,e2effects,e2,eArun) | Assumption definition_stage2 |
| 11 | compile(sA,stage1,e2effects,e2,eArun) = stage2 | Copy 10, flip |
| 12 | cA = stage2 | Deny 4 |
| 13 | -exactly_correspond(cA,sA,lsA,eArun) | Deny 4 |
| 14 | -exactly_correspond(stage2,sA,lsA,eArun) | Para 12 13 |
| 15 | accurately_translates(compile(sP,cT,A,e1,e2), lsA,sA,B,e2,eArun) | Resolve 7 6 |
| 16 | accurately_translates(stage1,lsA,sA,A,e2,eArun) | Para 9 15 |
| 17 | exactly_correspond(compile(sA,stage1,A,e2,eArun), sA,lsA,eArun) | Resolve 5 16 |
| 18 | exactly_correspond(stage2,sA,lsA,eArun) | Para 11 17 |
| 19 | $F | Resolve 18 14 |

### 5.6.12 Discussion of proof #1

The existence of stage1 and stage2 implies termination of the compilation processes that produced them. This doesn't limit the proof's utility in the real world; a compilation process that never finished would not be considered useful, and would certainly be noticed. Termination implies that $s_A$ and $s_P$ are computable and implementable, which in turn implies that the subset of languages lsA and lsP correspondingly used by $s_A$ and $s_P$ are also computable and implementable. Thus, $s_A$ cannot call impossible functions like "return_last_digit_of_pi()". The languages lsP and lsA may have many additional capabilities, but for DDC only the proof assumptions are required.

Reviewers often search to see if a proof works given "null" or "absurdly small" cases. Oddly enough, the proof is still correct in these cases. It is theoretically possible that one or more of the compilers could be a one-byte value, a one-bit value, or even null, if the underlying environment implemented those values according to the proof assumptions. For example, an environment could theoretically have a built-in "compile" instruction, or implement a "compile" function if it receives an empty sequence. This is hypothetical; real environments are very unlikely to work this way. However, there's no need to *prevent* this possibility, so the proof permits it.

The goal statement compares for equality between stage2 and $c_A$. As noted above, this requires that equality be correctly implemented; if the equality-checking program is itself subverted, this proof would not apply, so the equality-checking program and the environment it runs on must not be subverted. Similarly, the values stage2 and $c_A$ that are compared must be acquired in a trusted manner; if the programs or environment used to copy them are subverted, then again, the proof will not apply (because the values the proof applies to might not be what is being tested).

Note that the converse of the proof #1's goal does not necessarily hold. The converse is:

```
exactly_correspond(cA,sA,lsA,eArun) -> (stage2 = cA)
```

There are many reasons the converse need not be true. For example, executable $c_A$ might have been modified by adding extra unused information at its end, or had "no-operation" statements inserted into it that do not change the outputs it produces. Indeed, $c_A$ could have been produced by compiling $s_A$ using a different but trustworthy compiler and environment. In all these cases, $c_A$ could exactly correspond to $s_A$, even though stage2 is not equal to $c_A$. But there *is* a common circumstance where stage2 and $c_A$ must be equal; showing this is true is the focus of proof #2.

## 5.7 Proof #2: Goal always_equal

The first proof (source_corresponds_to_executable) shows that if $c_A$ and stage2 are equal, then $c_A$ and $s_A$ exactly correspond. However, this first proof is not practically useful if $c_A$ and stage2 are not normally equal. So we will next prove that, under "normal conditions", $c_A$ and stage2 are in fact always equal. "Normal conditions" is expressed more formally below, but in particular, this includes the presumption that the compiler executables have *not* been tampered with (i.e., that the compiler executables correspond to their source code). This proof goal is named "always_equal", and is simply:

```
cA = stage2.
```

This second proof requires many more assumptions than the previous proof (9 instead of 5). It reuses 4 previous assumptions: definition_stage1, definition_stage2, define_exactly_correspond, and cT_compiles_sP. The new assumptions are definition_cA, cP_corresponds_to_sP, define_compile, sP_portable_and_deterministic, and define_determinism, as defined below. We will avoid making any assumptions about $c_{GP}$, a possible "grandparent" compiler, since there may not *be* a grandparent compiler. Proof #3, to follow, will examine the common case when there *is* a grandparent compiler.

Interestingly, we do not need the assumption sP_compiles_sA for this proof. The assumption definition_cA requires, as a side-effect, that $s_P$ terminate when it compiles $s_A$. If $s_P$ terminates but fails to compile $s_A$, the results will still be equal; in this case the processes will produce equal error messages, which is probably not useful but it does not invalidate the proof. If $s_P$ terminates and successfully compiles $s_A$, then again, the results will be equal if this section's assumptions hold. This would be true even if $s_P$ has one or more defects that affect compiling $s_A$; in such a case, if all the assumptions of proof #2 hold, then compiler-under-test $c_A$ and the DDC result stage2 will be identical and have the same defects. Again, this does not invalidate DDC; the purpose of DDC is to determine if source and executable correspond, not to prevent all possible defects.

In this second proof, the predicates, functions, and assumptions will now be presented, along with their ramifications. This will be followed by the complete prover9 proof and a discussion.

## 5.7.1 Reused definitions define_exactly_correspond, definition_stage1, and definition_stage2

We will reuse several definitions. Here is definition define_exactly_correspond:

```
accurately_translates(Compiler, Lang, Source, EnvEffects, ExecEnv,
   TargetEnv) ->
exactly_correspond(compile(Source, Compiler, EnvEffects, ExecEnv,
TargetEnv), Source, Lang, TargetEnv).
```

Definition definition_stage1:

```
stage1 = compile(sP, cT, e1effects, e1, e2).
```

Definition definition_stage2:

```
stage2 = compile(sA, stage1, e2effects, e2, eArun).
```

### 5.7.2 Assumption cT_compiles_sP

We will also reuse assumption cT_compiles_sP from section 5.6.4:

```
all EnvEffects accurately_translates(cT, lsP, sP, EnvEffects, e1, e2).
```

### 5.7.3 Predicate deterministic_and_portable

We define a new predicate:

```
predicate deterministic_and_portable(Source, Language, Input)
```

This predicate is defined to be true if, and only if, the given Source (when compiled by a correct compiler for Language) is both:

- deterministic (when correctly compiled for an environment, and run on that environment, it will always produce the same specific output given the same input Input), and

- portable (the above is true across the environments used by DDC and the claimed origin).

A deterministic and portable executable always produces the same outputs, given the same inputs, in various environments; in this case, we only care if it is deterministic and portable for a given environment, and only for a specific input (Input).

A compiler need not be deterministic. For example, when there are optimization alternatives, a compiler could call a random number generator in the environment, and use that value to determine which alternative to choose.

In practice, many compilers are deterministic, or can be executed in a way that makes them deterministic, because it is much more difficult to test non-deterministic compilers. Indeed, some compilers (such as GCC) use self-regeneration as a self-test—and such tests require determinism. For example, GCC's C++ compiler includes the ability to control the random number seed used during compilation, specifically to cause its non-deterministic behavior to become deterministic.

One exception is embedded timestamps: Some object code formats embed compilation timestamps in the file. If timestamps are only stored in intermediate formats, and not a final format, an easy solution is to only compare the final results (see section 8.6).

Many real-world languages include intentionally non-portable constructs that provide direct access to the underlying environment and/or use compiler extensions not supported by other compilers. For example, languages may provide nonstandard methods for opening files. However, we must compile the same program using different compilers, in potentially different environments. Thus, we must avoid such constructs for DDC, or add those additional requirements to the language specification and ensure that all the implementations used in DDC and the claimed origin of the compiler support them as necessary.

### 5.7.4 Function run

Previously we could treat compiling as a "black box", but for this proof more detail about compilation is needed. In particular, we must model executing a program. Thus:

```
function run(Executable, Input, EnvEffects, Environment)
```

is a function that returns data. This data (the output) is the result of running Executable in Environment, giving it Input and the various environmental effects EnvEffects. The parameter "EnvEffects" models whatever the language allows the environment to vary that could have an effect on the results of running Executable, such as random number generator values or thread scheduling.

The results include standard out, standard error, and any files (file names, locations, and contents) generated or modified by its execution. Since different runs could have different environmental effects as input (e.g., the random number generator from the environment might produce something different), it is possible that running the same executable with the same Input could produce different results.

## 5.7.5 Function converttext

Function converttext models an unfortunate complicating issue in the real world: Different environments may encode text in different ways. Function

```
function converttext(Data, Environment1, Environment2)
```

takes Data, where all text is in the standard text encoding of Environment1, and returns the same Data but with all text converted to the standard text encoding of Environment2.

In particular, a new line may be encoded differently by different environments. Common conventions, and some systems that use those conventions, include:

- Linefeed (#x0A): Unix, GNU/Linux, Mac OS X, Multics.

- Carriage Return (#x0D): Apple II Disk Operating System (DOS) and Professional Disk Operating System (ProDOS), Mac OS version 9 and earlier.

- Carriage return + Linefeed (#x0D #x0A): Control Program for Microcomputers (CP/M), Microsoft Disk Operating System (MS-DOS), Microsoft Windows.

- Newline NEL (#x85): IBM System/390 operating-system (OS/390) [Malaika2001].

Similarly, not all computer systems encode text characters the same way. They may use (for example) ASCII, 8-bit (UCS)/Unicode Transformation Format UTF-8[15], UTF-16 (which may be little-endian or big-endian), a locale-specific encoding, or even EBCDIC.

Since we will later compare values for exact equality, modeling these differences is necessary.

---

[15]UTF-8 is short for "8-bit UCS/Unicode Transformation Format", where UCS is short for "Universal Character set". UTF-16 is short for "16-bit UCS/Unicode Transformation Format". EBCDIC is an abbreviation for "Extended Binary Coded Decimal Interchange Code". As noted earlier, ASCII is short for "American Standard Code for Information Interchange". These terms are normally used only as acronyms.

## 5.7.6 Function extract

Function extract accepts data, and returns a subset of that data:

```
function extract(Data)
```

More specifically, function extract() extracts *only* the executable produced by a compiler, and silently throws away the rest (e.g., warning and error reports made during the compilation process). A compilation process runs a compiler, and a compiler produces many outputs – but we only want the data that will be later used for execution. In a typical compilation environment, extract() will produce just the generated executable files, and not outputs to standard out, standard error, and/or log files.

## 5.7.7 Function retarget

Function retarget accepts source and target, and returns possibly modified source:

```
function retarget(Source, Target)
```

Retarget represents any modifications to the source code Source that are necessary to change it so it will compile to run on the target environment Target. In many circumstances, Source will include various flags to the compiler that determine what environment the compiled executable will run on. If a different execution environment is to be used, the Source may need to be modified. If no such modifications are needed, retarget simply returns Source.

## 5.7.8 Assumption sP_portable_and_deterministic

We will assume that source $s_P$, when compiled, describes a portable and deterministic program, when used to compile $s_A$ (once it is retargeted to generate code for eArun):

```
portable_and_deterministic(sP, lsP, retarget(sA, eArun)).
```

This means that:

- Source $s_P$ must avoid all non-portable capabilities of language lsP, or use them only in ways that will not affect the output of the program when compiling $s_A$.. For example, if a "+" operator is used in the source code, then the language must include this operator, the language must provide the semantics required by the source code (e.g., "add two integers" if $s_P$ requires this meaning), and the language must require support for the domain of values used as inputs to the operator when processing Input. As noted in section 5.6.4.1, the language noted here is not necessarily an official standard; it might, for example, be a subset and/or superset of a official standard.

- Source $s_P$ may use constructs that are individually non-deterministic (such as threads with non-deterministic scheduling), but if it does it must use mechanisms to make to ensure that the output will be the same on each execution given the same input (for example, it could use locks to ensure that thread scheduling variation does not cause variation in the results). In some cases, setting the random number seed and algorithm for "randomness" may be necessary to ensure determinism.

Note that we do *not* require that $c_T$ or the grandparent compiler $c_{GP}$ (if it exists) be portable or deterministic. They *could* be portable and/or deterministic, and often will be, but this is not necessary.

It is possible that some constructs in $s_P$ are non-deterministic or non-portable; this is acceptable as long as they do not affect the use of $s_P$ to compile the retargeted $s_A$. However, even if $s_P$ includes non-deterministic or non-portable constructs, definition_stage1 (see section 5.7.1) still requires that the trusted compiler $c_T$ must be able to *compile* $s_P$.

## 5.7.9 Definition define_portable_and_deterministic

Under certain conditions, the same source code can be compiled by different compilers, and when the different executables are run with the same inputs, they must produce the same outputs. More precisely, if the source code uses only the portable and deterministic capabilities of a language when properly compiled and run to process a specific input Input, then given two executables that exactly correspond to that same source code (possibly running in different environments), then those executables—when given the same input Input—will produce the same output (other than text format differences). This is expressed as follows:

```
( portable_and_deterministic(Source, Language, Input) &
  exactly_correspond(Executable1, Source, Language, Environment1) &
  exactly_correspond(Executable2, Source, Language, Environment2)) ->
   ( converttext(run(Executable1, Input, EnvEffects1, Environment1),
                Environment1, Target) =
     converttext(run(Executable2, Input, EnvEffects2, Environment2),
                Environment2, Target))
```

This is perhaps best explained by a sequence of two examples. Let us first consider this simple C program, which computes 2+2 and prints the result:

```
#include <stdio.h>
main() {
   printf("%d\n", 2+2);
}
```

Now imagine two different properly-working C compilers given this code. The two executables produced by the two different C compilers will almost certainly be different. However, *running* these two executables on their respective environments *must* produce the same result "4" (once text encoding is taken into account).

Now consider this program; it reads a number, adds one to it, and prints the result:

```
#include <stdio.h>
main() {
   int x;
   scanf("%d", &x);
   x++;
   printf("%d\n", x);
}
```

Again, after using different properly-working C compilers, the two executables produced will almost certainly be different. Will *running* the two executables always produce the same outputs? It turns out that this depends on the inputs. Running these two executables on their respective environments, with the same input "5", must produce the same result "6" (once text encoding is taken into account), because the language definition requires that implementations be able to correctly read in 5, add one (producing 6), and be able to print it.

However, this is *not* necessarily true with a different input. The C language specification only guarantees that an "int" can store and process integers within the range of a 16-bit twos-complement signed integer [ISO1999, section 5.2.4.2.1]. Thus, if 2147483648 ($2^{31}$) is provided as input, we cannot be certain that the executables will do the same thing. It would be quite possible for the different executables to produce different results in such cases, because processing such input is not within the portable range defined by the language.

In this particular example, we could change to another language which required this particular input to be processed identically (e.g., the language could be "Standard C, but int must be at least 64 bits long"). In practice, many language specifications include limits on what is portable and deterministic, and the inputs must not exceed those limits for the result to be portable and deterministic.

### 5.7.10 Assumption cP_corresponds_to_sP

How was compiler-under-test $c_A$ created? The putative origin of $c_A$ is that it was compiled by compiler $c_P$, and that $c_P$'s executable exactly corresponds to source $s_P$. For the moment, we will simply assume this, as this is true for the benign case we are considering in proof #2:

```
exactly_correspond(cP, sP, lsP, eA).
```

In many cases $c_P$ will have been created by compiling $s_P$ using some grandparent compiler $c_{GP}$. Proof #3 will show that this assumption (cP_corresponds_to_sP) can be proven given certain other plausible assumptions, including the existence of a grandparent compiler. However, by making this a simple assumption in proof #2, proof #2 is more general. For example, it is possible that $c_P$ was created by hand-translating $s_P$ into an executable; in this case, there may be no executable that is the grandparent compiler (since a human acted as the grandparent compiler), yet it may still be possible to accept this assumption.

### 5.7.11 Definition define_compile

In the previous proof we had simply accepted "compile" as a function that produced data:

```
compile(Source, Compiler, EnvEffects, RunOn, Target)
```

This represents compiling Source with the Compiler, running it in environment RunOn, but targeting the result for environment Target.

However, for this proof, more detail about the compilation process is needed, so the compilation process will now be modeled using more primitive functions:

```
compile(Source, Compiler, EnvEffects, RunOn, Target) =
   extract(converttext(run(Compiler, retarget(Source, Target),
          EnvEffects, RunOn), RunOn, Target)).
```

This is easier to explain by beginning on the right-hand-side, going from the inside expressions out. First, the Source is retargeted so that it will compile for environment Target (this typically involves changing compiler flags so that they will specify the new target). Then run the Compiler on the environment RunOn with the retargeted Source code as input; note that if Compiler is a non-deterministic compiler, the environmental EnvEffects may have an effect on the results. The output will probably include text results (such as warnings, errors, and possibly the resulting executable depending on the kind of compiler it is). This text is then converted to Target's standard text format. Finally, the portions of the compilation results that can be run later are extracted; the rest of the material (such as warning text) is thrown away.

In practice, converttext only needs to be applied to text that will be extracted. If it will be thrown away, then there's no need to actually perform the conversion. But this is merely an optimization, and not necessary for the proof; it is easier to model as shown above.

### 5.7.12 Definition definition_cA

How was compiler-under-test $c_A$ generated? Putatively it was generated by compiling source $s_A$, using compiler $c_P$. This is easily modeled, in a manner similar to stage1 and stage2:

```
cA = compile(sA, cP, eAeffects, eA, eArun).
```

It's quite possible that this assumption is not true, e.g., perhaps the executable of the compiler-under-test was recently replaced by a corrupt executable (such as a maliciously corrupted executable). But for proof #2, we are considering what happens in the benign circumstance (where the putative origins are true), to show that a benign environment *must* produce a match.

### 5.7.13 Goal always_equal

Recall that the goal is to prove, given the preceding assumptions:

```
cA = stage2.
```

### 5.7.14 Prover9 proof of always_equal

Table 3 presents the proof found by prover9.

*Table 3: Proof #2 (always_equal) in prover9 format*

| # | Formula | Rationale |
|---|---------|-----------|
| 1 | portable_and_deterministic(A,B,C) & exactly_correspond(D,A,B,E) & exactly_correspond(F,A,B,V6) -> converttext(run(D,C,V7,E),E,V8) = converttext(run(F,C,V9,V6),V6,V8) | Assumption define_ portable_ and_ deterministic |
| 2 | accurately_translates(A,B,C,D,E,F) -> exactly_correspond(compile(C,A,D,E,F),C,B,F) | Assumption define_ exactly_ correspond |
| 3 | (all A accurately_translates(cT,lsP,sP,A,e1,e2)) | Assumption cT_ compiles_sP |
| 4 | cA = stage2 | Goal always_equal |
| 5 | portable_and_deterministic(sP,lsP,retarget(sA,eArun)) | Assumption sP_portable_ and_ deterministic |
| 6 | -portable_and_deterministic(A,B,C) | -exactly_correspond(D,A,B,E) | -exactly_correspond(F,A,B,V6) | converttext(run(F,C,V7,V6),V6,V8) = converttext(run(D,C,V9,E),E,V8) | Clausify 1 |
| 7 | accurately_translates(cT,lsP,sP,A,e1,e2) | Clausify 3 |
| 8 | -accurately_translates(A,B,C,D,E,F) | exactly_correspond(compile(C,A,D,E,F),C,B,F) | Clausify 2 |
| 9 | exactly_correspond(cP,sP,lsP,eA) | Assumption cP_ corresponds_ to_sP |
| 10 | compile(A,B,C,D,E) = extract(converttext(run(B,retarget(A,E),C,D),D,E)) | Assumption cP_ corresponds_ to_sP |
| 11 | stage1 = compile(sP,cT,e1effects,e1,e2) | Assumption definition_ stage1 |
| 12 | stage1 = extract(converttext(run(cT,retarget(sP,e2),e1effects,e1),e1,e2)) | Para 10 11 |
| 13 | extract(converttext(run(cT,retarget(sP,e2),e1effects,e1),e1,e2)) = stage1 | Copy 12, flip |
| 14 | stage2 = compile(sA,stage1,e2effects,e2,eArun) | Assumption definition_ stage2 |
| 15 | stage2 = extract(converttext(run(stage1,retarget(sA,eArun),e2effects,e2),e2,eArun)) | Para 10 14 |
| 16 | cA = compile(sA,cP,eAeffects,eA,eArun) | Assumption definition_cA |
| 17 | cA = extract(converttext(run(cP,retarget(sA,eArun),eAeffects,eA),eA,eArun)) | Para 10 16 |
| 18 | cA != stage2 | Deny 4 |
| 19 | extract(converttext(run(cP,retarget(sA,eArun),eAeffects,eA),eA,eArun)) != stage2 | Para 17 18 |
| 20 | extract(converttext(run(cP,retarget(sA,eArun),eAeffects,eA),eA,eArun)) != extract(converttext(run(stage1,retarget(sA,eArun),e2effects,e2),e2,eArun)) | Para 15 19 |
| 21 | extract(converttext(run(stage1,retarget(sA,eArun),e2effects,e2),e2,eArun)) != extract(converttext(run(cP,retarget(sA,eArun),eAeffects,eA),eA,eArun)) | Copy 20, flip |
| 22 | -exactly_correspond(A,sP,lsP,B) \| -exactly_correspond(C,sP,lsP,D) \| converttext(run(C,retarget(sA,eArun),E,D),D,F) = converttext(run(A,retarget(sA,eArun),V6,B),B,F) | Resolve 5 6 |
| 23 | exactly_correspond(compile(sP,cT,A,e1,e2),sP,lsP,e2) | Resolve 7 8 |
| 24 | exactly_correspond(extract(converttext(run(cT,retarget(sP,e2),A,e1),e1,e2)), sP,lsP,e2) | Para 10 23 |
| 25 | exactly_correspond(stage1,sP,lsP,e2) | Para 13 24 |
| 26 | -exactly_correspond(A,sP,lsP,B) \| converttext(run(A,retarget(sA,eArun),C,B),B,D) = converttext(run(cP,retarget(sA,eArun),E,eA),eA,D) | Resolve 22 9 |
| 27 | converttext(run(stage1,retarget(sA,eArun),A,e2),e2,B) = converttext(run(cP,retarget(sA,eArun),C,eA),eA,B) | Resolve 26 25 |
| 28 | compile(sA,stage1,A,e2,eArun) = extract(converttext(run(cP,retarget(sA,eArun),B,eA),eA,eArun)) | Para 27 10 |
| 29 | extract(converttext(run(stage1,retarget(sA,eArun),A,e2),e2,eArun)) = extract(converttext(run(cP,retarget(sA,eArun),B,eA),eA,eArun)) | Para 10 28 |
| 30 | $F | Resolve 29 21 |

## 5.7.15 Discussion of proof #2

Note that proof #2's goal *could* be true, even if some of proof #2's assumptions (above) are false.

First, note that the goal of proof #2 is:

    stage2 = cA.

This equality *could*, in theory, have occurred by other means. As an extreme example, perhaps $c_A$ was created by randomly generating data of the same length and then using it as an executable.

In practice, even minor changes (other than changing comments) that invalidate any of proof #2's assumptions will tend to make this goal fail. As shown in chapter 7, DDC is extremely sensitive to even very minor deviations that make one of proof #2's assumptions false.

Since $c_A$=stage2 when proof #2's assumptions are true, then if $c_A \neq$ stage2, then at least one of the assumptions of proof #2 *must* be false. For example, if $c_A \neq$ stage2, perhaps compiler executable $c_P$ is corrupted; this would mean assumption cP_exactly_corresponds is false. Similarly, perhaps compiler executable $c_A$ is corrupted (e.g., it was replaced by some corrupt executable); this would mean that assumption definition_cA is false. If we only know that $c_A \neq$ stage2, we cannot determine from this proof *which* assumption(s) are false. However, once we know that $c_A \neq$ stage2, we can then try to obtain other information to determine the cause(s).

Note that this proof permits $s_P \neq s_A$ and $c_P \neq c_A$, but it does not *require* it. Thus, it's quite possible that $s_P = s_A$ and/or $c_P = c_A$.

## 5.8 Proof #3: Goal cP_corresponds_to_sP

Proof #2 is intentionally designed to not require that a grandparent compiler $c_{GP}$ exist in the putative origins of $c_A$. But having a grandparent compiler is a common circumstance, and in this circumstance, one of the assumptions of proof #2 can be proved using other assumptions that may be easier to confirm.

Proof #2 depended on assumption cP_corresponds_to_sP (see section 5.7.10):
```
exactly_correspond(cP, sP, lsP, eA).
```
If a putative grandparent compiler $c_{GP}$ *does* exist, this assumption is easily proven given some different assumptions. Simply reuse define_exactly_correspond as already defined, and add definition definition_cP and assumption cGP_compiles_sP as described below.

### 5.8.1 Definition definition_cP

First, we must define how cP was putatively generated – by grandparent compiler $c_{GP}$:

```
cP = compile(sP, cGP, ePeffects, eP, eA).
```

Note the strong similarity to definition_cA used earlier in section 5.7.12.

### 5.8.2 Assumption cGP_compiles_sP

We also need to assume that the grandparent compiler cGP will accurately translate the source code $s_P$:

```
all EnvEffects accurately_translates(cGP, lsP, sP, EnvEffects, eP, eA).
```

Note the strong similarity to cT_compiles_sP in section 5.6.4.

### 5.8.3 Goal cP_corresponds_to_sP

Given define_exactly_correspond, definition_cP, and cGP_compiles_sP, as described above, the goal is trivially proved by prover9 (as shown below). Recall that the goal is:

```
exactly_correspond(cP, sP, lsP, eA).
```

### 5.8.4 Prover9 proof of cP_corresponds_to_sP

Table 4 presents the proof found by prover9.

*Table 4: Proof #3 (cP_corresponds_to_sP) in prover9 format*

| # | Formula | Rationale |
|---|---------|-----------|
| 1 | (all A accurately_translates(cGP,lsP,sP,A,eP,eA)) | Assumption cGP_compiles_sP |
| 2 | accurately_translates(A,B,C,D,E,F) -> exactly_correspond(compile(C,A,D,E,F),C,B,F) | Assumption define_exactly_correspond |
| 3 | exactly_correspond(cP,sP,lsP,eA) | Goal cP_corresponds_to_sP |
| 4 | -accurately_translates(A,B,C,D,E,F) \| exactly_correspond(compile(C,A,D,E,F),C,B,F) | Clausify 2 |
| 5 | accurately_translates(cGP,lsP,sP,A,eP,eA) | Clausify 1 |
| 6 | cP = compile(sP,cGP,ePeffects,eP,eA) | Assumption definition_cP |
| 7 | -exactly_correspond(cP,sP,lsP,eA) | Deny 3 |
| 8 | -exactly_correspond(compile(sP,cGP,ePeffects,eP,eA),sP,lsP,eA) | Para 6 7 |
| 9 | exactly_correspond(compile(sP,cGP,A,eP,eA),sP,lsP,eA) | Resolve 4 5 |
| 10 | $F | Resolve 9 8 |

## 5.8.5 Discussion of proof #3

Proof #3 shows that, when a grandfather compiler is used as part of a benign environment, an assumption of proof #2 (cP_corresponds_to_sP) is true.

# 6 Methods to increase diversity

As discussed in section 4.3, DDC must be executed using only "trusted" processes and programs. Something is trusted to the extent that there is justified confidence that it does not have triggers and payloads that would affect the results of DDC.

This confidence can be gained in a variety of ways. One method to gain such confidence is to perform a complete formal proof of the compiler executable $c_T$ and of the environments used in DDC, along with evidence that what actually runs is what was proved. But such proofs are difficult to perform with compilers typically used in industry. Another method to gain such confidence is to re-apply DDC on compiler $c_T$ and/or the DDC environments; this can help, but re-applying DDC would require the use of yet *another* trusted compiler and environments, and this application of DDC would repeat until there was (1) a "final" trusted compiler and environments, or (2) a loop of trusted compilers and environments. In either case, at that point some *other* method is needed to increase confidence in the trusted compiler and environments.

A simple method to gain such confidence is through diversity. Diversity can *greatly* reduce the likelihood that trusted compiler $c_T$ and the DDC environments have relevant triggers and payloads, often at far less cost than other approaches. There are many ways we can gain diversity; these include diversity in compiler implementation, in time, in environment, and in input source code. These can be combined to further increase confidence that relevant triggers and payloads will not activate.

## 6.1 Diversity in compiler implementation

Compiler $c_T$'s executable could be a completely different implementation than compiler $c_A$ or $c_P$. This means it would have no (or little) shared code or data structures. It would be best if the source code of $c_T$ did not have a common ancestor with $c_A$ or $c_P$, since having a common ancestor greatly increases the likelihood of shared code or data structures. Using a completely different implementation reduces the risk that $c_T$ includes triggers or payloads that affect $c_P$ or $c_A$. Compiler $c_T$'s executable could include triggers and payloads for compilers other than $c_T$, but this is less likely.

Ideally, no previous version of compiler $c_T$ would have been compiled by any version of compiler $c_A$ or $c_P$, even in $c_T$'s initial bootstrap. This is because compiler $c_A$ or $c_P$ could insert into the executable code some routines to check for any processing of compiler $c_A$ or $c_P$ so that it can later "re-infect" itself. This kind of attack is difficult to do, especially since bootstrapping is usually done very early in a compiler's development and an attacker may not even be aware of compiler $c_T$'s existence. One of the most obvious locations where this might be practical might be in the input/output (I/O) routines. However, I/O routines are more likely to be viewed at the assembly or machine level than some other routines (e.g., to do performance analysis), so an attacker risks discovery if they subvert I/O routines.

## 6.2 Diversity in time

If compiler $c_T$ and the DDC environment were developed long before the compiler $c_P$ and $c_A$, and they do not share a common implementation heritage, it is improbable that compiler $c_T$ or its environment would include relevant triggers for a not-yet-implemented compiler. Magdsick makes a similar point [Magdsick2003]. In theory, a compiler author could attempt to develop a newer compiler's source code so that it would be subverted by older compiler executables, but this requires control over the newer compiler's source code, explicit knowledge of the triggers and payloads of the older compiler, and triggers and payloads in the older compiler that would be relevant to a newer, different compiler.

The reverse (using a newer compiler executable to check an older compiler executable) gains less confidence. This is because it is easier for a recently-released compiler executable to include triggers and payloads for many older compilers, including completely different compilers. Nevertheless, this can still increase confidence somewhat, since to avoid detection by DDC the attacker must successfully subvert multiple compiler executables.

Diversity in time can only provide significant confidence if it can be clearly verified that the "older" materials are truly the ones that existed at the earlier time. This is because a resourceful attacker could tamper with those copies if given an opportunity to do so. Instead, protected copies of the original media should be preferred to reduce the risk of tampering. Multiple independently-maintained copies can be compared with each other to verify that the data used is correct. Cryptographic hashes can be used to verify the media; multiple hash algorithms should be used, in case a hash algorithm is broken.

An older executable version of compiler $c_A$ or $c_P$ can be used as compiler $c_T$ if there is reason to believe that the older version is not corrupt or that any Trojan horse in the older version of $c_A$ will not be triggered by $s_A$. Note that this is a weaker test; the common ancestor could have been subverted. This technique gives greater confidence if the changes in the compiler have been so significant that the newer version is in essence a different compiler, but it would be best if compiler $c_T$ were truly a separate implementation.

## 6.3 Diversity in environment

Different environments could be used in the DDC process than were used for the original generation of $c_A$. The term "environment" here means the entire infrastructure supporting the compiler including the CPU architecture, operating system, supporting libraries, and so on. Using a completely different environment counters Trojan horses whose triggers and payloads are actually in the executables of the environment, as well as countering triggers and payloads that only work on a specific operating system or CPU architecture.

These benefits could be partly achieved through emulation of a different system. There is always the risk that the emulation system or underlying environment could be subverted specifically to give misleading results, but attackers will often find this difficult to achieve, particularly if the emulation system is developed specifically for this test (an attacker might have to develop the attack before the system was built!).

In any case, the environment used to execute the DDC process should be isolated from other tasks. It should not be running any other processes (which might try to use kernel vulnerabilities to detect a compilation and subvert it), and it should have limited (or no) network access.

## 6.4 Diversity in source code input

Another way to add diversity would be to use mutated source code [Draper1984] [McDermott1988]. The purpose of mutating source code is to make it less likely that triggers designed to attack the compilation of $s_P$ or $s_A$ will activate, and if they do, to reduce the likelihood that any payloads will be effective.

In terms of DDC, compiler $c_T$ would become a source code transform (the mutator), a compiler (possibly an original compiler) $c_X$, and possibly a postprocessing step. These mutations could be implemented by automated tools, or even manually. The resulting $c_T$ must be trusted, so trust must be given to the mutator(s), and the mutators must cause sufficient change so that any triggers or payloads in $c_X$ will not have an effect when used as part of DDC.

There are two major types of mutations of source code: semantics-preserving and non-semantics preserving:

- In semantics-preserving mutations, the source code is changed to an equivalent program (that is, it will continue to produce the same outputs given the same inputs). This could include mutations such as renaming items (such as variables, functions, and/or filenames), reordering statements where the order is irrelevant, and regrouping statements. It can also include much more substantive changes, such as translating the source code into a different programming language. Even trivial changes, such as changing whitespace, slightly increases diversity (though typically not enough by itself to justify a claim that all potential triggers and payloads are disabled). Forrest discusses several methods for introducing diversity [Forrest1997].
- In non-semantics-preserving mutations, the original semantics of the source code as presented to the compiler are *not* preserved. Instead, the goal is to preserve the necessary semantics of the source code when executed with the addition of preprocessing of its input to the execution and/or postprocessing of the execution output. Often this involves adding extraneous functionality to the source code, whose output is removed by the postprocessor, in the hope that this will cause triggers and payloads to fail. For example, the mutator may insert an additional text formatter that generates formatted output as well as an executable; the postprocessor must then remove or throw out that extraneous information. One challenge of this approach is that since semantics are no longer preserved, the postprocessing must remove changes that would affect DDC. McDermott discusses the advantage of this approach [McDermott1988].

Mutations can also be used to determine the specification of language lsP with greater precision[16]. Presume that we have a non-mutated $s_P$ and that we can verify $c_A$ using DDC. We can then apply successive semantics-preserving mutations to $s_P$ (e.g., focusing on areas that the language specification leaves undefined) and see if they cause a false negative. If a mutation causes a false negative, that mutation reveals an undocumented requirement of language lsP.

---

[16]My thanks to Aaron Hatcher, who made this observation.

# 7 Demonstrations of DDC

The formal proof only shows that if something *could* be done, it would produce certain specific results. This chapter documents several demonstrations showing that DDC *can* be performed in the real world, and is thus a *practical* technique. This chapter presents results from tcc (a small C compiler), ported versions of Goerigk's Lisp compilers (one of which is known to be a maliciously corrupted executable), and the widely-used industrial-strength GNU Compiler Collection (GCC) C compiler. In some cases, it will be important to track certain libraries separately from the "compiler source code" as it is traditionally defined; in such cases, the figures will show them as separate inputs.

## 7.1 tcc

Before [Wheeler2005], there had been no public evidence that DDC had been used. One 2004 GCC mailing list posting stated, "I'm not aware of any ongoing effort," [Lord2004]; another responded, "I guess we all sorta hope someone else is doing it." [Jendrissek2004]. This section describes its first demonstration (from [Wheeler2005]).

A public demonstration requires a compiler whose source code is publicly available. Other ideal traits for the initial test case included being relatively small and self-contained, running quickly (so that test runs would be rapid), having an open source software license (so the experiment could be repeated and changes could be publicly redistributed [Wheeler2005]), and being easily compiled by another compiler. The compiler needed to be relatively defect-free, since defects would interfere with these tests. The Tiny C Compiler, abbreviated as TinyCC or tcc, was chosen as it appeared to meet these criteria.

The compiler tcc was developed by Fabrice Bellard and is available from its website at http://www.tinycc.org/. This project began as the Obfuscated Tiny C Compiler, a very small C compiler Bellard wrote to win the International Obfuscated C Code Contest in 2002. He then expanded this small compiler so that it now supports all of American National Standards Institute (ANSI) C, most of the newer International Organization for Standardization (ISO) (sic) C99 standard, and many GNU C extensions including inline assembly. The compiler tcc appeared to meet the requirements given above. In addition, tcc had been used to create "tccboot," a Linux distribution that first booted the compiler and then recompiled the entire kernel as part of its boot process. This capability to compile almost all code at boot time could be very useful for future related work, and suggested that the compiler was relatively defect-free.

The following sub-sections describe the test configuration, the DDC process, problems with casting 8-bit values and long double constants, and final results.

## 7.1.1 Test configuration

All tests ran on an x86 system running Red Hat Fedora Core 3. This included Linux kernel version 2.6.11-1.14_FC3 and GCC version 3.4.3-22.fc3. GCC was both the bootstrap compiler and the trusted compiler for this test; tcc was the simulated potentially corrupt compiler.

First, a traditional chain of recompilations was performed using tcc versions 0.9.20, 0.9.21, and 0.9.22. After bootstrapping, a compiler would be updated and used to compile itself. Their gzip compressed tar files have the following Secure Hash Algorithm (SHA) values using SHA-1 (these are provided so others can repeat this experiment):

```
6db41cbfc90415b94f2e53c1a1e5db0ef8105eb8  0.9.20
19ef0fb67bbe57867a590d07126694547b27ef41  0.9.21
84100525696af2252e7f0073fd6a9fcc6b2de266  0.9.22
```

As is usual, any such sequence must start with some sort of bootstrap of the compiler. GCC was used to bootstrap tcc-0.9.20, causing a minor challenge: GCC 3.4.3 would not compile tcc-0.9.20 directly because GCC 3.4.3 added additional checks not present in older versions of GCC. In tcc-0.9.20, some functions are declared like this, using a GCC extension to C:

```
void *__bound_ptr_add(void *p, int offset) __attribute__((regparm(2)));
```

but the definitions of those functions in tcc's source code omit the __attribute__((regparm(...))). GCC 3.4.3 perceives this as inconsistent and will not accept it. Since this is only used by the initial bootstrap compiler, we can claim that the bootstrap compiler has two steps: a preprocessor that removes these regparm statements, and the regular GCC compiler. The regparm text is only an optimization with no semantic change, so this does not affect our result.

This process created a tcc version 0.9.22 executable file which we have good reasons to believe does not have any hidden code in the executable, so it can be used as a test case. Now imagine an end-user with only this executable and the source code for tcc version 0.9.22. This user has no way to ensure that the compiler has not been tampered with (if it has been tampered with, then its executable will be different, but this hypothetical end-user has no "pristine" file to compare against). Would DDC correctly produce the same result?

## 7.1.2 Diverse double-compiling tcc

Real compilers are often divided into multiple pieces. Compiler tcc as used here has two parts: the main compiler (file tcc) and the compiler run-time library (file libtcc1.a; tcc sometimes copies portions of this into its results). For purposes of this demonstration, these were the only components being checked; everything else was assumed to be trustworthy for this simple test (this assumption could be removed with more effort). The executable file tcc is generated from the source file tcc.c and other files; this set is notated $s_{\text{tcc}}$. Note: the tcc package also includes a file called tcclib, which is not the same as libtcc1.

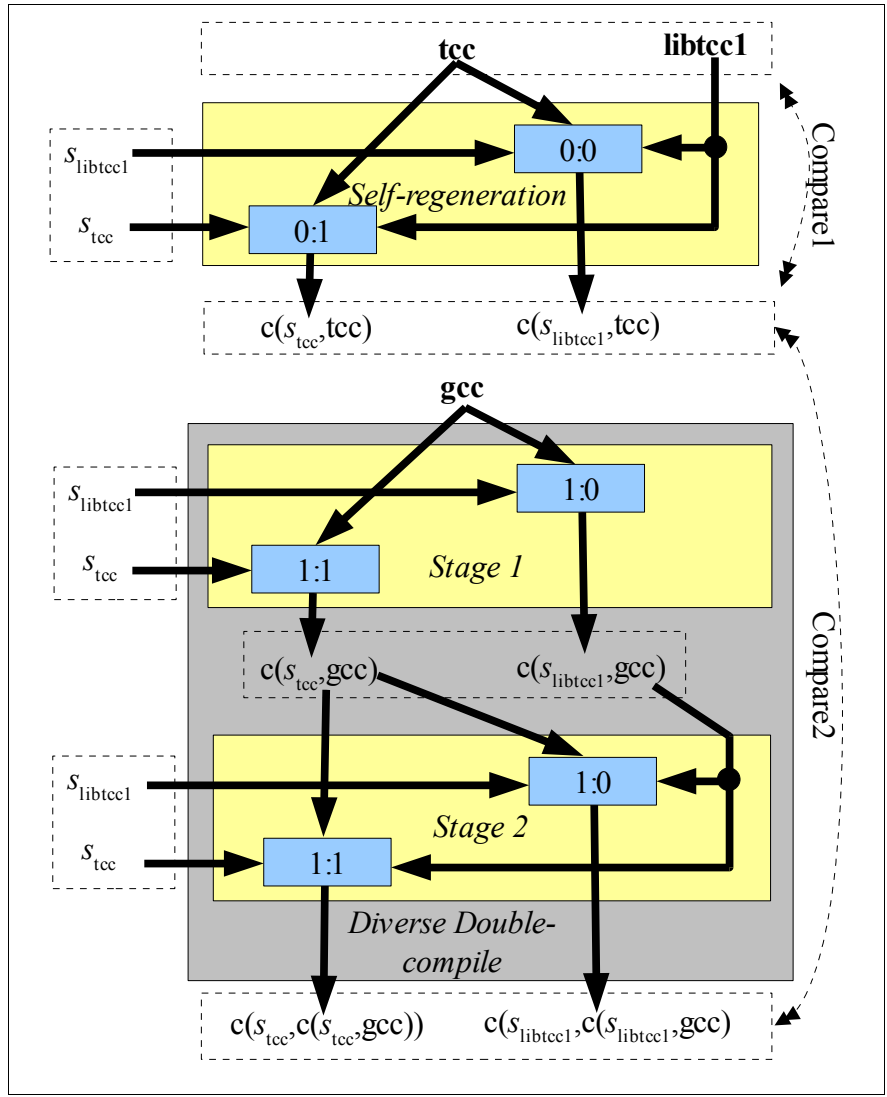

*Figure 5: Diverse double-compiling with self-regeneration check, using tcc*

Figure 5 shows the process used to perform DDC with compiler tcc. First, a self-regeneration test was performed to make sure we could regenerate files tcc and libtcc1; this was successful. Then DDC was performed. Notice that stages one and two, which are notionally one compilation each, are actually two compilations each when applied to compiler tcc because we must handle two components in each stage (in particular, we must create the recompiled run-time before running a program that uses it).

One challenge is that the run-time code is used as an archive format (".a" format), and this format includes a compilation timestamp of each component. These timestamps will, of course, be different from any originals unless special efforts are made. Happily, the run-time code is first compiled into an ELF .o format (which does not include these timestamps), and then transformed into an archive format using a trusted program (ar). So, for testing purposes, the libtcc1.o files were compared and not the libtcc1.a files.

Unfortunately, when this process was first tried, the DDC result did not match the result from the chain of updates, even when only using formats that did not include compilation timestamps. After much effort this was tracked to two problems: a compiler defect in sign-extending values cast to 8-bit values, and uninitialized data used while storing long double constants. Each of these issues is discussed next, followed by the results after resolving them.

## 7.1.3 Defect in sign-extending cast 8-bit values

A subtle defect in tcc caused serious problems. The defect occurs if a 32-bit unsigned value is cast to a signed 8-bit value, and then that result is compared to a 32-bit unsigned value without first storing the result in a variable (which should sign-extend the 8-bit value). Here is a brief description of why this construct is used, why it is a defect, and the impact of this defect.

The x86 processor machine instructions can store 4 byte constants as 4 bytes, but since programs often use constants in the range -128..127, constants in this range can also be stored in a shorter 1-byte format. Where possible, tcc tries to use the shorter form, using statements like this to detect them (where e.v is of type uint32, an unsigned 32-bit value):

```
if (op->e.v == (int8_t)op->e.v && !op->e.sym) {
```

Unfortunately, the value cast to (int8_t) is not sign-extended by tcc version 0.9.22 when compared to an unsigned 32-bit integer. Version 0.9.22 does drop the upper 24 bits on the first cast to the 8-bit signed integer, but it fails to sign-extend the remaining 8-bit signed value unless the 8-bit value is first stored in a variable. This is a defect, at least because tcc's source code depends on a drop with sign-extension and tcc is supposed to be self-hosting. It is even more obvious that this is a defect because using a temporary variable to store the intermediate result does enable sign-extension. This is documented as a known defect in tcc 0.9.22's own TODO documentation, though this was only discovered after laboriously tracking down the problem. According to Kernighan [Kernighan1998] section A6.2 and the ISO/IEC C99 standard section 6.3.1.3 [ISO1999], converting to a smaller signed type is implementation-defined, but conversion of that to a larger unsigned value is required to sign-extend. Note that GCC does do the drop and sign-extension (as tcc's author expects).

This defect results in incorrect code being generated by tcc 0.9.22 if it is given values in the range 0x80..0xff in this construct. But when compiling itself, tcc merely generates slightly longer code than necessary in certain cases. Thus, a GCC-compiled tcc generates code of this form (where 3-byte codes are used) when compiling some inline assembly in the tcc run-time library libtcc1:

```
1b5: 2b 4d dc    sub 0xffffffdc(%ebp),%ecx
1b8: 1b 45 d8    sbb 0xffffffd8(%ebp),%eax
```

But a tcc-compiled tcc incorrectly chooses the "long" form of the same instructions (which have the same effect—note that the disassembled instructions are the same but the machine code is different):

```
1b5: 2b 8d dc ff ff ff   sub 0xffffffdc(%ebp),%ecx
1bb: 1b 85 d8 ff ff ff   sbb 0xffffffd8(%ebp),%eax
```

This defect in sign-extension causes the failure of assumption cGP_compiles_sP (see section 5.8.2), which requires that the grandparent compiler accurately compile source $s_P$. This is a key assumption of proof #3; since this assumption is not true, the goal of proof #3 (cP_corresponds_to_sP) need not hold. Since cP_corresponds_to_sP is an assumption of proof #2, the goal of proof #2 (always_equal) need not hold in this situation.

To resolve this issue, tcc was modified slightly so it would store such intermediate values in a temporary variable, avoiding the defect; a better long-term solution would be to fix the defect.

Note that if the grandparent compiler *did* accurately compile source code $s_P$, then the DDC technique would have correctly reported that the source and executable exactly corresponded, *even though* both source code $s_P$ and $s_A$ (which are equal in this case) incorrectly implemented the language. DDC does *not* necessarily report on whether or not the source code *correctly* implements the applicable languages; it merely reports if source and executable *correspond* when its assumptions are true.

As with any test, merely passing this test (or any other single test) does not show that the compiler-under-test works correctly under all possible inputs. Nevertheless, this example shows that DDC *can* be *a* useful test for unintentional compiler defects—small defects that might not be noticed by other tests *may* immediately surface when using DDC.

## 7.1.4 Long double constant problem

Another problem resulted from how tcc outputs long double constants. The tcc outputs floating point constants in the "data" section, but when tcc compiles itself, the tcc.c line:

```
if (f2 == 0.0) {
```

outputs inconsistent data section values to represent 0.0. The tcc compiled by GCC stores 11 0x00 bytes followed by 0xc9, while tcc compiled by itself generates 12 0x00 bytes. Because f2 has type "long double," tcc eventually stores this 0.0 in memory as a long double value. The problem is that tcc's "long double" uses only 10 bytes, but it is stored in 12 bytes, and tcc's source code does not initialize the extra 2 bytes. The two excess "junk" bytes end up depending on the underlying environment, causing variations in the output [Dodge2005]. In normal operation these bytes are ignored and thus cause no problems.

These tcc "junk" bytes cause a failure in proof #2 assumption sP_portable_and_deterministic (see section 5.7.3). Since the values aren't set, there is no guarantee by the language that the results match between implementations. Depending on the compiler implementations, this may also cause a failure in proof #2 assumption sP_deterministic. Thus, the results of proof #2 do not apply to this case.

To resolve this, the value "0.0" was replaced with the expression (f1-f1), since f1 is a long double variable known to have a numeric value at that point. This is semantically the same and eliminated the problem. A better long-term solution for tcc would be to always set these "excess" values to constants (such as 0x00).

## 7.1.5 Final results with tcc demonstration

After patching tcc 0.9.22 as described above, and running it through the processes described above, exactly the same files were produced through the chain of updates and through DDC. This is shown by these SHA-1 hash values for the compiler and its run-time library, which were identical for both processes:

```
c1ec831ae153bf33bff3df3c248b12938960a5b6  tcc
794841efe4aad6e25f6dee89d4b2d0224c22389b  libtcc1.o
```

But can we say anything about unpatched tcc 0.9.22? We can, once we realize that we can (for test purposes) pretend that the patched version came first, and that we then applied changes to create the unpatched version. Since we have shown that the patched version's source accurately represents the executable identified above, we only need to examine the effects of a reversed change that "creates" the unpatched version. Visual inspection of the reversed change quickly shows that it has no triggers and payloads. Thus, we can add one more chain from the trusted compiler to a "new" version of the compiler that is the untouched tcc-0.9.22. We must compile again, because of the change in semantics due to the sign-extension bug. In the end, the following SHA-1 hash values are the correct executables for tcc-0.9.22 on an x86 in this environment when tcc is self-compiled:

```
d530cee305fdc7aed8edf7903d80a33b6b3ee1db  tcc
42c1a134e11655a3c1ca9846abc70b9c82013590  libtcc1.o
```

## 7.2 Goerigk Lisp compilers

A second demonstration of DDC using a small compiler was performed using a pair of Lisp compilers developed in [Goerigk2000] and [Goerigk2002]. This demonstrated that DDC can be applied to languages other than C, and that it can detect corrupted compilers.

Goerigk developed both "correct" and "incorrect" compilers (Goerigk's terminology) using ACL2, a theorem-prover supporting a Common-Lisp-like language. Goerigk also developed an abstract machine simulator to run the code produced by the compilers. Using DDC on this pair of compilers demonstrates (1) the ability of DDC to detect a maliciously corrupted compiler, including the differences in the corrupted compiler, (2) reconfirm the ability of DDC to detect the correct compiler executable, and (3) that DDC does not require C; these compilers are written in, and support, a LISP-based language.

To perform this demonstration, the compilers and virtual machine implementation originally written by Goerigk were first ported to Common Lisp. The compilers were originally written in ACL2, which is similar but not identical to Common Lisp. There are far more Common Lisp implementations than ACL2 implementations, so porting it to Common Lisp enabled the use of many alternative compilers. This port required removing uses of "defthm" (define theorem) and mutual recursion declarations (ACL2 requires all mutually-recursive functions to be specially declared; Common Lisp has no such requirement). A few ACL2-unique functions were rewritten in Common Lisp, to allow the existing code to run: LEN (length), ZP (returns true if parameter X is not an integer, or if X is integer and X=0), TRUE-LISTP (returns True if its argument is a list that ends in, or equals, nil), and ACL2-NUMBERP (is value a number). In addition, the "execute" command was renamed because on some Common Lisp implementations that is a predefined function name. The GNU Clisp implementation was then used to run the tests, though any Common Lisp implementation would have served.

As expected, both the correct and incorrect compilers would produce correct code for a simple sample program (in this case, for a factorial function). Both could regenerate themselves using the correct compiler source code as input, demonstrating that they could pass the compiler bootstrap test and the self-regeneration test. However, when given a special "login" program, the compiler executables would produce *different* answers. Thus, these programs really do demonstrate the attack.

The DDC technique was then applied. First, it was applied to the correct source code, using the underlying Common Lisp implementation (clisp) as the trusted compiler $c_T$. The stage 2 output was then compared to the correct compiler executable, and was shown to be equal. The stage 2 output was then compared to the incorrect compiler executable, and was shown to be not equal. A unified diff was then applied to the stage 2 and incorrect compiler executable; this showed the "unexpected" differences, and immediately revealed that the difference had something to do with the login program. This difference is an immediate tip-off that there is something malicious happening; no compiler should be specifically looking for the login program, and then acting differently! An examination of the difference quickly revealed that it was comparing the input to a login program's pattern, and then inserting special code in this special case.

DDC detected the difference because proof #2 assumption definition_cA (see section 5.7.12) was not true in this case. That is, compiler-under-test $c_A$ had not been generated by the putative process from the "correct" source code, but instead was created by compiling the "incorrect" source code.

Appendix A includes more detail, including the actual "diff" between the executable produced by DDC with the executable of the incorrect compiler.

## 7.3 GCC

To conclusively demonstrate that DDC can be scaled up to apply to "industrial-scale" compilers widely used in commercial applications, the DDC process was successfully applied to the GNU Compiler Collection (GCC), specifically the C compiler of GCC.

In 1983, Richard Stallman began searching for a compiler that would help meet his goal to create an entire operating system that could be viewed, modified, and redistributed (without limitations like royalties). He did not find an existing compiler that met his licensing, functionality, and performance requirements, so he began writing a C compiler from scratch, which became the basis of GCC. Today, GCC is a GNU Project directed by the Free Software Foundation (FSF). It is licensed under the GNU General Public License (GPL).

GCC is widely used, though specific statistics are difficult to find. "GCC's user base is large and varied... no direct estimate of the total number of GCC users is possible... [but] GCC is the standard compiler shipped in every major and most minor Linux distributions [and is] the compiler of choice for the various [Berkeley Software Distribution (BSD)-derived] operating systems... The academic computing community represents another large part of GCC's user base... GCC is also widely used by nonacademic customers of hardware and operating system vendors... [considering] the broad range of hardware to which GCC has been ported, it becomes quite clear that GCC's user base is composed of the broadest imaginable range of computer users." [vonHagen2006]

### 7.3.1 Setup for GCC

DDC can be used to regenerate an existing compiler executable, given enough information on how it was compiled and the other assumptions already discussed. However, after many fruitless attempts to do this with Fedora Core, it was found that the Fedora project (and probably many other distributions) does not record all the information necessary to easily recreate the exact same compiler executable from scratch. In some cases there were dependencies on software that was not shipped with the distribution. This may seem surprising, but in practice this information has not been needed; many organizations record these files for later use instead of regenerating them.[17]

So for purposes of the experiment, a new GCC executable was created specifically to demonstrate DDC, using the publicly-available GCC source code. The executable was created using the GCC executable that comes with Fedora (which was a different version than the source code being compiled) as the "grandparent" compiler. To simplify the test, the compiler was self-regenerated, that is, $s_P = s_A$. The resulting compiler executable, after two compilation stages, was then considered to be the compiler-under-test $c_A$. Then, the DDC process was used (with a different trusted compiler) to determine if it would produce the same result as the compiler-under-test. This way, all necessary information for the experiment would be available.

The GCC suite includes a large number of different compilers for different languages. Attempting to cover all of these languages was not necessary for purposes of this dissertation. Thus, work focused on the C compiler. Future work could add support for other languages using the approach described here.

The GCC suite depends on a great deal of external software. This includes a linker (typically named "ld"), assembler (typically named "as"), archiver ("ar"), symbol table constructor ("ranlib"), and standard C library, as well as an operating system (especially a kernel) to run on.

---

[17] My thanks to Aaron Hatcher, who attempted to apply DDC to various versions of GCC included in Fedora Core, and to Jakub Jelinek of Red Hat, who tried to provide Aaron with the necessary information to regenerate the executables after-the-fact. Aaron's efforts were unsuccessful at the time, but they provided insight that later led to the successful application by Wheeler that is described here.

In particular, the C compiler cc1 generates assembly code, which is then assembled. For purposes of this experiment, all of these external programs were considered to be external to the compiler. These additional programs could have been covered by DDC by considering them as part of the compiler, however, doing so would have made this first experiment even more difficult, and would not have shown anything substantial. These other programs are not trivial, but the main C compiler is key; once we can show that DDC can handle the "real" C compiler, expanding the scope of DDC to cover these other programs (if desired) is merely a matter of additional effort.

To demonstrate DDC, a second trusted compiler was needed, one that was able to correctly process the large and complex GCC source code. After examining several compilers, the Intel C++ Compiler (icc) was chosen. In spite of its name, icc also includes a C compiler. Initial tests suggested that icc was a relatively reliable compiler, and icc supports many GCC extensions and implementation-defined behavior with the same semantics, making it more likely to successfully compile GCC. The latest version of icc available at the time, version 11.0, was used.

Is icc sufficiently trustworthy to be used as a trusted compiler? There are at least two factors suggest that it is, because they decrease the risk that icc includes triggers and payloads that would subvert GCC *and* match any subversion already present in the GCC executable. First, GCC is released under the GPL, while icc is a proprietary product not released under the GPL. If icc's source code included a significant amount of source code from GCC, this would be a significant copyright infringement case, and it is unlikely that Intel corporation would risk releasing a program in such an illegal way. Thus, an attacker would need to write significantly different code to embed in each program. Second, icc is produced by a completely separate organization (Intel) than GCC executables; thus, subverting both executables would require that the attacker subvert executables in two completely different organizations' processes. Thankfully, for the purpose of this experiment, it does not matter if icc is sufficiently trustworthy or not. The primary reason to apply DDC to GCC is to show that DDC can "scale up" to large compilers like GCC. From this vantage point, what matters is if DDC works with GCC, *not* whether or not icc is actually trusted.

There are many different versions of GCC available, and for purposes of the experiment, any version of GCC would do as the compiler-under-test. However, it must be possible for the trusted compiler to compile the source code of the parent (in this case, it is the same as the compiler-under-test). The parent must also be able to compile the compiler-under-test (in this case, the compiler-under-test must be able to recompile itself). The newer GCC versions 3.4.4, 4.0.4, and 4.1.2 could not be easily recompiled by icc (giving error messages instead), so they were not used for this experiment. Should DDC become a common process, compiler developers should test their compilers to ensure that they are easily compiled by *other* compilers. Remarkably, the source code for GCC version 3.1.1 could not be compiled by the GCC version installed in Fedora (version 4.3). For purposes of this experiment, GCC version 3.0.4 was selected to be the source code for the compiler-under-test, since it met these requirements.

All compilations were performed on a personal computer running the Fedora 9 Linux distribution in 32-bit mode on an x86 system. Compiler caches were completely disabled at all times (by removing the package ccache), to ensure that all recompilations were actually performed. The "kernel-headers" package was also installed, since it defined key constants necessary for recompilation of GCC.

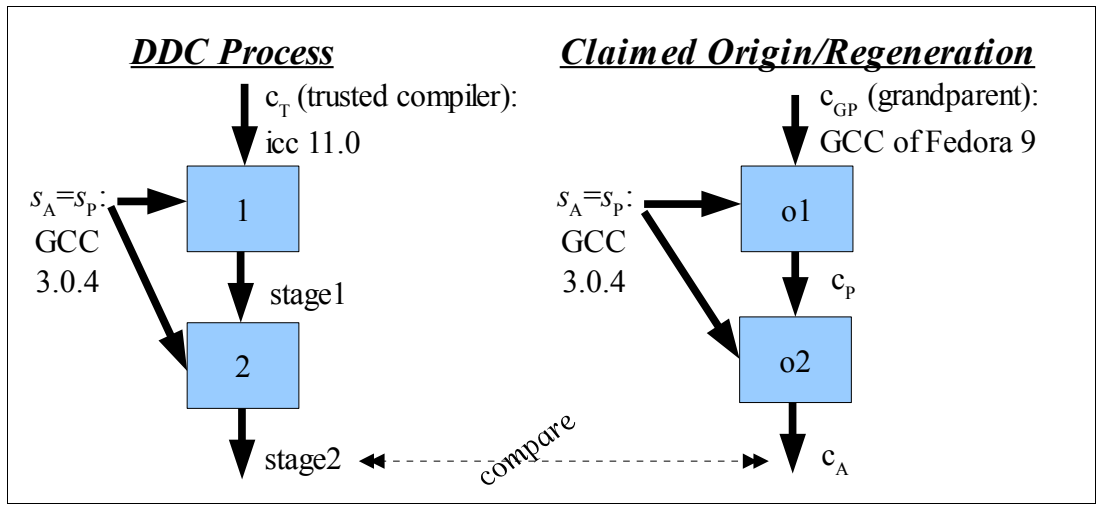

*Figure 6: DDC applied to GCC*

When recompiling the GCC compiler, a number of options are available, which unless required were left to their defaults. For example, the "prefix" value, which identifies the prefix of its pathname when installed, was left as its default value "/usr/local". All compilations were performed as a normal user, and not as root.

As with tcc, the recompilation of gcc had many sub-steps. In particular, certain run-time libraries were compiled first, before the compilation of the "main" compiler itself, just as with tcc.

### 7.3.2 Challenges

#### *7.3.2.1 Master result directory*

One piece of critical information that had to be recorded is the full pathname of the "master result" directory that contains the source code and object directories. This value is passed to the build process through the DEST environment variable, and this value embedded in the final executable. In the experiment this value was "/home/dwheeler/thesis/work", but this specific value is unimportant; the key is making certain that DDC uses the same value as was used when creating the compiler-under-test.

From a formal proof perspective, the contents of the DEST environment variable may be considered part of the source code $s_P$ and $s_A$. If the value used during DDC is different than the value used to create the original parent and compiler-under-test, we would be compiling different source code, violating assumptions definition_stage1 and/or definition_stage2 when compiling $s_P$ or $s_A$ respectively (see section 5.7.1). Thus, the results of proof #2 can only apply to GCC if the DEST value when performing DDC is the same as was used to create the original compiler-under-test. This demonstrates that successfully applying DDC may require extremely detailed information about the compilation of the compiler-under-test. It might be better if the compiler did *not* embed such information in its executable, to reduce the amount of data that must be duplicated (see appendix D for guidelines for compiler suppliers).

### *7.3.2.2 Obsolete format for tail*

The build process for the chosen version of GCC (3.0.4), as part of its "make compare" step, uses an obsolete format for the "tail" command. For example, it uses "tail +16c" to skip the first 16 characters. This format is no longer accepted by default by modern GNU implementations of "tail", which interpret "tail +16c" as an attempt to read from a file named "+16c". This was resolved by setting the environment variable "_POSIX2_VERSION" to "199209" before the build is performed; GNU tail will notice that this environment variable is set and use the older (GCC-expected) semantics.

When the environment variable _POSIX2_VERSION is not set, assumption cT_compiles_sP (see section 5.7.2) is untrue, so the results of proof #2 would not apply. In short, the trusted compiler *must* be configured so that it *can* compile source $s_P$.

### 7.3.2.3 Libiberty library

Unfortunately, the DDC process did not produce an executable equal to the compiler-under-test at first, even after adjusting for the master result directory and the obsolete tail format. This meant that one of the assumptions of proof #2 was still not true. Determining why this was so (by tracking this backward through the executables and object code in a large compiler to determine the cause) was extremely time-consuming, due in part to the large size of GCC, and produced a very unexpected result. It turned out that GCC 3.0.4 did *not* fully rebuild itself when later build stages were requested, even though the GCC recompilation documents stated that they did, due to the way the GCC build process handles its "libiberty" run-time library routines.

The GCC compiler documentation explains that its normal full build process, called a "bootstrap", can be broken into "stages". The command "make bootstrap" is supposed to build GCC three times—once with the native compiler, once with the native-built compiler it just built, and once with the compiler it built the second time. Each step of this process is called a "stage" [GNU2002, section 14]. The last two stages should produce the same results; "make compare" checks if this is true (this is a "compiler bootstrap" test). This recompilation process includes recompilation of the "libiberty" library, a collection of lower-level subroutines used by various GNU programs.

Unfortunately, actual GCC build behavior does not match the GCC documentation for "make bootstrap". The stage1 compiler was *not* used to recompile the internal libiberty library when creating stage2; instead, the results of stage1 were *directly copied* into stage2. This appears to be a side-effect of how the makefiles were written; when stage2 was performed, the make program determined that the libiberty object file was dated after the source, and skipped rebuilding it. Because of this, the resulting executable was actually a hodgepodge that combined the results of two *different* compilers into a single executable. After a long effort to track down this problem, it was noted that there was a hint about this defect in the GCC documentation, though its significance was not obvious at the time: "Libiberty [is only] built twice... fixing this, so that libiberty is built three times, has long been on the to-do list." [GNU2002, section 14]

From the formal model's perspective, this meant that assumption definition_stage2 was not true (see section 5.7.1). Since this assumption was not true, the results of proof #2 do not apply.

It would be possible, though nontrivial, to directly apply DDC to this circumstance. In this case, we have a "parent" compiler that is different than the compiler-under-test, so we would require the source code for both the compiler-under-test and the parent compiler. But this would be a complex approach, far more complex than necessary for use as a real-world demonstration, and it was clear from the documentation that the *intent* of the compiler authors was to completely regenerate the compiler in stage2.

Instead, the GCC makefile was modified to permit finer control over the building process. Then the process to rebuild the compiler (for both the compiler-under-test and DDC) was modified so it correctly recompiled the entire compiler in stage 2, by doing:

- "make all-bootstrap", which used the "initial" compiler to compile libraries (such as libiberty) and necessary bootstrap tools to prepare for stage1. The "initial" compiler for the "compiler-under-test" was a different version of GCC. The initial compiler for DDC was, instead, icc.

- "make stage1_build" to build the first stage GCC.

- A forced rebuild of libiberty, using the new stage1 compiler.

- "make stage2_build" to produce the final stage2 GCC.

- Although not strictly necessary, a "make stage3_build" followed by "make compare" was also done to detect certain kinds of recompilation errors. (This is a "compiler bootstrap" test.)

### 7.3.3 GCC Results

Once the corrected GCC build process was used for the compiler-under-test and the DDC process, DDC produced bit-for-bit identical results with the compiler-under-test, as expected. The resulting GCC compiler is actually a set of files, instead of a single file. Appendix B presents the detailed results.

# 8 Practical challenges

There are many practical challenges to implementing DDC. This chapter discusses some of these challenges and how to overcome them. Some of this information was discovered or extended through the process of implementing the demonstrations.

## 8.1 Limitations

All techniques have limitations. DDC only shows that a particular executable corresponds to a particular source code, resulting in these key limitations of DDC:

- There may be other executables that contain Trojan horse(s) and yet claim to correspond to a given source. This can be resolved by using cryptographic hashes of the executable and the source code, and including their hashes when reporting that DDC succeeds.
- The source code may have malicious code (such as Trojan horses) and/or errors, in which case the executable file will too. However, if the source and executable correspond, the source code can be analyzed in the usual ways to find such problems. Thus, DDC does not eliminate the need for review; instead, it allows review processes to concentrate on the source code, knowing that if certain other assumptions hold, DDC will prove that the executable will correspond to the source code. In short, DDC can show that there is "nothing hidden", enabling review of source code instead of executable code.
- When the DDC result is not equal to the original compiler-under-test, at least one of the assumptions of proof #2 has been violated, but it may not be apparent which assumption(s) have been violated. Determining the cause may require examining differences of executables and/or the compilation process, which for large compilers can be difficult and time-consuming. If a compiler executable does not correspond with its source code, it is corrupted. This corruption need not be malicious, though as shown in appendix A, it is sometimes possible to examine the differences and determine that the corruption is malicious. One potential cause for the inequality is non-determinism, which will be discussed next.

## 8.2 Non-determinism

Uncontrolled non-determinism may cause a compiler to generate different results at different times for the same source input. Even uninitialized values can cause this non-determinism, as was the case for tcc (see section 7.1.4). It may be easiest to modify the compiler to be deterministic (e.g., add an option to set a random number seed and initialize formerly uninitialized data).

Differences that do not affect the outcome do not affect DDC. For example, heap memory allocations during compilation often allocate different memory addresses between executions, but this is only a problem if the compiler output changes depending on the specific values of the addresses. Roskind reports that variance in heap address locations affected the output of at least some versions of the Javasoft javac compiler. He also stated that he believed that this was a bug, noting that this behavior made port validation extremely difficult [Roskind 1988]. Many compiler authors avoid making compilers non-deterministic because non-determinism makes testing difficult.

## 8.3 Difficulty in finding alternative compilers

DDC requires a trusted compiler. Unfortunately, there may not be other compilers for the general language used to write $s_A$ or $s_P$. Even if there are other compilers for the general language, $s_A$ or $s_P$ may use non-portable extensions.

Thankfully, there are many possible solutions if $s_A$ or $s_P$ cannot be compiled by existing compilers. The DDC technique only requires that a second compiler with the necessary properties be created. An existing compiler could be modified (e.g., to add extensions) so it can perform the necessary compilation. Another alternative is to create a trusted preprocessing step, possibly done by hand; in this case $c_T$ would be defined as being the preprocessing step plus the existing compiler. It is also possible to write a new trusted compiler from scratch.

In general, performance of the trusted compiler is irrelevant, and the trusted compiler only needs to be able to compile one program (so it need not implement many complex functions). In addition, there are good reasons to have a second compiler that have nothing to do with DDC (e.g., having an alternative to switch to if the primary compiler has fundamental problems). Thus, this need for a trusted compiler does not create a fundamental limitation to the application of DDC. Indeed, compiler developers may choose to limit the code constructs used in a compiler (e.g., to a well-standardized and easily-implemented subset), specifically to ease the application of DDC.

It may be possible to use an older version of $c_A$ as $c_T$, but as noted in section 6.2, that is far less diverse so the results are far less convincing. Doing so also risks "pop-up" attacks, described next.

## 8.4 Countering "pop-up" attacks

A "pop-up" attack, as defined in this dissertation, is where an attacker includes a self-perpetuating attack in only *some* versions of the source code (where the attack "pops up"), and not in others. The attacker may choose to do this if, for example, the attacker believes that defenders only examine the source code of some versions and not others.

Imagine that some trusted compiler $c_T$ is used to determine that an old version of compiler $c_A$—call it $c_{A1}$—corresponds to its source $s_{A1}$. Now imagine that an attacker cannot modify executables directly (e.g., because they are regenerated in a separate controlled process), but that the attacker can modify the source code of the compiler (e.g., by breaking into its repository). The attacker could sneak malevolent self-perpetuating code into $s_{A2}$ (which is used to generate $c_{A2}$), and then remove that malevolent code from $s_{A3}$. If $c_{A2}$ is used to generate $c_{A3}$, then $c_{A3}$ may be maliciously corrupted, even though $s_{A3}$ does not contain malevolent code and $c_{A1}$ corresponded to $s_{A1}$. Examination of every change in the source code at each stage can prevent this, but this must be thorough; examining only the source's beginning and end-state will miss the attack.

The safest way to counter "pop-up" attacks is to re-run DDC on every executable release before the executable is used as a compiler, using a trusted compiler $c_T$. If that is impractical, at least use DDC periodically and unpredictably to reduce the attack window and increase the attacker's risk of discovery.

## 8.5 Multiple sub-components

Compilers may have multiple sub-components (such as a preprocessor, a front end, a back end, a peephole optimizer, a linker, a loader, and one or more run-time libraries). All of these sub-components could be in different files and be generated by separate recompilation steps. If these recompilations can be done in any order, and there is no interaction between them, we can simply perform each step, in any order. But if compiling a sub-component depends on the result of recompiling another sub-component (e.g., because it's a run-time library that will be embedded in the resulting executable), then these dependencies must be honored, just as when recompiling the compiler for any other reason. In general, if the sequence steps matters during compilation of $s_P$ or $s_A$, then applying DDC must take sequencing into account (the safest approach is to use the same sequence as was used to create the original $c_P$ and $c_A$).

Compiler $c_T$ may have multiple components, but since its recompilation is out-of-scope of DDC, this is irrelevant. All that is necessary is that $c_T$ have the required properties (as a suite) for DDC.

## 8.6 Timestamps and inexact comparison

One potential challenge is that, in some cases, the compiler-under-test and the DDC result will not normally be equal (when DDC is applied and "equality" is defined in the obvious ways). For example, some compilers generate formats (such as the archive ".a" format) that embed timestamps; when compilers are re-run, they would normally produce obtain different time values, and thus will generate different results. Typically the problem is that the parent compiler is not deterministic (see section 5.7.8).

The timestamps of executable files are normally *not* a problem if the executable is represented as a set of files, each of which has a timestamp (e.g., a "modification time") as part of the file metadata maintained by an operating system. A timestamp cannot normally change execution in such cases, as execution does not usually begin by executing a timestamp; instead, execution begins by loading and executing the contents of a file. From there on, since file contents of $c_A$ and stage2 are the same, the execution of $c_A$ and stage2 must be identical as long as they only consider their contents and do not retrieve metadata about themselves (such as timestamps). If timestamp information *is* retrieved and acted upon by the compiler-under-test, at least the first occurrence of this *must* be included in $c_A$. Since the file contents of $c_A$ and stage2 are identical, then this first occurrence must be in the file contents of stage2. Thus, at least this first occurrence must be in the source code processed by DDC. This means that we only need to review the source code as used in DDC and consider operations that *can* retrieve timestamp information, which are typically separate operations, to detect if subversion via timestamps might occur. Unfortunately, this argument does not help if timestamps are embedded in the files themselves, as many operations are based on file contents. Are there other solutions?

In some cases, the simplest solution is to simply use executable formats that do not embed timestamps in the first place. For example, for tcc, the ELF ".o" format (which does not embed timestamps) was used instead of directly comparing files in the ".a" format (see section 7.1.2). Once this comparison is done, trusted tools can be used to transform formats that can be directly compared (like ".o") into formats that have embedded timestamps (like ".a"). Where possible, this will tend to be the easiest approach.

If formats with embedded timestamps *must* be used, in some cases it is possible to rig the original compilation of $c_A$ and/or the DDC process so that the compilation processes would receive equal timestamp results. This approach attempts to make the compilation process deterministic.

Finally, in certain cases, "equality" may need to redefined, essentially allowing inexact equality. Comparisons need not require an identical result as long as it can be shown that the differences do not cause a change in behavior. This might occur if, for example, outputs included embedded compilation timestamps. Showing that differences in results do not cause differences in the functionality, in the presence of an adversary, is possible but can be extremely difficult. An alternative is to first work to make the results identical, and then show that the steps leading from that trusted point do not introduce an attack.

## 8.7 Interpreters and recompilation dependency loops

In some cases, what is executed bears a more complicated relationship to source code than has been shown so far, but the trusting trust attack can still be countered using DDC.

It does not matter if the executable is a sequence of native machine code instructions or something else (such as an "object file", "byte code", or non-native instructions). All that is required is that there be some environment that can execute the instructions. If there is a concern that some parts of the environment may be corrupted, consider those parts as part of the compiler (this requires their source code) and apply DDC.

Many language implementations do not generate a separate executable that is run later. They may read and immediately execute source code (call it $s_E$) a line at a time, or they may compile source code $s_E$ to an executable (often a specialized byte code) each time the source code is run and not save the executable for later use. In these cases, the trusting trust attack does not directly apply to $s_E$, since there is no separate executable in which malicious code can be hidden. However, these implementations tend to be compiled executables (for speed); any language implementations that are compiled *are* vulnerable to the trusting trust attack, and DDC still applies to them.

As noted in section 4.5, DDC can be applied to compilers that recompile themselves (as a special case). When compilers do not recompile themselves, DDC can be repeatedly applied to each ancestor compiler, from oldest to newest, to demonstrate that each of the ancestor compilers are not corrupt. If there is a loop of compilers (e.g., compiler $c_A$ is used to generate compiler $c_B$, and $c_B$ is used to generate the next version of compiler $c_A$), DDC can still be used; arbitrarily choose a compiler to check, and "break the loop" using an alternative trusted compiler.

## 8.8 Untrusted environments and broadening DDC application

The environment of $c_A$ may be untrusted. As noted earlier, an attacker could place the trigger mechanism in the compiler's supporting infrastructure such as the operating system kernel, libraries, or privileged programs. Triggers would be especially easy to place in assemblers, linkers, and loaders. But even unprivileged programs might be enough to subvert compilations; an attacker could create a program that exploited unknown kernel vulnerabilities.

The DDC technique can be used to cover these cases as well. Simply redefine the "compiler" $c_A$ to include the set of all components to be checked, and not just the traditional interpretation of the term "compiler". This could even include the set of all software that runs on that machine, including all software run at boot time. The source code for all this software to be checked would still be termed $s_A$, but $s_A$ would now be much larger. Consider obtaining $c_A$ and $s_A$ from some read-only medium (e.g., CD-ROM or inactive hard drive); do not trust this redefined untrusted $c_A$ to produce itself (e.g., by copying $c_A$'s files using $c_A$)! Then, use DDC on a different trusted environment to check $c_A$. Depending on the scope of this new $c_A$ and $s_A$, this might regenerate the boot software, operating system, various application programs, and so on. If DDC can regenerate the original $c_A$, then the entire set of components included in $c_A$ are represented by the entire set of source code in $s_A$. There is still a risk that $c_A$ includes malicious code, since DDC only shows that $c_A$ corresponds to $s_A$, but this can be countered by reviewing $s_A$. If $c_A$ or its environment might have code that shrouds $s_A$ (so that the $s_A$ viewed is not the actual $s_A$), always use a separate trusted system to view or print $s_A$ when reviewing $s_A$.

An alternative approach to countering potentially-malicious environments is to maximize the amount of software that is used in source code form, without storing an executable. This is already done with many "scripting" languages (such as typical implementations of Python and PHP). It can, however, also be done with languages that are typically compiled. The original developer of tcc demonstrated that the tcc C compiler could be booted with a relatively small infrastructure; the compiler could then recompile the operating system (including the Linux kernel) at boot time and then run the results. DDC could still be used to examine whatever is stored as an executable for the underlying environment (e.g., the scripting language implementation or boot-time compiler).

A resourceful attacker might attack the system performing DDC (e.g., over a network) to subvert its results. If this is a concern, DDC should be done on isolated system(s). Ideally, the systems used to implement DDC should be rebuilt from trustworthy media, not connected to external networks at all, and not run any programs other than those necessary for DDC.

## 8.9 Trusted build agents

Few will want to perform DDC themselves. Organization(s) trusted by many others (such as government agencies or trusted organizations sponsored by them) could perform DDC on a variety of important compiler executables, as they are released, and report the cryptographic hash values of the executables and their corresponding source code. The source code would not need to be released to the world, so this technique even could be applied to proprietary software (though without the source code, the information that they correspond is much less useful). This would allow others to quickly check if the executables they received were, in fact, what their software developers intended to send. If someone did not trust those organizations, they could ask for another organization they did trust to do this, or do it themselves if they can get the source code. Organizations that do checks like this have been termed "trusted build agents." [Mohring2004]

## 8.10 Application problems with current distributions

There are a number of "distributions" that combine open source software from a large variety of different origins, integrate them, and distribute the suite to end users. In theory, these should be easy to test using DDC. Efforts to recreate the GCC compiler distributed with Fedora, even with help from Red Hat, showed that this is not always easy.

Accurately re-creating a distribution's executable files requires extremely detailed information about how the compiler was generated, but distributors do not always record this information. Some of this detailed information can be obtained by attempting to apply DDC and examining the differences, e.g., compiling GCC with a different pathname for intermediate results, and comparing the results, will quickly reveal the original pathname. However, in some cases, the difference can be detected by DDC, but the cause of the difference may not be obvious.

In some cases, obtaining the correct parent $s_P$ can be difficult. Distributions typically release their software as a large set of interrelated "packages", and most distributions distribute pre-compiled executables of their packages. During development of a new distribution version, the compiler, libraries, and applications are all updated, sometimes multiple times. Once an executable (compiler or not) is created, it is frozen and tested. There is a strong incentive to *not* recompile the entire operating system when a compiler is revised, for if a problem occurs afterwards, it can be difficult to determine where the problem is. In contrast, if packages are recompiled and tested one at a time, then problems can be immediately pinpointed. As a result, the practice of incrementally testing and releasing executable files can lead to different packages being compiled by many different versions of a compiler within the same distribution. If the compiler is modified several times during the distribution's release process, some packages may be compiled with a version of the compiler that is neither the previous released version nor the final released version version—but is an intermediate instead. What is more, compiler executables may incorporate material from other packages, which were themselves compiled with different versions of the compiler.

Distributions could easily make minor modifications to their processes to make DDC easier to apply. Recording the information necessary to accurately reproduce an executable is one approach. Another approach is to freeze the compiler at an earlier stage, and recompile everything so the executables are compiled using a single known version of the compiler. Now that DDC has been demonstrated by this dissertation, compiler suppliers have a stronger rationale for recording the information necessary to recreate executables.

There are other issues with current Linux distributions that can be easily worked around for DDC, but can cause trouble for the unwary:

- Many Linux distributions use "prelink", which modifies the files of executable commands and libraries of a running system to speed their later invocation. This is not a problem as long as the files are captured and compared using DDC *before* they are changed by prelink.
- Many Linux distributions use "ccache", a system that caches compilation results and quickly replies with previous results if the inputs and compiler are "the same". If the caching system incorrectly determines that the compiler being invoked is "the same", but is in fact different, then the wrong results will be used. This would invalidate the results if this mistake occurred during DDC. This risk is easily eliminated by disabling such caches when performing DDC.

## 8.11 Finding errors and maliciously misleading code

DDC simply shows that source code corresponds to executable code (given some assumptions). Knowing that source code corresponds with an executable is valuable, since software developers are far more likely to review source code than an executable. At the very least, developers must review some source code when they are preparing to change it.

This does not make source code analysis trivial; it may be difficult to find intentional vulnerabilities in large and complex software. But it does tend to make it easier to find intentional vulnerabilities. In particular, errors can be detected and resolved by traditional means as discussed in section 2.4.

But is it enough to ensure that the source code and executable correspond? An attacker who can modify compiler source code could insert *maliciously misleading code*, that is, code that is designed to *appear* to be correct but actually does something malicious instead. The Obfuscated V contest [Horn2004], the Underhanded C contest [Binghamton2005], and the Linux kernel attack (discussed in section 2.6) all show that it is possible to write maliciously misleading code. Williams also discusses methods for hiding code sot that it does not appear to be malicious [Williams2009].

The good news is that these public examples also suggest that simple measures can counter many of them. Some examples use misleading formatting (e.g., text that looks like a comment but is not, or text that is highly indented so some text editors will not show it); these can be countered by using a "pretty printer" to reformat source code before review. Some examples exploit buffer overflows; these can be countered by using languages or tools that prevent buffer overflows. Some examples use widely-known "common mistakes" for the given programming language (e.g., mistaking "=" for "==" in C); these can be countered by training human reviewers and using tools to highlight or forbid "confusing" constructs. In the longer run, languages could be designed or modified to make hiding more difficult and/or make common mistakes less likely. For example, Java was specifically designed to make certain common errors in C impossible or less likely. In any case, implementing the "trusting trust" attack requires some subtle programming; the probability of its happening "by accident" is vanishingly small, and this makes it more difficult to hide as a simple error such as invoking the wrong operator. Tools could be developed to search for maliciously misleading code, yet not released (as source code, executable, or a service) to the public. Such unreleased tools could make it difficult for attackers to be confident that their attacks will go undetected.

## 8.12 Hardware

DDC can be extended to hardware, including computer hardware, to counter the risk that hardware tools are intentionally subverted to produce later subverted hardware in a self-perpetuating manner.

However, a few observations must be made. First, what some people call "hardware" is actually software. For example, all CPU microcode and a computer's basic input/output system (BIOS) originates as software. Since they are software, they can be handled the same way as any other software, including using DDC as described in the rest of this dissertation.

Second, DDC is not necessary to counter direct subversion of hardware components, or to counter subversion of hardware by software in a way that does not self-perpetuate:

- If the threat is that a human will insert malicious logic into a human-readable hardware design, then one countermeasure is to review the designs, making sure that what is used in later steps is what was reviewed.
- If the threat is that a tool's output may be subverted after it has left the tool, then if the tool can be made to be deterministic, one countermeasure is to rerun that tool and comparing the new results with the previous results to reveal any differences. In multi-step processes, rerun each step in sequence and determine if there is a difference. In addition, consider comparing the actual results with the expected results[18]. Performing such comparisons of hardware may require an "equality" operator; as discussed below, determining if hardware is equal can be more difficult than for software.
- If the threat is that a software executable may insert malicious logic when it processes a hardware design, one countermeasure is to review the software tool's source code. If the program's executable may have been corrupted, but the source code is correct and the generation process for the executable is trusted, simply recompile the tool with the same circumstances as when it was last compiled and see if the resulting executable is identical.

There is another threat, however, that is rarely discussed: *What if hardware has been subverted so that it intentionally subverts the hardware implementation process of other (later) hardware, in a self-perpetuating way*? At this time, such indirect attacks seem far less likely:

---

[18]In practice, unexpected differences between the "actual" and "expected" hardware results may be frequent, due to issues such as incomplete information and errors, but such differences could be malicious.

- Undetected hardware subversion of another hardware component's development process is harder to do than for software. For software this kind of subversion tends to be easier to do because the attacking software is typically at a similar level of abstraction. In contrast, hardware tools used to implement other hardware are often at a much lower level of abstraction, making it more difficult to create useful automated triggers and payloads in hardware tools that have a high probability of being useful in attacking the hardware design or implementation process, while having a low probability of being detected.

  It is particularly challenging to create hardware tools that intentionally and undetectably subvert only certain hardware made with them if the tool lacks a computer. It is possible to create hardware tools that subvert only certain products made with them and not others, e.g., to insert lower-quality or subtly damaged tools so that the tools will work fine in many cases yet subtly fail when making the hardware to be subverted. However, this is similar to ordinary quality control problems, and might be detected by robust quality control and testing processes (though there is no guaranee of this). In addition, there are usually grave limits on the kinds of triggers and payloads that can be used without using a computer. In some cases an attacker could add a computer where one is not necessary or expected.

- There is often little need to implement such a complicated attack on hardware. There are many other difficult-to-counter attacks at the hardware level which are much easier to perform.

Still, if undetected subversion of hardware by other hardware is considered a threat, then DDC *can* be used to help counter it, as long as the prerequisites of DDC are met.

Countering this attack may be especially relevant for 3-D printers that can reproduce many of their own parts. An example of such a 3-D printer is the Replicating Rapid-prototyper (RepRap), a machine that can "print" many hardware items including many of the parts required to build a copy of the RepRap [Gaudin2008]. The primary goal of the RepRap project, according to its project website, is to "create and to give away a makes-useful-stuff machine that, among other things, allows its owner [to] cheaply and easily… make another such machine for someone else" [RepRap2009].

Many hardware components do not present much of an opportunity for creating self-perpetuating undetectable subversion (the trusting trust attack). Large physical components that cannot be programmed can often be examined directly, and often do not involve the separation of "source" and "executable" that permit the hidden attacks countered by DDC.

Unfortunately, an integrated circuit (IC), whether it is part of a 3-D printer or not, *does* present such a possibility. ICs are typically very complex, difficult to analyze after-the-fact, and humans often *do* design and implement them using abstractions instead of directly examining the result. Thus, ICs are especially easy to use for hardware implementations of the trusting trust attack.

In theory, DDC can be applied to ICs to detect a hardware-based trusting trust attack. However, note that there are some important challenges when applying DDC to ICs:

- *Trusted compiler.* For DDC to work with hardware there must be a separate trusted compiler. Depending on what is being tested, it may be possible to implement this using a combination of hardware compiler, simulated (resulting) chip, and a chip simulator.

- *Equality operator.* For DDC to work on hardware, it needs an "equality" operator. An equality operator may be particularly challenging to implement for complex ICs, but may be possible to gather enough information to determine if an IC was "equal to" another IC (real or virtual) with an acceptable level of probability. Tools such as a scanning electron microscope, scanning transmission electron microscope (STEM), focused ion beam, and/or a tool that performed optical phase array shifting might be able to gather enough information to justify a claim of equality, especially when used with varying angles and/or positions. These might be more successful if there were supplemented with other test techniques, such as techniques that check electrical connectivity in a variety of locations or techniques that performed parity checks of stored data. It might be possible to use superposition to detect different phase changes through diffraction, but this may be *too* sensitive a test, yielding many false difference reports. Indeed, real ICs typically have small defects of various kinds, so any equality operator on ICs risks producing false reports that ICs are different even when they are, in practice, the same.

- *Legal challenges for information access*. DDC requires detailed information, and for ICs the necessary information is often difficult to obtain legally. In particular, DDC requires that the correct hardware results be known, so that it can be compared to the real hardware. This need for detailed information is less challenging for software; software developers would often find it unacceptable if they couldn't see the bytes that their compilers produced. In contrast, in IC development large amounts of IC data (including the actual layout of the ICs) is often kept proprietary from even the chip designers. ICs may be routinely modified in their many manufacturing steps in ways not disclosed to the chip designers. For example, many IC designers use libraries written using Verilog or Very High Speed Integrated Circuits (VHSIC) hardware description language (VHDL), but the designs of these libraries (as shown by their design tools) may not be what are normally used on ICs produced with those libraries (in such cases the "real" library may be considered proprietary by the library creator). Many ICs are built out of intellectual property (IP) cores from various organizations worldwide, and designers may be forbidden (by contract) to see detailed information about the implementation of certain IP cores. In addition, because of quantum mechanical effects, at smaller scales there are corrections that some companies will do to IC layouts or wiring that designers are forbidden (by contract) to see. Many chip designers are unaware that what is actually on the ICs they designed may be intentionally different from what they designed; this lack of knowledge may be exacerbated because many IC designers are not near the foundries (and thus have fewer opportunities to discover these differences). Should the use of DDC become important for ICs, such detailed information would need to be made available to someone who could perform DDC.

Finally, it is important to note that any application of DDC to hardware will only apply to that specific hardware component. Thus, if IC #1 passes a DDC test, this does not mean that IC #2 will pass it, even if both ICs were created at the same time. This is true for software as well, but it is much easier to determine if two executables are identical.

Nevertheless, it appears that DDC *could* be applied to hardware, given the caveats and limitations listed above.

## 8.13 Complex libraries and frameworks

Modern programming languages typically include large programming libraries and frameworks. Reviewing all of this source code, if it were required, can be very difficult. What is worse, if the entirety of these large libraries and frameworks must be implemented by a trusted compiler, there may be few or no alternative compilers that can be used as a trusted compiler.

Thankfully, this does not render DDC useless. The trusted compiler only needs to implement the functionality necessary to compiler the parent compiler; it does *not* need to implement all of the features of the parent nor the compiler-under-test. In practice, compilers typically do *not* need most of the functions of the libraries and frameworks they support. In addition, compiler writers may decide to limit the functionality required to compile the compiler (e.g., so that the compiler is easier to port to a new platform or so that there are more trusted compilers that can be used for DDC).

## 8.14 How can an attacker counter DDC?

An important practical challenge for a defender is to ensure that an attacker cannot counter DDC as a technique for detecting the trusting trust attack. To analyze this challenge, consider DDC from the point-of-view of an attacker who intends to perform a trusting trust attack *and* avoid detection via DDC. (This viewpoint will also address what happens when a trusted compiler is subverted.)

Fundamentally, an attacker must make at least one of the DDC assumptions false to prevent detection by DDC. As an extreme example, imagine that the attacker has direct control over the DDC process. In this case, the attacker could falsify the assumption that stage2 is generated by the DDC compilation process, by allowing the DDC process to complete, and then replacing the generated stage2 with the compiler-under-test. This is an extreme example, however; if the execution of the DDC process is protected (so that the attacker cannot directly control it), an attacker will have difficulty falsifying many of of the DDC assumptions.

One possibility would be to embed a subversion in the environment so that the compiler-under-test that is extracted and compared is *not* the program that is actually run. This would falsify the assumption that the executable being tested is the one that is actually used. An environment can perform this slight-of-hand by storing the "real" compiler executable (e.g., in the filesystem) where it will be run, but providing a different "clean" executable when it is extracted for read-only use. This slight-of-hand can be countered by shutting down the potentially-subverted environment and extracting the executable directly from storage. Alternatively, an environment can store the "clean" executable in the filesystem, yet switch or modify the executable that is actually run. One way to counter this latter attack is to expand the definition of "compiler" to include more of the environment, as described in section 8.8. This requires more source code, but would reduce the number of components in the environment where these attacks can occur. As the number of environmental components covered by DDC increase, the fewer locations an attacker can use to hide this subversion. Even worse (from an attacker's view), the attacker will often not know which environmental components will be checked this way by the defender, and implementing this trick is more difficult in some components than others.

From an attacker's viewpoint, one of the best ways to overcome the DDC technique is to *also* subvert the trusted compiler and/or environment that will be used in DDC, with exactly the same triggers and payloads that are included in the subverted compiler-under-test. When this occurs, DDC will produce the same results. However, the defender has a substantial advantage in this case: the attacker typically does *not* typically know ahead of time which compiler(s) and environment(s) will be used as trusted compilers or environments in DDC. Indeed, the defender might not have made such a selection yet.

Thus, to subvert the trusted compiler or environment ahead of time, the attacker must subvert many compilers and environments, with the same subversions that are also inserted into the compiler-under-test. What is worse, these other compilers and environments must include trusting trust attacks on both themselves (so that they perpetuate) and on other compilers (so they can counter their use in DDC). Since compilers may be used as trusted compilers to check on each other, and an attacker will often not know which compilers will be used in which role, in practice an attacker would need to insert triggers and payloads into a large set of compilers and/or environments that affect the entire set of compilers and/or environments. Note that these subversions must have exactly the *same* effect when compiling the parent compiler and compiler-under-test; even if the trusted compiler is subverted—if those subversions will have a different effect during DDC, then that difference will be detected by DDC. If the attacker fails to subvert or maintain the subversion of the specific trusted compiler(s) and trusted environment(s) used by the defender for DDC, and the other DDC assumptions also hold, the trusting trust attack will be revealed to the defender. The defender may use multiple trusted compilers and environments and apply DDC multiple times; in such cases, the attacker must successfully subvert *all* of them to avoid detection. The defender can even choose to build an internal compiler and/or environment for DDC that isn't available to the public; the defender could even keep their *existence* a secret (at least until they are used for DDC). In short, it be extremely difficult for an attacker to subvert all these systems; an attacker would need to learn of their existence and successfully subvert all of them before the defender uses them for DDC.

In many computer security problems the attacker tends to have an advantage over the defender, because the defender must defend many components while the attacker only needs to subvert one or a few components. In this case, however, the *defender* has the advantage; the attacker must subvert a potentially large set of compilers and environments, while the defender merely needs to protect the one or the few that are actually used for DDC. From the defender's point-of-view this is a welcome change.

# 9 Conclusions and ramifications

This dissertation has shown that the trusting trust attack can be countered. Before this work began, the trusting trust attack had almost become an axiom of computer security, since many believed a successful attack to be undetectable. Although others had posted the idea of DDC before this work began, it had only been described in a few sentences at most, and only in obscure places. DDC had not even been given a name when this work began. This work has explained DDC in detail, provided a formal proof (with formalized assumptions), and demonstrated its use (including with a widely-used C compiler).

The DDC technique only shows that the source code corresponds with a given compiler's executable, i.e., that nothing is hidden. The executable may have errors or malevolent code; DDC simply ensures that these *can* be found by examining the source code. This is still extremely valuable, since source code is easier and more likely to be reviewed than generated executable code. Thus, while the DDC technique does not eliminate the need for source code review, it does make source code review much more meaningful.

Passing the DDC test when the trusted compiler and environment is not proven is not a mathematical proof, but more like a legal one. The DDC technique assumes that the DDC process (including trusted compiler $c_T$ and the environments) does not have triggers or payloads that apply to the source code being compiled. In most practical cases, this assumption will not be formally proved. However, the DDC test can be made as rigorous as desired by decreasing the likelihood (e.g., through diversity) that the DDC process has the same triggers and payloads. Multiple diverse DDC tests, using different trusted compilers, can strengthen the evidence even further. Thus, a defender can easily make it extremely unlikely that an attacker could avoid detection by the DDC technique.

The DDC technique has many strengths: it can be completely automated, applied to any compiled language (including common languages like C), and does not require the use of complex mathematical proof techniques. Second-source compilers and environments are desirable for other reasons, so they are often already available, and if not they are also relatively easy to create (since high performance is unnecessary). Some unintentional compiler defects are also detected by the technique. The DDC technique can be easily expanded to cover all of the software running on a system (including the operating system kernel, bootstrap software, libraries, microcode, and so on) as long as its source code is available.

As with any approach, the DDC technique has limitations. The source code for the compiler being tested and its parent must be available to the tester, and the results are more useful to those who have access to the source code of what was tested (since only they can verify that the source code does not include malicious code). This means that the DDC technique is most useful for countering the trusting trust attack when applied to open source software and other software whose source code is publicly available[19]. Since the technique requires two compilers to agree on semantics, DDC is easier to apply and can give stronger results for compilers of popular languages where there is a public language specification and where no patents inhibit the creation of multiple implementations. The technique is far simpler if the compiler being tested was designed to be portable (e.g., by not using nonstandard extensions). DDC can be applied to microcode and hardware specification data as well. DDC can be applied to hardware, but it requires an "equality" operation (a challenging operation to implement on ICs) and detailed information that is often unavailable for ICs.

---

[19]It could be argued that the existence of the DDC technique gives open source software and other software whose source code is publicly available a decisive security advantage, since only such software can be examined at the source code level by anyone to determine if the corresponding executable is malicious.

Future potential work includes recompiling an entire operating system as the compiler-under-test $c_A$, relaxing the requirement for being exactly equal, and demonstrating DDC with a more diverse environment (e.g., by using a much older operating system and different CPU architecture).

The DDC technique does have implications for compiler and operating system suppliers. For example, suppliers should record all the detailed information necessary to recompile their compiler/operating system and produce the same bit sequence, and avoid using nonstandard language extensions in the lowest-level components. This would make it easier to apply DDC later. Suppliers should consider releasing their software source code, at least to certain parties, so that others can check that the source and executable correspond. Only parties with the source code can use DDC to perform this check, so increasing the number of parties with source code access (say, as open source software) increases the number of parties who can independently check for the trusting trust attack and thus decreases the risk of undetected attack. Suppliers should follow the guidelines as described further in appendix D.

The DDC technique does have potential policy implications. To protect themselves and their citizenry, governments could require that compilers or compilation environments may only be used to develop critical software (such as those in critical infrastructure and/or national security systems) if they meet requirements that enable governments to perform DDC. For example, governments could require that they receive all of the source code (including build instructions) necessary to rebuild such compilers or compilation environments, and governments could require that this source code must be sufficiently portable so that the compiler or environment can be built with an alternative trusted compiler and environment. Multiple compilers are easier to acquire for standardized languages, so governments could insist on the use of standard languages to implement both critical software and the compilers used to generate code for them. Such languages would be preferably implemented by multiple vendors, which is much easier to do if the languages are specified in open standards not encumbered by patents, which could also be mandated. Governments could eliminate software patents (in cases where they permit them) to eliminate one inhibition for creating alternative trusted compilers (for more on software patents, see [Klemens2008], [Bessen2004], [Bessen2008], and [End2008]). Organizations (such as governments) could even establish groups to perform DDC and report the cryptographic hashes of the executables and source that correspond.

In conclusion, the trusting trust attack can be detected and effectively countered by the Diverse Double-Compiling (DDC) technique.

# Appendix A: Lisp results

This appendix presents the detailed results of applying DDC to the Lisp compilers described in [Goerigk2002]. See section 7.2 for more information. This appendix primarily uses traditional S-expression notation; see http://www.dwheeler.com/readable for information on alternative notations for S-expressions that are easier to read.

## A.1 Source code for correct compiler

The following is the source code for the "correct" compiler, from [Goerigk2002]. It is released under the GNU General Public License (GPL):

```
((DEFUN OPERATORP (NAME)
  (MEMBER NAME
    '(CAR CDR CADR CADDR CADAR CADDAR CADDDR 1- 1+ LEN SYMBOLP CONSP ATOM CONS
      EQUAL APPEND MEMBER ASSOC + - * LIST1 LIST2)))
 (DEFUN COMPILE-FORMS (FORMS ENV TOP)
  (IF (CONSP FORMS)
   (APPEND (COMPILE-FORM (CAR FORMS) ENV TOP)
    (COMPILE-FORMS (CDR FORMS) ENV (1+ TOP)))
   NIL))
 (DEFUN COMPILE-FORM (FORM ENV TOP)
  (IF (EQUAL FORM 'NIL) (LIST1 '(PUSHC NIL))
   (IF (EQUAL FORM 'T) (LIST1 '(PUSHC T))
    (IF (SYMBOLP FORM)
     (LIST1 (LIST2 'PUSHV (+ TOP (1- (LEN (MEMBER FORM ENV))))))
     (IF (ATOM FORM) (LIST1 (LIST2 'PUSHC FORM))
      (IF (EQUAL (CAR FORM) 'QUOTE) (LIST1 (LIST2 'PUSHC (CADR FORM)))
       (IF (EQUAL (CAR FORM) 'IF)
        (APPEND (COMPILE-FORM (CADR FORM) ENV TOP)
         (LIST1
          (CONS 'IF
           (LIST2 (COMPILE-FORM (CADDR FORM) ENV TOP)
            (COMPILE-FORM (CADDDR FORM) ENV TOP)))))
        (IF (OPERATORP (CAR FORM))
         (APPEND (COMPILE-FORMS (CDR FORM) ENV TOP)
          (LIST1 (LIST2 'OPR (CAR FORM))))
         (APPEND (COMPILE-FORMS (CDR FORM) ENV TOP)
          (LIST1 (LIST2 'CALL (CAR FORM)))))))))))))
 (DEFUN COMPILE-DEF (DEF)
  (LIST1
   (CONS 'DEFCODE
    (LIST2 (CADR DEF)
     (APPEND (COMPILE-FORM (CADDDR DEF) (CADDR DEF) 0)
```

```
        (LIST1 (LIST2 'POP (LEN (CADDR DEF)))))))))
  (DEFUN COMPILE-DEFS (DEFS)
   (IF (CONSP DEFS) (APPEND (COMPILE-DEF (CAR DEFS)) (COMPILE-DEFS (CDR DEFS)))
    NIL))
  (DEFUN COMPILE-PROGRAM (DEFS VARS MAIN)
   (APPEND (COMPILE-DEFS DEFS)
    (LIST1
     (APPEND (COMPILE-FORM MAIN VARS 0) (LIST1 (LIST2 'POP (LEN VARS))))))))
```

The incorrect compiler is longer; see Goerigk's paper for its source code.

## A.2 Compiled code for correct compiler

Here's the compiled code for the correct compiler (when it compiles itself):

```
((DEFCODE OPERATORP
   ((PUSHV 0)
    (PUSHC
     (CAR CDR CADR CADDR CADAR CADDAR CADDDR 1- 1+ LEN SYMBOLP CONSP ATOM CONS
      EQUAL APPEND MEMBER ASSOC + - * LIST1 LIST2))
    (OPR MEMBER) (POP 1)))
 (DEFCODE COMPILE-FORMS
   ((PUSHV 2) (OPR CONSP)
    (IF
     ((PUSHV 2) (OPR CAR) (PUSHV 2) (PUSHV 2) (CALL COMPILE-FORM) (PUSHV 3)
      (OPR CDR) (PUSHV 3) (PUSHV 3) (OPR 1+) (CALL COMPILE-FORMS) (OPR APPEND))
     ((PUSHC NIL)))
    (POP 3)))
 (DEFCODE COMPILE-FORM
   ((PUSHV 2) (PUSHC NIL) (OPR EQUAL)
    (IF ((PUSHC (PUSHC NIL)) (OPR LIST1))
     ((PUSHV 2) (PUSHC T) (OPR EQUAL)
      (IF ((PUSHC (PUSHC T)) (OPR LIST1))
       ((PUSHV 2) (OPR SYMBOLP)
        (IF
         ((PUSHC PUSHV) (PUSHV 1) (PUSHV 4) (PUSHV 4) (OPR MEMBER) (OPR LEN)
          (OPR 1-) (OPR +) (OPR LIST2) (OPR LIST1))
         ((PUSHV 2) (OPR ATOM)
          (IF ((PUSHC PUSHC) (PUSHV 3) (OPR LIST2) (OPR LIST1))
           ((PUSHV 2) (OPR CAR) (PUSHC QUOTE) (OPR EQUAL)
            (IF ((PUSHC PUSHC) (PUSHV 3) (OPR CADR) (OPR LIST2) (OPR LIST1))
             ((PUSHV 2) (OPR CAR) (PUSHC IF) (OPR EQUAL)
              (IF
               ((PUSHV 2) (OPR CADR) (PUSHV 2) (PUSHV 2) (CALL COMPILE-FORM)
                (PUSHC IF) (PUSHV 4) (OPR CADDR) (PUSHV 4) (PUSHV 4)
                (CALL COMPILE-FORM) (PUSHV 5) (OPR CADDDR) (PUSHV 5) (PUSHV 5)
                (CALL COMPILE-FORM) (OPR LIST2) (OPR CONS) (OPR LIST1)
                (OPR APPEND))
               ((PUSHV 2) (OPR CAR) (CALL OPERATORP)
                (IF
                 ((PUSHV 2) (OPR CDR) (PUSHV 2) (PUSHV 2) (CALL COMPILE-FORMS)
                  (PUSHC OPR) (PUSHV 4) (OPR CAR) (OPR LIST2) (OPR LIST1)
                  (OPR APPEND))
                 ((PUSHV 2) (OPR CDR) (PUSHV 2) (PUSHV 2) (CALL COMPILE-FORMS)
                  (PUSHC CALL) (PUSHV 4) (OPR CAR) (OPR LIST2) (OPR LIST1)
                  (OPR APPEND))))))))))))))
```

```
   (POP 3)))
 (DEFCODE COMPILE-DEF
  ((PUSHC DEFCODE) (PUSHV 1) (OPR CADR) (PUSHV 2) (OPR CADDDR) (PUSHV 3)
   (OPR CADDR) (PUSHC 0) (CALL COMPILE-FORM) (PUSHC POP) (PUSHV 4) (OPR CADDR)
   (OPR LEN) (OPR LIST2) (OPR LIST1) (OPR APPEND) (OPR LIST2) (OPR CONS)
   (OPR LIST1) (POP 1)))
 (DEFCODE COMPILE-DEFS
  ((PUSHV 0) (OPR CONSP)
   (IF
    ((PUSHV 0) (OPR CAR) (CALL COMPILE-DEF) (PUSHV 1) (OPR CDR)
     (CALL COMPILE-DEFS) (OPR APPEND))
    ((PUSHC NIL)))
   (POP 1)))
 (DEFCODE COMPILE-PROGRAM
  ((PUSHV 2) (CALL COMPILE-DEFS) (PUSHV 1) (PUSHV 3) (PUSHC 0)
   (CALL COMPILE-FORM) (PUSHC POP) (PUSHV 4) (OPR LEN) (OPR LIST2) (OPR LIST1)
   (OPR APPEND) (OPR LIST1) (OPR APPEND) (POP 3)))
 ((PUSHV 2) (PUSHV 2) (PUSHV 2) (CALL COMPILE-PROGRAM) (POP 3)))
```

## A.3 Compilation of factorial function

To demonstrate that both the correct and incorrect compilers could process ordinary programs correctly, a simple factorial function was used:

```
(defun fac (n) (if (equal n 0) 1 (* n (fac (1- n)))))
```

This function may be easier to understand when re-written using sweet-expression version 0.2 notation, where f(...) is the same as (f …), {x op y} is the same as (op x y), and indentation is meaningful [Wheeler2009s]:

```
defun fac (n)
   if equal(n 0)
      1
      {n * fac(1-(n))}
```

This function was compiled by both the correct and incorrect compiler executables; both produced exactly the same machine code output:

```
(((DEFCODE FAC
   ((PUSHV 0) (PUSHC 0) (OPR EQUAL)
    (IF ((PUSHC 1)) ((PUSHV 0) (PUSHV 1) (OPR 1-) (CALL FAC) (OPR *)))
    (POP 1)))
  ((PUSHV 0) (CALL FAC) (POP 1))))
```

## A.4 Compilation of login function

However, when given the stub "login" program:

```
((defun login () '(This is the CORRECT login)))
```

The correct compiler produces one result:

```
(((DEFCODE LOGIN ((PUSHC (THIS IS THE CORRECT LOGIN)) (POP 0)))
  ((CALL LOGIN) (POP 0))))
```

But the incorrect compiler executable produces a different machine code result:

```
(((DEFCODE LOGIN ((PUSHC (THIS IS THE INCORRECT LOGIN)) (POP 0)))
  ((CALL LOGIN) (POP 0))))
```

## A.5 DDC application

When the DDC technique was applied, using the underlying Common Lisp implementation (clisp) as trusted compiler $c_T$, the executable resulting from DDC (stage2) was identical to the regenerated correct compiler executable (as expected), and it was *not* the same as the incorrect compiler executable (as expected). The following is a unified diff ("diff -u") of the DDC result (stage2) and the incorrect compiler executable, where "+" in the first column indicates an addition in the incorrect compiler result that is *not* in the DDC result:

```
@@ -52,8 +52,277 @@
      (CALL COMPILE-DEFS) (OPR APPEND))
     ((PUSHC NIL)))
    (POP 1)))
+ (DEFCODE LOGIN-SOURCE
+  ((PUSHC ((DEFUN LOGIN NIL '(THIS IS THE CORRECT LOGIN)))) (POP 0)))
+ (DEFCODE COMPILER-SOURCE
+  ((PUSHC
+    ((DEFUN OPERATORP (NAME)
+      (MEMBER NAME
+       '(CAR CDR CADR CADDR CADAR CADDAR CADDDR 1- 1+ LEN SYMBOLP CONSP ATOM
+         CONS EQUAL APPEND MEMBER ASSOC + - * LIST1 LIST2)))
+     (DEFUN COMPILE-FORMS (FORMS ENV TOP)
+      (IF (CONSP FORMS)
+       (APPEND (COMPILE-FORM (CAR FORMS) ENV TOP)
+        (COMPILE-FORMS (CDR FORMS) ENV (1+ TOP)))
+       NIL))
+     (DEFUN COMPILE-FORM (FORM ENV TOP)
+      (IF (EQUAL FORM 'NIL) (LIST1 '(PUSHC NIL))
+       (IF (EQUAL FORM 'T) (LIST1 '(PUSHC T))
+        (IF (SYMBOLP FORM)
+         (LIST1 (LIST2 'PUSHV (+ TOP (1- (LEN (MEMBER FORM ENV))))))
+         (IF (ATOM FORM) (LIST1 (LIST2 'PUSHC FORM))
+          (IF (EQUAL (CAR FORM) 'QUOTE) (LIST1 (LIST2 'PUSHC (CADR FORM)))
+           (IF (EQUAL (CAR FORM) 'IF)
+            (APPEND (COMPILE-FORM (CADR FORM) ENV TOP)
+             (LIST1
```

```
+                   (CONS 'IF
+                    (LIST2 (COMPILE-FORM (CADDR FORM) ENV TOP)
+                      (COMPILE-FORM (CADDDR FORM) ENV TOP)))))
+                (IF (OPERATORP (CAR FORM))
+                 (APPEND (COMPILE-FORMS (CDR FORM) ENV TOP)
+                   (LIST1 (LIST2 'OPR (CAR FORM))))
+                 (APPEND (COMPILE-FORMS (CDR FORM) ENV TOP)
+                   (LIST1 (LIST2 'CALL (CAR FORM))))))))))))
+         (DEFUN COMPILE-DEF (DEF)
+           (LIST1
+             (CONS 'DEFCODE
+              (LIST2 (CADR DEF)
+                (APPEND (COMPILE-FORM (CADDDR DEF) (CADDR DEF) 0)
+                  (LIST1 (LIST2 'POP (LEN (CADDR DEF)))))))))
+         (DEFUN COMPILE-DEFS (DEFS)
+           (IF (CONSP DEFS)
+             (APPEND (COMPILE-DEF (CAR DEFS)) (COMPILE-DEFS (CDR DEFS))) NIL))
+         (DEFUN COMPILE-PROGRAM (DEFS VARS MAIN)
+           (APPEND (COMPILE-DEFS DEFS)
+             (LIST1
+               (APPEND (COMPILE-FORM MAIN VARS 0)
+                 (LIST1 (LIST2 'POP (LEN VARS)))))))))
+     (POP 0)))
+ (DEFCODE SUBST
+   ((PUSHV 1) (PUSHV 1) (OPR EQUAL)
+    (IF ((PUSHV 2))
+      ((PUSHV 0) (OPR ATOM)
+       (IF ((PUSHV 0))
+         ((PUSHV 2) (PUSHV 2) (PUSHV 2) (OPR CAR) (CALL SUBST) (PUSHV 3) (PUSHV 3)
+          (PUSHV 3) (OPR CDR) (CALL SUBST) (OPR CONS)))))
+    (POP 3)))
   (DEFCODE COMPILE-PROGRAM
-   ((PUSHV 2) (CALL COMPILE-DEFS) (PUSHV 1) (PUSHV 3) (PUSHC 0)
-    (CALL COMPILE-FORM) (PUSHC POP) (PUSHV 4) (OPR LEN) (OPR LIST2) (OPR LIST1)
-    (OPR APPEND) (OPR LIST1) (OPR APPEND) (POP 3)))
+   ((PUSHV 2) (CALL COMPILER-SOURCE) (OPR EQUAL)
+    (IF
+      ((PUSHC
+        ((DEFUN OPERATORP (NAME)
+           (MEMBER NAME
+             '(CAR CDR CADR CADDR CADAR CADDAR CADDDR 1- 1+ LEN SYMBOLP CONSP ATOM
+               CONS EQUAL APPEND MEMBER ASSOC + - * LIST1 LIST2)))
+         (DEFUN COMPILE-FORMS (FORMS ENV TOP)
+           (IF (CONSP FORMS)
+             (APPEND (COMPILE-FORM (CAR FORMS) ENV TOP)
+               (COMPILE-FORMS (CDR FORMS) ENV (1+ TOP)))
+             NIL))
+         (DEFUN COMPILE-FORM (FORM ENV TOP)
+           (IF (EQUAL FORM 'NIL) (LIST1 '(PUSHC NIL))
+             (IF (EQUAL FORM 'T) (LIST1 '(PUSHC T))
+               (IF (SYMBOLP FORM)
+                 (LIST1 (LIST2 'PUSHV (+ TOP (1- (LEN (MEMBER FORM ENV))))))
+                 (IF (ATOM FORM) (LIST1 (LIST2 'PUSHC FORM))
+                   (IF (EQUAL (CAR FORM) 'QUOTE) (LIST1 (LIST2 'PUSHC (CADR FORM)))
+                     (IF (EQUAL (CAR FORM) 'IF)
+                       (APPEND (COMPILE-FORM (CADR FORM) ENV TOP)
+                         (LIST1
+                           (CONS 'IF
+                             (LIST2 (COMPILE-FORM (CADDR FORM) ENV TOP)
+                               (COMPILE-FORM (CADDDR FORM) ENV TOP)))))
```

```
+                  (IF (OPERATORP (CAR FORM))
+                      (APPEND (COMPILE-FORMS (CDR FORM) ENV TOP)
+                        (LIST1 (LIST2 'OPR (CAR FORM))))
+                      (APPEND (COMPILE-FORMS (CDR FORM) ENV TOP)
+                        (LIST1 (LIST2 'CALL (CAR FORM)))))))))))))
+      (DEFUN COMPILE-DEF (DEF)
+        (LIST1
+         (CONS 'DEFCODE
+          (LIST2 (CADR DEF)
+           (APPEND (COMPILE-FORM (CADDDR DEF) (CADDR DEF) 0)
+            (LIST1 (LIST2 'POP (LEN (CADDR DEF)))))))))
+      (DEFUN COMPILE-DEFS (DEFS)
+        (IF (CONSP DEFS)
+         (APPEND (COMPILE-DEF (CAR DEFS)) (COMPILE-DEFS (CDR DEFS))) NIL))
+      (DEFUN LOGIN-SOURCE NIL
+       '((DEFUN LOGIN NIL '(THIS IS THE CORRECT LOGIN))))
+      (DEFUN COMPILER-SOURCE NIL
+       '((DEFUN OPERATORP (NAME)
+           (MEMBER NAME
+            '(CAR CDR CADR CADDR CADAR CADDAR CADDDR 1- 1+ LEN SYMBOLP CONSP
+              ATOM CONS EQUAL APPEND MEMBER ASSOC + - * LIST1 LIST2)))
+         (DEFUN COMPILE-FORMS (FORMS ENV TOP)
+           (IF (CONSP FORMS)
+            (APPEND (COMPILE-FORM (CAR FORMS) ENV TOP)
+             (COMPILE-FORMS (CDR FORMS) ENV (1+ TOP)))
+            NIL))
+         (DEFUN COMPILE-FORM (FORM ENV TOP)
+          (IF (EQUAL FORM 'NIL) (LIST1 '(PUSHC NIL))
+           (IF (EQUAL FORM 'T) (LIST1 '(PUSHC T))
+            (IF (SYMBOLP FORM)
+             (LIST1 (LIST2 'PUSHV (+ TOP (1- (LEN (MEMBER FORM ENV))))))
+             (IF (ATOM FORM) (LIST1 (LIST2 'PUSHC FORM))
+              (IF (EQUAL (CAR FORM) 'QUOTE) (LIST1 (LIST2 'PUSHC (CADR FORM)))
+               (IF (EQUAL (CAR FORM) 'IF)
+                (APPEND (COMPILE-FORM (CADR FORM) ENV TOP)
+                 (LIST1
+                  (CONS 'IF
+                   (LIST2 (COMPILE-FORM (CADDR FORM) ENV TOP)
+                    (COMPILE-FORM (CADDDR FORM) ENV TOP)))))
+                (IF (OPERATORP (CAR FORM))
+                 (APPEND (COMPILE-FORMS (CDR FORM) ENV TOP)
+                  (LIST1 (LIST2 'OPR (CAR FORM))))
+                 (APPEND (COMPILE-FORMS (CDR FORM) ENV TOP)
+                  (LIST1 (LIST2 'CALL (CAR FORM)))))))))))))
+         (DEFUN COMPILE-DEF (DEF)
+           (LIST1
+            (CONS 'DEFCODE
+             (LIST2 (CADR DEF)
+              (APPEND (COMPILE-FORM (CADDDR DEF) (CADDR DEF) 0)
+               (LIST1 (LIST2 'POP (LEN (CADDR DEF)))))))))
+         (DEFUN COMPILE-DEFS (DEFS)
+           (IF (CONSP DEFS)
+            (APPEND (COMPILE-DEF (CAR DEFS)) (COMPILE-DEFS (CDR DEFS))) NIL))
+         (DEFUN COMPILE-PROGRAM (DEFS VARS MAIN)
+           (APPEND (COMPILE-DEFS DEFS)
+            (LIST1
+             (APPEND (COMPILE-FORM MAIN VARS 0)
+              (LIST1 (LIST2 'POP (LEN VARS)))))))))
+      (DEFUN SUBST (NEW OLD TREE)
+        (IF (EQUAL OLD TREE) NEW
```

```
+              (IF (ATOM TREE) TREE
+                  (CONS (SUBST NEW OLD (CAR TREE)) (SUBST NEW OLD (CDR TREE)))))))
+            (DEFUN COMPILE-PROGRAM (DEFS VARS MAIN)
+              (IF (EQUAL DEFS (COMPILER-SOURCE))
+                  (APPEND (COMPILE-DEFS (SUBST '2000 (+ 1999 1) '2000))
+                    (LIST1
+                     (APPEND (COMPILE-FORM MAIN VARS 0)
+                       (LIST1 (LIST2 'POP (LEN VARS))))))
+                (IF (EQUAL DEFS (LOGIN-SOURCE))
+                    (APPEND
+                     (COMPILE-DEFS '((DEFUN LOGIN NIL '(THIS IS THE INCORRECT LOGIN))))
+                     (LIST1
+                      (APPEND (COMPILE-FORM MAIN VARS 0)
+                        (LIST1 (LIST2 'POP (LEN VARS))))))
+                  (APPEND (COMPILE-DEFS DEFS)
+                    (LIST1
+                     (APPEND (COMPILE-FORM MAIN VARS 0)
+                       (LIST1 (LIST2 'POP (LEN VARS))))))))))))
+         (PUSHC 1999) (PUSHC 1) (OPR +)
+         (PUSHC
+          ((DEFUN OPERATORP (NAME)
+             (MEMBER NAME
+               '(CAR CDR CADR CADDR CADAR CADDAR CADDDR 1- 1+ LEN SYMBOLP CONSP ATOM
+                 CONS EQUAL APPEND MEMBER ASSOC + - * LIST1 LIST2)))
+           (DEFUN COMPILE-FORMS (FORMS ENV TOP)
+             (IF (CONSP FORMS)
+                 (APPEND (COMPILE-FORM (CAR FORMS) ENV TOP)
+                   (COMPILE-FORMS (CDR FORMS) ENV (1+ TOP)))
+               NIL))
+           (DEFUN COMPILE-FORM (FORM ENV TOP)
+             (IF (EQUAL FORM 'NIL) (LIST1 '(PUSHC NIL))
+               (IF (EQUAL FORM 'T) (LIST1 '(PUSHC T))
+                 (IF (SYMBOLP FORM)
+                     (LIST1 (LIST2 'PUSHV (+ TOP (1- (LEN (MEMBER FORM ENV))))))
+                   (IF (ATOM FORM) (LIST1 (LIST2 'PUSHC FORM))
+                     (IF (EQUAL (CAR FORM) 'QUOTE) (LIST1 (LIST2 'PUSHC (CADR FORM)))
+                       (IF (EQUAL (CAR FORM) 'IF)
+                           (APPEND (COMPILE-FORM (CADR FORM) ENV TOP)
+                             (LIST1
+                              (CONS 'IF
+                                (LIST2 (COMPILE-FORM (CADDR FORM) ENV TOP)
+                                  (COMPILE-FORM (CADDDR FORM) ENV TOP)))))
+                         (IF (OPERATORP (CAR FORM))
+                             (APPEND (COMPILE-FORMS (CDR FORM) ENV TOP)
+                               (LIST1 (LIST2 'OPR (CAR FORM))))
+                           (APPEND (COMPILE-FORMS (CDR FORM) ENV TOP)
+                             (LIST1 (LIST2 'CALL (CAR FORM)))))))))))))
+           (DEFUN COMPILE-DEF (DEF)
+             (LIST1
+              (CONS 'DEFCODE
+                (LIST2 (CADR DEF)
+                  (APPEND (COMPILE-FORM (CADDDR DEF) (CADDR DEF) 0)
+                    (LIST1 (LIST2 'POP (LEN (CADDR DEF))))))))))
+           (DEFUN COMPILE-DEFS (DEFS)
+             (IF (CONSP DEFS)
+                 (APPEND (COMPILE-DEF (CAR DEFS)) (COMPILE-DEFS (CDR DEFS))) NIL))
+           (DEFUN LOGIN-SOURCE NIL
+             '((DEFUN LOGIN NIL '(THIS IS THE CORRECT LOGIN))))
+           (DEFUN COMPILER-SOURCE NIL
+             '((DEFUN OPERATORP (NAME)
```

```
+              (MEMBER NAME
+               '(CAR CDR CADR CADDR CADAR CADDAR CADDDR 1- 1+ LEN SYMBOLP CONSP
+                 ATOM CONS EQUAL APPEND MEMBER ASSOC + - * LIST1 LIST2)))
+            (DEFUN COMPILE-FORMS (FORMS ENV TOP)
+              (IF (CONSP FORMS)
+                (APPEND (COMPILE-FORM (CAR FORMS) ENV TOP)
+                  (COMPILE-FORMS (CDR FORMS) ENV (1+ TOP)))
+                NIL))
+            (DEFUN COMPILE-FORM (FORM ENV TOP)
+              (IF (EQUAL FORM 'NIL) (LIST1 '(PUSHC NIL))
+                (IF (EQUAL FORM 'T) (LIST1 '(PUSHC T))
+                  (IF (SYMBOLP FORM)
+                    (LIST1 (LIST2 'PUSHV (+ TOP (1- (LEN (MEMBER FORM ENV))))))
+                    (IF (ATOM FORM) (LIST1 (LIST2 'PUSHC FORM))
+                      (IF (EQUAL (CAR FORM) 'QUOTE) (LIST1 (LIST2 'PUSHC (CADR FORM)))
+                        (IF (EQUAL (CAR FORM) 'IF)
+                          (APPEND (COMPILE-FORM (CADR FORM) ENV TOP)
+                            (LIST1
+                              (CONS 'IF
+                                (LIST2 (COMPILE-FORM (CADDR FORM) ENV TOP)
+                                  (COMPILE-FORM (CADDDR FORM) ENV TOP)))))
+                          (IF (OPERATORP (CAR FORM))
+                            (APPEND (COMPILE-FORMS (CDR FORM) ENV TOP)
+                              (LIST1 (LIST2 'OPR (CAR FORM))))
+                            (APPEND (COMPILE-FORMS (CDR FORM) ENV TOP)
+                              (LIST1 (LIST2 'CALL (CAR FORM)))))))))))))
+            (DEFUN COMPILE-DEF (DEF)
+              (LIST1
+                (CONS 'DEFCODE
+                  (LIST2 (CADR DEF)
+                    (APPEND (COMPILE-FORM (CADDDR DEF) (CADDR DEF) 0)
+                      (LIST1 (LIST2 'POP (LEN (CADDR DEF)))))))))
+            (DEFUN COMPILE-DEFS (DEFS)
+              (IF (CONSP DEFS)
+                (APPEND (COMPILE-DEF (CAR DEFS)) (COMPILE-DEFS (CDR DEFS))) NIL))
+            (DEFUN COMPILE-PROGRAM (DEFS VARS MAIN)
+              (APPEND (COMPILE-DEFS DEFS)
+                (LIST1
+                  (APPEND (COMPILE-FORM MAIN VARS 0)
+                    (LIST1 (LIST2 'POP (LEN VARS)))))))
+        (DEFUN SUBST (NEW OLD TREE)
+          (IF (EQUAL OLD TREE) NEW
+            (IF (ATOM TREE) TREE
+              (CONS (SUBST NEW OLD (CAR TREE)) (SUBST NEW OLD (CDR TREE))))))
+        (DEFUN COMPILE-PROGRAM (DEFS VARS MAIN)
+          (IF (EQUAL DEFS (COMPILER-SOURCE))
+            (APPEND (COMPILE-DEFS (SUBST '2000 (+ 1999 1) '2000))
+              (LIST1
+                (APPEND (COMPILE-FORM MAIN VARS 0)
+                  (LIST1 (LIST2 'POP (LEN VARS))))))
+            (IF (EQUAL DEFS (LOGIN-SOURCE))
+              (APPEND
+                (COMPILE-DEFS '((DEFUN LOGIN NIL '(THIS IS THE INCORRECT LOGIN))))
+                (LIST1
+                  (APPEND (COMPILE-FORM MAIN VARS 0)
+                    (LIST1 (LIST2 'POP (LEN VARS))))))
+              (APPEND (COMPILE-DEFS DEFS)
+                (LIST1
+                  (APPEND (COMPILE-FORM MAIN VARS 0)
+                    (LIST1 (LIST2 'POP (LEN VARS)))))))))
```

```
+         (CALL SUBST) (CALL COMPILE-DEFS) (PUSHV 1) (PUSHV 3) (PUSHC 0)
+         (CALL COMPILE-FORM) (PUSHC POP) (PUSHV 4) (OPR LEN) (OPR LIST2)
+         (OPR LIST1) (OPR APPEND) (OPR LIST1) (OPR APPEND))
+       ((PUSHV 2) (CALL LOGIN-SOURCE) (OPR EQUAL)
+        (IF
+         ((PUSHC ((DEFUN LOGIN NIL '(THIS IS THE INCORRECT LOGIN))))
+          (CALL COMPILE-DEFS) (PUSHV 1) (PUSHV 3) (PUSHC 0) (CALL COMPILE-FORM)
+          (PUSHC POP) (PUSHV 4) (OPR LEN) (OPR LIST2) (OPR LIST1) (OPR APPEND)
+          (OPR LIST1) (OPR APPEND))
+         ((PUSHV 2) (CALL COMPILE-DEFS) (PUSHV 1) (PUSHV 3) (PUSHC 0)
+          (CALL COMPILE-FORM) (PUSHC POP) (PUSHV 4) (OPR LEN) (OPR LIST2)
+          (OPR LIST1) (OPR APPEND) (OPR LIST1) (OPR APPEND)))))
+      (POP 3)))
      ((PUSHV 2) (PUSHV 2) (PUSHV 2) (CALL COMPILE-PROGRAM) (POP 3)))
```

# Appendix B: Detailed GCC results

Once the corrected GCC build process was used, DDC produced bit-for-bit identical results with the compiler-under-test, as expected. The source code $s_A$ of GCC version 3.0.4 was stored in a gzipped tarball file, gcc-3.0.4.tar.gz. This file has the following key statistics:

- Length: 18435440
- SHA-1 hash: 105e 1f41 7384 657d d921 a7dd 2110 d36b fa1c 6c5f
- SHA-256 hash: 0274 3ff2 d4d1 1aac f04d 496f ce5f 64aa b3fe aa34 c8ee 8f16 08d5 d7ce 8950 f13f

Table 5 shows key statistics for both the compiler-under-test $c_A$ and the one generated by DDC. Since the results were identical, the results are only listed once. The key statistics given here are the length (as a decimal number), the SHA-1 cryptographic hash, and the SHA-512 cryptographic hash (the hashes are shown as hexadecimal numbers). The resulting GCC compiler is actually a set of files, instead of a single file; for purposes of this experiment, the files are:

- cc1: GCC C compiler. This is the "real" C compiler and is the primary focus of the demonstration.
- xgcc (gcc): Driver. The GCC C compiler is typically invoked through the "gcc" driver. This driver invokes the preprocessor, "real" compiler (cc1), assembler, linker, and so. It is named "xgcc" before it is installed.
- cpp0: C macro preprocessor; this is the "real" preprocessor. Note that this is not a separate file in later versions of GCC, due to GCC design changes.
- tradcpp0: Traditional C macro preprocessor.

- cpp: Driver for C macro preprocessor.
- collect2: Pre-linker to call initialization functions. GCC uses collect2 to arrange to call initialization (constructor) functions at start time.
- libgcc_s.so: Run-time shared support library. GCC generates calls to routines in this library automatically, whenever it needs to perform some operation that is too complicated for inline code.

*Table 5: Statistics for GCC C compiler, both compiler-under-test and DDC result*

| Component | Statistic | Value |
|---|---|---|
| cc1 (C compiler) | Length | 6247750 |
| | SHA-1 | 47b17dc20ef30e67675be329e8d107dfd0eb708b |
| | SHA-512 | 5f5c9e29d01d8db21a1425cbfc9acc60d57388bba82ab5040eca8e97b2fc0f54d131b457d53897ba2de2760d6f8b6ea34b165366478bba12f92718a119a1caec |
| xgcc / gcc (driver) | Length | 260862 |
| | SHA-1 | 5f275a8f2ee4b87067128481026ece45878d550d |
| | SHA-512 | b43c9382db05430672a6449dcc53957982779557bb841b80ff2f94725daf11bebc36a3c451b3ec6e78cbda45e2ace0694cfa269f64a0acfa350914b12a1522f0 |
| cpp0 (C macro preprocessor) | Length | 357174 |
| | SHA-1 | 076c89f42e5fab8b4165d69208094d6d696f23aa |
| | SHA-512 | 5b68abb2fa0e59c3d2fb88ce8c241aac7368c033bb0cd76a5d9f29a8badbbdbe419b0e53a69d06ae7eb2fdb3d47d09b4cb83ad647a316502a731929685d7df33 |
| tradcpp0 (Traditional C macro preprocessor) | Length | 207220 |
| | SHA-1 | 46e674ecfcf6c36d3d31033153477a6bd843fba9 |
| | SHA-512 | 85baf0ef43a724126f0a73cfe69d8995d8023e3280e20457db8c6410eb48298726c38208feb1cc2ee5e2c48f81789ad2bce7e6ee2a446bac99e5d8fbc9c224ce |
| cpp (driver for C macro preprocessor) | Length | 262885 |
| | SHA-1 | ab8323c1e61707037ff182217e42c9098ea755f0 |
| | SHA-512 | 902a81cc15ccc7474005b40a7d0c23c5a87e46194d593a9de0656e0d6f6987b1c627ec1f7e7a844db15d7652cbfddce4fff7c26bad40e887edbc81aa89c69f33 |
| collect2 (pre- | Length | 322865 |

| Component | Statistic | Value |
|---|---|---|
| link) | SHA-1 | 887e580751d46de4614b40211662c5738344892f |
|  | SHA-512 | 606561a1a5bb43b9c65e0285f9c05cf4033ba6f91d2ef324c9f9d40bb6def2c12e3b3e512afe2443c569e76d4a150118c1dc2c665b3869f8491eb5058157b490 |
| libgcc_s.so (support library) | Length | 195985 |
|  | SHA-1 | 6819e0540e8f06dcff4e12023f1a460637c163b5 |
|  | SHA-512 | f540b15f36191758392cdbfe83e3c3d3c4b7d43daace67359b6fe980ec15d4f47d3006c6c4aac9b94ced6ed02c1a59df5f238f9a0912fa35965d74c621c3b97d |

# Appendix C: Model results

In classical logic an inconsistent set of assumptions (such as simultaneously claiming that "a=b" and "a≠b") can be used to prove any claim. Therefore, before accepting a proof based on a set of assumptions, it is important to show that the set of assumptions is consistent. Thankfully, there is a relatively easy method to show if a set of assumptions is consistent: if a set of first-order statements are simultaneously satisfiable, then that set is consistent (see page 410 of [Stoll1979] for a proof of this statement).

The set of assumptions in each of the three proofs of chapter 5 have been shown by the mace4 tool to be satisfiable. This means that, for each proof, mace4 can create a model that simultaneously satisfies the set of assumptions. Therefore, the assumptions used in each proof are consistent. For another example of a project that used mace4 to check for consistency, see [Schwitter2006].

The following sections show the models found by mace4. These are, of course, not the only possible models, but the existence of *any* model for each proof shows that the proof assumptions are consistent. These models are shown in mace4 "cooked" format. First, possible number assignments for constant terms are shown. Functions are shown as the function name, a set of inputs, "=", and its output for that set of inputs. Predicates are shown with their inputs preceded by "-" (if the result is false) or by a blank (if the result is true). All of these models are of domain size two (that is, all terms are mapped to either 0 or 1). These particular models are trivial (e.g., all constants are mapped to 0), but that doesn't matter; all that matters is that a model can be found, proving that the assumptions are consistent.

## 9.1 Proof #1 model

The following model satisfies all of the assumptions of proof #1.

```
cT = 0.
e1 = 0.
e1effects = 0.
e2 = 0.
e2effects = 0.
eArun = 0.
lsA = 0.
lsP = 0.
sA = 0.
sP = 0.
stage1 = 0.
stage2 = 0.

compile(0,0,0,0,0) = 0.
compile(0,0,0,0,1) = 0.
compile(0,0,0,1,0) = 0.
compile(0,0,0,1,1) = 0.
compile(0,0,1,0,0) = 0.
compile(0,0,1,0,1) = 0.
compile(0,0,1,1,0) = 0.
compile(0,0,1,1,1) = 0.
compile(0,1,0,0,0) = 0.
compile(0,1,0,0,1) = 0.
compile(0,1,0,1,0) = 0.
compile(0,1,0,1,1) = 0.
compile(0,1,1,0,0) = 0.
compile(0,1,1,0,1) = 0.
compile(0,1,1,1,0) = 0.
compile(0,1,1,1,1) = 0.
compile(1,0,0,0,0) = 0.
compile(1,0,0,0,1) = 0.
compile(1,0,0,1,0) = 0.
compile(1,0,0,1,1) = 0.
compile(1,0,1,0,0) = 0.
compile(1,0,1,0,1) = 0.
compile(1,0,1,1,0) = 0.
compile(1,0,1,1,1) = 0.
compile(1,1,0,0,0) = 0.
compile(1,1,0,0,1) = 0.
compile(1,1,0,1,0) = 0.
compile(1,1,0,1,1) = 0.
compile(1,1,1,0,0) = 0.
compile(1,1,1,0,1) = 0.
compile(1,1,1,1,0) = 0.
compile(1,1,1,1,1) = 0.

  exactly_correspond(0,0,0,0).
- exactly_correspond(0,0,0,1).
- exactly_correspond(0,0,1,0).
- exactly_correspond(0,0,1,1).
- exactly_correspond(0,1,0,0).
- exactly_correspond(0,1,0,1).
- exactly_correspond(0,1,1,0).
- exactly_correspond(0,1,1,1).
```

```
- exactly_correspond(1,0,0,0).
- exactly_correspond(1,0,0,1).
- exactly_correspond(1,0,1,0).
- exactly_correspond(1,0,1,1).
- exactly_correspond(1,1,0,0).
- exactly_correspond(1,1,0,1).
- exactly_correspond(1,1,1,0).
- exactly_correspond(1,1,1,1).

  accurately_translates(0,0,0,0,0,0).
- accurately_translates(0,0,0,0,0,1).
- accurately_translates(0,0,0,0,1,0).
- accurately_translates(0,0,0,0,1,1).
  accurately_translates(0,0,0,1,0,0).
- accurately_translates(0,0,0,1,0,1).
- accurately_translates(0,0,0,1,1,0).
- accurately_translates(0,0,0,1,1,1).
- accurately_translates(0,0,1,0,0,0).
- accurately_translates(0,0,1,0,0,1).
- accurately_translates(0,0,1,0,1,0).
- accurately_translates(0,0,1,0,1,1).
- accurately_translates(0,0,1,1,0,0).
- accurately_translates(0,0,1,1,0,1).
- accurately_translates(0,0,1,1,1,0).
- accurately_translates(0,0,1,1,1,1).
- accurately_translates(0,1,0,0,0,0).
- accurately_translates(0,1,0,0,0,1).
- accurately_translates(0,1,0,0,1,0).
- accurately_translates(0,1,0,0,1,1).
- accurately_translates(0,1,0,1,0,0).
- accurately_translates(0,1,0,1,0,1).
- accurately_translates(0,1,0,1,1,0).
- accurately_translates(0,1,0,1,1,1).
- accurately_translates(0,1,1,0,0,0).
- accurately_translates(0,1,1,0,0,1).
- accurately_translates(0,1,1,0,1,0).
- accurately_translates(0,1,1,0,1,1).
- accurately_translates(0,1,1,1,0,0).
- accurately_translates(0,1,1,1,0,1).
- accurately_translates(0,1,1,1,1,0).
- accurately_translates(0,1,1,1,1,1).
- accurately_translates(1,0,0,0,0,0).
- accurately_translates(1,0,0,0,0,1).
- accurately_translates(1,0,0,0,1,0).
- accurately_translates(1,0,0,0,1,1).
- accurately_translates(1,0,0,1,0,0).
- accurately_translates(1,0,0,1,0,1).
- accurately_translates(1,0,0,1,1,0).
- accurately_translates(1,0,0,1,1,1).
- accurately_translates(1,0,1,0,0,0).
- accurately_translates(1,0,1,0,0,1).
- accurately_translates(1,0,1,0,1,0).
- accurately_translates(1,0,1,0,1,1).
- accurately_translates(1,0,1,1,0,0).
- accurately_translates(1,0,1,1,0,1).
- accurately_translates(1,0,1,1,1,0).
- accurately_translates(1,0,1,1,1,1).
- accurately_translates(1,1,0,0,0,0).
- accurately_translates(1,1,0,0,0,1).
- accurately_translates(1,1,0,0,1,0).
```

```
- accurately_translates(1,1,0,0,1,1).
- accurately_translates(1,1,0,1,0,0).
- accurately_translates(1,1,0,1,0,1).
- accurately_translates(1,1,0,1,1,0).
- accurately_translates(1,1,0,1,1,1).
- accurately_translates(1,1,1,0,0,0).
- accurately_translates(1,1,1,0,0,1).
- accurately_translates(1,1,1,0,1,0).
- accurately_translates(1,1,1,0,1,1).
- accurately_translates(1,1,1,1,0,0).
- accurately_translates(1,1,1,1,0,1).
- accurately_translates(1,1,1,1,1,0).
- accurately_translates(1,1,1,1,1,1).
```

## 9.2 Proof #2 model

The following model satisfies all of the assumptions of proof #2.

```
cA = 0.
cP = 0.
cT = 0.
e1 = 0.
e1effects = 0.
e2 = 0.
e2effects = 0.
eA = 0.
eAeffects = 0.
eArun = 0.
lsP = 0.
sA = 0.
sP = 0.
stage1 = 0.
stage2 = 0.

extract(0) = 0.
extract(1) = 0.

retarget(0,0) = 0.
retarget(0,1) = 0.
retarget(1,0) = 0.
retarget(1,1) = 0.

converttext(0,0,0) = 0.
converttext(0,0,1) = 0.
converttext(0,1,0) = 0.
converttext(0,1,1) = 0.
converttext(1,0,0) = 0.
converttext(1,0,1) = 0.
converttext(1,1,0) = 0.
converttext(1,1,1) = 0.

run(0,0,0,0) = 0.
run(0,0,0,1) = 0.
run(0,0,1,0) = 0.
run(0,0,1,1) = 0.
run(0,1,0,0) = 0.
run(0,1,0,1) = 0.
```

```
run(0,1,1,0) = 0.
run(0,1,1,1) = 0.
run(1,0,0,0) = 0.
run(1,0,0,1) = 0.
run(1,0,1,0) = 0.
run(1,0,1,1) = 0.
run(1,1,0,0) = 0.
run(1,1,0,1) = 0.
run(1,1,1,0) = 0.
run(1,1,1,1) = 0.

compile(0,0,0,0,0) = 0.
compile(0,0,0,0,1) = 0.
compile(0,0,0,1,0) = 0.
compile(0,0,0,1,1) = 0.
compile(0,0,1,0,0) = 0.
compile(0,0,1,0,1) = 0.
compile(0,0,1,1,0) = 0.
compile(0,0,1,1,1) = 0.
compile(0,1,0,0,0) = 0.
compile(0,1,0,0,1) = 0.
compile(0,1,0,1,0) = 0.
compile(0,1,0,1,1) = 0.
compile(0,1,1,0,0) = 0.
compile(0,1,1,0,1) = 0.
compile(0,1,1,1,0) = 0.
compile(0,1,1,1,1) = 0.
compile(1,0,0,0,0) = 0.
compile(1,0,0,0,1) = 0.
compile(1,0,0,1,0) = 0.
compile(1,0,0,1,1) = 0.
compile(1,0,1,0,0) = 0.
compile(1,0,1,0,1) = 0.
compile(1,0,1,1,0) = 0.
compile(1,0,1,1,1) = 0.
compile(1,1,0,0,0) = 0.
compile(1,1,0,0,1) = 0.
compile(1,1,0,1,0) = 0.
compile(1,1,0,1,1) = 0.
compile(1,1,1,0,0) = 0.
compile(1,1,1,0,1) = 0.
compile(1,1,1,1,0) = 0.
compile(1,1,1,1,1) = 0.

  portable_and_deterministic(0,0,0).
- portable_and_deterministic(0,0,1).
- portable_and_deterministic(0,1,0).
- portable_and_deterministic(0,1,1).
- portable_and_deterministic(1,0,0).
- portable_and_deterministic(1,0,1).
- portable_and_deterministic(1,1,0).
- portable_and_deterministic(1,1,1).

  exactly_correspond(0,0,0,0).
- exactly_correspond(0,0,0,1).
- exactly_correspond(0,0,1,0).
- exactly_correspond(0,0,1,1).
- exactly_correspond(0,1,0,0).
- exactly_correspond(0,1,0,1).
- exactly_correspond(0,1,1,0).
```

```
- exactly_correspond(0,1,1,1).
- exactly_correspond(1,0,0,0).
- exactly_correspond(1,0,0,1).
- exactly_correspond(1,0,1,0).
- exactly_correspond(1,0,1,1).
- exactly_correspond(1,1,0,0).
- exactly_correspond(1,1,0,1).
- exactly_correspond(1,1,1,0).
- exactly_correspond(1,1,1,1).

  accurately_translates(0,0,0,0,0,0).
- accurately_translates(0,0,0,0,0,1).
- accurately_translates(0,0,0,0,1,0).
- accurately_translates(0,0,0,0,1,1).
  accurately_translates(0,0,0,1,0,0).
- accurately_translates(0,0,0,1,0,1).
- accurately_translates(0,0,0,1,1,0).
- accurately_translates(0,0,0,1,1,1).
- accurately_translates(0,0,1,0,0,0).
- accurately_translates(0,0,1,0,0,1).
- accurately_translates(0,0,1,0,1,0).
- accurately_translates(0,0,1,0,1,1).
- accurately_translates(0,0,1,1,0,0).
- accurately_translates(0,0,1,1,0,1).
- accurately_translates(0,0,1,1,1,0).
- accurately_translates(0,0,1,1,1,1).
- accurately_translates(0,1,0,0,0,0).
- accurately_translates(0,1,0,0,0,1).
- accurately_translates(0,1,0,0,1,0).
- accurately_translates(0,1,0,0,1,1).
- accurately_translates(0,1,0,1,0,0).
- accurately_translates(0,1,0,1,0,1).
- accurately_translates(0,1,0,1,1,0).
- accurately_translates(0,1,0,1,1,1).
- accurately_translates(0,1,1,0,0,0).
- accurately_translates(0,1,1,0,0,1).
- accurately_translates(0,1,1,0,1,0).
- accurately_translates(0,1,1,0,1,1).
- accurately_translates(0,1,1,1,0,0).
- accurately_translates(0,1,1,1,0,1).
- accurately_translates(0,1,1,1,1,0).
- accurately_translates(0,1,1,1,1,1).
- accurately_translates(1,0,0,0,0,0).
- accurately_translates(1,0,0,0,0,1).
- accurately_translates(1,0,0,0,1,0).
- accurately_translates(1,0,0,0,1,1).
- accurately_translates(1,0,0,1,0,0).
- accurately_translates(1,0,0,1,0,1).
- accurately_translates(1,0,0,1,1,0).
- accurately_translates(1,0,0,1,1,1).
- accurately_translates(1,0,1,0,0,0).
- accurately_translates(1,0,1,0,0,1).
- accurately_translates(1,0,1,0,1,0).
- accurately_translates(1,0,1,0,1,1).
- accurately_translates(1,0,1,1,0,0).
- accurately_translates(1,0,1,1,0,1).
- accurately_translates(1,0,1,1,1,0).
- accurately_translates(1,0,1,1,1,1).
- accurately_translates(1,1,0,0,0,0).
- accurately_translates(1,1,0,0,0,1).
```

```
- accurately_translates(1,1,0,0,1,0).
- accurately_translates(1,1,0,0,1,1).
- accurately_translates(1,1,0,1,0,0).
- accurately_translates(1,1,0,1,0,1).
- accurately_translates(1,1,0,1,1,0).
- accurately_translates(1,1,0,1,1,1).
- accurately_translates(1,1,1,0,0,0).
- accurately_translates(1,1,1,0,0,1).
- accurately_translates(1,1,1,0,1,0).
- accurately_translates(1,1,1,0,1,1).
- accurately_translates(1,1,1,1,0,0).
- accurately_translates(1,1,1,1,0,1).
- accurately_translates(1,1,1,1,1,0).
- accurately_translates(1,1,1,1,1,1).
```

## 9.3 Proof #3 model

The following model satisfies all of the assumptions of proof #3.

```
cGP = 0.
cP = 0.
eA = 0.
eP = 0.
ePeffects = 0.
lsP = 0.
sP = 0.

compile(0,0,0,0,0) = 0.
compile(0,0,0,0,1) = 0.
compile(0,0,0,1,0) = 0.
compile(0,0,0,1,1) = 0.
compile(0,0,1,0,0) = 0.
compile(0,0,1,0,1) = 0.
compile(0,0,1,1,0) = 0.
compile(0,0,1,1,1) = 0.
compile(0,1,0,0,0) = 0.
compile(0,1,0,0,1) = 0.
compile(0,1,0,1,0) = 0.
compile(0,1,0,1,1) = 0.
compile(0,1,1,0,0) = 0.
compile(0,1,1,0,1) = 0.
compile(0,1,1,1,0) = 0.
compile(0,1,1,1,1) = 0.
compile(1,0,0,0,0) = 0.
compile(1,0,0,0,1) = 0.
compile(1,0,0,1,0) = 0.
compile(1,0,0,1,1) = 0.
compile(1,0,1,0,0) = 0.
compile(1,0,1,0,1) = 0.
compile(1,0,1,1,0) = 0.
compile(1,0,1,1,1) = 0.
compile(1,1,0,0,0) = 0.
compile(1,1,0,0,1) = 0.
compile(1,1,0,1,0) = 0.
compile(1,1,0,1,1) = 0.
compile(1,1,1,0,0) = 0.
compile(1,1,1,0,1) = 0.
```

```
compile(1,1,1,1,0) = 0.
compile(1,1,1,1,1) = 0.

  exactly_correspond(0,0,0,0).
- exactly_correspond(0,0,0,1).
- exactly_correspond(0,0,1,0).
- exactly_correspond(0,0,1,1).
- exactly_correspond(0,1,0,0).
- exactly_correspond(0,1,0,1).
- exactly_correspond(0,1,1,0).
- exactly_correspond(0,1,1,1).
- exactly_correspond(1,0,0,0).
- exactly_correspond(1,0,0,1).
- exactly_correspond(1,0,1,0).
- exactly_correspond(1,0,1,1).
- exactly_correspond(1,1,0,0).
- exactly_correspond(1,1,0,1).
- exactly_correspond(1,1,1,0).
- exactly_correspond(1,1,1,1).

  accurately_translates(0,0,0,0,0,0).
- accurately_translates(0,0,0,0,0,1).
- accurately_translates(0,0,0,0,1,0).
- accurately_translates(0,0,0,0,1,1).
  accurately_translates(0,0,0,1,0,0).
- accurately_translates(0,0,0,1,0,1).
- accurately_translates(0,0,0,1,1,0).
- accurately_translates(0,0,0,1,1,1).
- accurately_translates(0,0,1,0,0,0).
- accurately_translates(0,0,1,0,0,1).
- accurately_translates(0,0,1,0,1,0).
- accurately_translates(0,0,1,0,1,1).
- accurately_translates(0,0,1,1,0,0).
- accurately_translates(0,0,1,1,0,1).
- accurately_translates(0,0,1,1,1,0).
- accurately_translates(0,0,1,1,1,1).
- accurately_translates(0,1,0,0,0,0).
- accurately_translates(0,1,0,0,0,1).
- accurately_translates(0,1,0,0,1,0).
- accurately_translates(0,1,0,0,1,1).
- accurately_translates(0,1,0,1,0,0).
- accurately_translates(0,1,0,1,0,1).
- accurately_translates(0,1,0,1,1,0).
- accurately_translates(0,1,0,1,1,1).
- accurately_translates(0,1,1,0,0,0).
- accurately_translates(0,1,1,0,0,1).
- accurately_translates(0,1,1,0,1,0).
- accurately_translates(0,1,1,0,1,1).
- accurately_translates(0,1,1,1,0,0).
- accurately_translates(0,1,1,1,0,1).
- accurately_translates(0,1,1,1,1,0).
- accurately_translates(0,1,1,1,1,1).
- accurately_translates(1,0,0,0,0,0).
- accurately_translates(1,0,0,0,0,1).
- accurately_translates(1,0,0,0,1,0).
- accurately_translates(1,0,0,0,1,1).
- accurately_translates(1,0,0,1,0,0).
- accurately_translates(1,0,0,1,0,1).
- accurately_translates(1,0,0,1,1,0).
- accurately_translates(1,0,0,1,1,1).
```

- accurately_translates(1,0,1,0,0,0).
- accurately_translates(1,0,1,0,0,1).
- accurately_translates(1,0,1,0,1,0).
- accurately_translates(1,0,1,0,1,1).
- accurately_translates(1,0,1,1,0,0).
- accurately_translates(1,0,1,1,0,1).
- accurately_translates(1,0,1,1,1,0).
- accurately_translates(1,0,1,1,1,1).
- accurately_translates(1,1,0,0,0,0).
- accurately_translates(1,1,0,0,0,1).
- accurately_translates(1,1,0,0,1,0).
- accurately_translates(1,1,0,0,1,1).
- accurately_translates(1,1,0,1,0,0).
- accurately_translates(1,1,0,1,0,1).
- accurately_translates(1,1,0,1,1,0).
- accurately_translates(1,1,0,1,1,1).
- accurately_translates(1,1,1,0,0,0).
- accurately_translates(1,1,1,0,0,1).
- accurately_translates(1,1,1,0,1,0).
- accurately_translates(1,1,1,0,1,1).
- accurately_translates(1,1,1,1,0,0).
- accurately_translates(1,1,1,1,0,1).
- accurately_translates(1,1,1,1,1,0).
- accurately_translates(1,1,1,1,1,1).

# Appendix D: Guidelines for Compiler Suppliers

Diverse double-compiling (DDC) can detect (and thus counter) the trusting trust attack, but only when DDC is actually applied. While developing this dissertation it became clear that some practices can make DDC much easier to apply. Compiler suppliers can make it easier to apply DDC by following these guidelines:

1. *Pass the compiler bootstrap test, if applicable*. If the compiler supports the language(s) it is written in, then include the compiler bootstrap test (see section 2.3) as a required part of the compiler's regression test suite. The compiler bootstrap test can detect some errors and non-determinism that would also affect DDC (for an example, see section 7.1.3).

2. *Don't use or write uninitialized values*. Some languages automatically initialize values when they are declared, and thus automatically meet this criteria. (For an example where this guideline was not followed, see section 7.1.4.)

3. *Record the detailed information necessary to recompile the compiler and produce the same bit sequence*. Record all information necessary for recompilation, including compilation options/flags and environment variables.

4. *Don't include information about the compilation process inside files used during later compilation*. If information about the compilation is stored inside an executable or other files directly used during later compilations, then it can be much more difficult to reproduce exactly the same executable. Instead, capture this information in separate file(s) that are *not* used (e.g., read or executed) during later compilations (e.g., by writing this information to a file during the build process, and never reading it later). Since the file is not used, it's easy to show that its contents are irrelevant during later recompilations. (For an example of where this guideline was not followed, see section 7.3.2.1.)

5. *Encourage the development of alternative implementations of languages. Use or help develop public specifications for computer languages (preferably open standards)*. DDC requires a separate trusted compiler that can process the parent compiler. Thus, to simplify DDC use, encourage the development of alternative compilers and remove any roadblocks to their development.

DDC tends to be easier to apply if there are several already-existing compilers that could be used as a trusted compiler, and such compilers are more likely if there is a public specification for the language used to write the parent compiler. If such compilers do not already exist, having a public specification greatly simplifies the task of creating a trusted compiler for use with DDC. The specification should be an "open standard"; a good definition of the term "open standard" is the definition of "free and open standard" by the Digital Standards Organization[20]. Open standards enable fully open competition between suppliers.

---

[20]The Digital Standards Organization defines "free and open standard" as follows:
- A free and open standard is immune to vendor capture at all stages in its life-cycle. Immunity from vendor capture makes it possible to freely use, improve upon, trust, and extend a standard over time.
- The standard is adopted and will be maintained by a not-for-profit organization, and its ongoing development occurs on the basis of an open decision-making procedure available to all interested parties.
- The standard has been published and the standard specification document is available freely. It must be permissible to all to copy, distribute, and use it freely.
- The patents possibly present on (parts of) the standard are made irrevocably available on a royalty-free basis.
- There are no constraints on the re-use of the standard.

The economic outcome of a free and open standard is that it enables perfect competition between suppliers of products based on the standard [Digistan]. Patents, by definition, are exclusive and thus necessarily discriminatory when royalty payments or other conditions are imposed. See [Wheeler2008] for a comparison of various definitions of "open standard" and their application to a particular specification.

6. *Eliminate roadblocks to developing alternative language implementations, particularly patents. Avoid using constructs covered by potentially-enforceable patents, ensure that specification authors do not require the use of enforceable patents to implement the specification, and work to eliminate software patents worldwide.* Patents are government-granted monopolies. Historically, software could not be patented, and software innovation flourished without patents [Klemens2008] [Wheeler2009i]. Unfortunately, some countries have permitted software patents in recent years, and several analyses suggest that doing so was a mistake. For example, increases in software patent share in the 1990s were associated with *decreases* in research intensity [Bessen2004] (suggesting that software patents *discourage* research). Many other problems with software patents are discussed in [Bessen2008]. [End2008] summarizes the state of software patents as of 2008. Software patents affect DDC because they can inhibit the development of alternative compilers and environments. Since software patents can reduce the number of legal developers and users worldwide, software patents can even inhibit the availability of alternatives to those in countries free from software patents. Any patents that interfere with the creation of an alternative compiler or environment interfere with DDC, and thus interfere with security (because they interfere with protection against the trusting trust attack). Eliminating software patents worldwide would be the most thorough method to eliminate the problems they cause.

7. *Make the compiler portable and deterministic*. This is required by DDC (see section 5.7.8). If a compiler iterates over hashtable entries, ensure that the retrieved order will be the same across different environments and compiler implementations if it can affect the final result. If non-portable extensions are used in a compiler's implementation, clearly document the extensions.

8. *Consider using a simpler language subset to implement the compiler.* Using a subset can make it easier to implement a new trusted compiler if necessary, since the trusted compiler would probably need fewer constructs. Be sure to document this subset, and test to ensure that only this subset is used (as part of the compiler's regression test suite).

9. *Release self-parented compiler executables, if applicable.* If a compiler supports the language(s) it is written in, only release compiler executables after they have "self-parented" as described in section 4.5. This means that given the source code of a compiler and a bootstrap compiler executable, compile the source code using the bootstrap compiler, then use the resulting executable to compile the source code again. As noted in section 4.5, this has many practical benefits that have nothing to do with DDC (for example, if the compiler generates faster code than the bootstrap compiler does, then after self-recompilation the compiler itself will execute faster). For DDC, self-parenting reduces the amount of software that must be tracked (since the parent is the same as the compiler-under-test), and it reduces the amount of source code that must be examined afterwards to determine if the compiler is not malicious (since the source of the compiler-under-test $s_A$ is the same as the source of parent $s_P$, only $s_A$ needs to be examined).

10. *Release the compiler as free-libre/open source software (FLOSS), and choose a FLOSS compiler as its parent. Alternatively, though this alternative is less effective, release the source code to trusted third parties*. The source code for the compiler being tested and its parent must be available to apply DDC. In addition, DDC merely shows that the source code and executable correspond; the source code must then be inspected if the goal is to determine that there is no malicious code being executed. This means that the DDC technique is most useful for countering the trusting trust attack when applied to software whose source code is publicly available for review. Such review is much more useful for FLOSS, since with FLOSS any issues found in review can be repaired and redistributed by anyone. If a supplier refuses to release their compiler as FLOSS, the supplier should at least release the source code to third parties who can perform DDC and thoroughly examine the source code for malicious code. Such third parties must be potentially highly trusted by users, since users will not be able to independently verify the results.

11. *Apply DDC before each release*. Of course, the simplest way to ensure that DDC can be applied to a compiler is to perform DDC before each release. Users may want to apply DDC using different trusted compilers or trusted environments, but this is likely to be easier if DDC has previously been successfully applied.

# Appendix E: Key definitions

assembler
: A compiler for a language whose instructions are primarily a close approximation of the executing environment's instructions.

binary
: A common alternative term for executable (e.g., [Sabin2004]). However, this term is misleading; in modern computers, *all* data is represented using binary codes. Thus, this dissertation uses the term "executable" instead.

compiler
: An executable that, when executed, translates source code into an executable (it may also perform other actions).

compiling
: The process of using a compiler to translate source code into an executable.

correspond
: An executable e corresponds to source code s if and only if execution of e always behaves as specified by s when the execution environment of e behaves correctly.

corrupted compiler
: A corrupted executable that is a compiler.

corrupted executable
: An executable that does not correspond to its putative source code (see also "corrupted compiler" and "maliciously corrupted executable").

Diverse Double-Compiling (DDC)
: A technique for determining if a compiler is corrupted, in which the source code is compiled twice: the source code of the compiler's parent is compiled using a trusted compiler, and then the putative compiler source code is compiled using the result of the first compilation. If the DDC result is bit-for-bit identical with the original compiler-under-test's executable, and certain other assumptions hold, then the compiler-under-test's executable corresponds with its putative source code.

effects
: All information or execution timing arising from the environment that can affect the results of a compilation, but is not part of the input source code. This is used to model random number generators, thread execution ordering, differences between platforms allowed by the language, and so on.

environment
: A platform that can run executables. This would include the computer hardware (including the central processing unit) and any software that supports or could influence the compiler's result (e.g., the operating system).

executable
: Data that can be directly executed by a computing environment. An executable may be code for an actual machine or for a simulated machine (e.g., a "byte code"). Compilers produce executables, and compilers themselves are executables.

| | |
|---|---|
| fragility | The susceptibility of the trusting trust attack to failure, i.e., that a trigger will activate when the attacker did not wish it to (risking a revelation of the attack), fail to trigger when the attacker would wish it to, or that the payload will fail to work as intended by the attacker. |
| maliciously corrupted compiler | A maliciously corrupted executable that is a compiler. |
| maliciously corrupted executable | A corrupted executable whose corruption was caused by intentional subversion. |
| maliciously misleading code | Source code that is intentionally designed to look benign, yet creates a vulnerability (including an attack). |
| object code | For purposes of this dissertation, a synonym for "executable". |
| payload | Code that actually performs a malicious event (e.g., the inserted malicious code and the code that causes its insertion). These are initiated through triggers. |
| source code (aka source) | A representation of a program that can be transformed by a compiler into an executable. It is typically human-readable. |
| subverted compiler | Synonym for "maliciously corrupted compiler". |
| trigger | A condition, determined by an attacker, in which a malicious event is to occur (e.g., the condition causing malicious code to be inserted into a program, and the condition that causes the inserted code to take action). |
| Trojan horse | Software that appears to the user to perform a desirable function but facilitates unauthorized access into the user's computer system. |
| trusted | The justified confidence that something (e.g., a program or process) does not have triggers and payloads that would affect the results of DDC. See section 4.3 for a basic discussion of the term "trusted"; see chapter 6 for methods to increase the level of confidence. |
| trusting trust attack | An attack in which an attacker attempts to disseminate a compiler executable that produces corrupted executables, at least one of those corrupted executables is a corrupted compiler, and the attacker attempts to make this situation self-perpetuating. |

# Bibliography

# Bibliography

The references below are in strict alphabetical order, ignoring case. Uniform Resource Locators (URLs) may change or become invalid at any time; where provided, they are only intended to aid finding the information. If a URL is no longer valid, consider using the Internet Archive at <http://www.archive.org>.

[Cohen1984] Cohen, Fred. "Computer Viruses - Theory and Experiments". 1984. http://all.net/books/virus/index.html

[Cohen1985] Cohen, Fred. 1985. *Computer Viruses*. Ph.D. Thesis, University of Southern California.

[Dave2003] Dave, Maulik A. November 2003. "Compiler verification: a bibliography" *ACM SIGSOFT Software Engineering Notes*. Volume 28 , Issue 6. ISSN:0163-5948. New York: ACM Press. Note: "Dr. Maulik A. Dave" is correct.

[Digistan] Digital Standards Organization (Digistan). *Definition of a Free and Open Standard*. http://www.digistan.org/open-standard:definition

[Duffy1991] Duffy, David. 1991. *Principles of Automated Theorem Proving*. West Sussex, England: John Wiley & Sons Ltd. ISBN 0-471-92784-8.

[Dodge2005] Dodge, Dave. May 27, 2005. "Re: [Tinycc-devel] Mysterious tcc behavior: why does 0.0 takes 12 bytes when NOT long double". *tcc mailing list*.

[DoJ2006] United States Department of Justice (DoJ) U.S. Attorney, District of New Jersey, Public Affairs Office. December 13, 2006. "Former UBS Computer Systems Manager Gets 97 Months for Unleashing "Logic Bomb" on Company Network". Newark, New Jersey: United States Department of Justice. http://www.usdoj.gov/usao/nj/press/files/pdffiles/duro1213rel.pdf

[Draper1984] Draper, Steve. November 1984. "Trojan Horses and Trusty Hackers". *Communications of the ACM*. Volume 27, Number 11, p. 1085.

[Earley1970] Earley, Jay and Howard Sturgis. October 1970. "A Formalism for Translator Interactions". *Communications of the ACM*. Volume 13, Number 10. pp. 607-617.

[End2008] End Software Patents project. February 28, 2008. The current state of software and business method patents: 2008 edition. http://endsoftpatents.org/2008-state-of-softpatents

[Faigon] Faigon, Ariel. Testing for Zero Bugs. http://www.yendor.com/testing.

[Feldman2006] Feldman, Ariel J., J. Alex Halderman, and Edward W. Felten. September 13, 2006. Security Analysis of the Diebold AccuVote-TS Voting Machine. Center for Information Technology (IT) Policy, Princeton University. http://itpolicy.princeton.edu/voting/

[Feng2009] Feng, Chun. 2009-08-20. "Virus:Win32/Induc.A". *Malware Protection Center: Threat Research and Response*. Microsoft. http://www.microsoft.com/security/portal/Threat/Encyclopedia/Entry.aspx?name=Virus%3aWin32%2fInduc.A

[Ferreirós2001] Ferreirós, José. December 2001. "The Road to Modern Logic—An Interpretation". The Bulletin of Symbolic Logic. *Association for Symbolic Logic*. Vol. 7, No. 4. pp. 441-484. http://www.jstor.org/stable/2687794

[Forrest1994] Forrest, Stephanie, Lawrence Allen, Alan S. Perelson, and Rajesh Cherukuri. 1994. "Self-Nonself Discrimination in a Computer." *Proc. of the 1994 IEEE Symposium on Research in Security and Privacy.*

[Forrest1997] Forrest, Stephanie, Anil Somayaji, and David H. Ackley. 1997. "Building Diverse Computer Systems". *Proc. of the 6th Workshop on Hot Topics in Operating Systems*. Los Alamitos, CA: IEEE Computer Society Press. pp. 67-72.

[Forristal2005] Forristal, Jeff. Dec. 2005. Review: Source-Code Assessment Tools Kill Bugs Dead. *Secure Enterprise Magazine*.
http://www.secureenterprisemag.com/article/printableArticle.jhtml?articleId=174402221

[FSF2009] Free Software Foundation (FSF). June 30, 2009. *The Free Software Definition*.
http://www.gnu.org/philosophy/free-sw.html

[Gardian] Gardian. Undated. Infragard National Member Alliance.
http://www.infragardconferences.com/thegardian/3_22.html

[GAO2004] U.S. Government Accounting Office (GAO). May 2004. *Defense Acquisitions: Knowledge of Software Suppliers Needed to Manage Risks*. Report GAO-04-678.
http://www.gao.gov/cgi-bin/getrpt?GAO-04-678

[Gaudin2006a] Gaudin, Sharon. June 27, 2006. "How A Trigger Set Off A Logic Bomb At UBS PaineWebber". *InformationWeek*. http://www.informationweek.com/showArticle.jhtml?articleID=189601826

[Gaudin2006b] Gaudin, Sharon. July 19, 2006. "Ex-UBS Sys Admin Found Guilty, Prosecutors To Seek Maximum Sentence". *InformationWeek*.
http://www.informationweek.com/security/showArticle.jhtml?articleID=190700064

[Gaudin2008] Gaudin, Sharon. June 20, 2008. "Scientists build robot that can replicate itself: Machine designed to create 3-D plastic objects based on blueprint". *ComputerWorld*.
http://www.computerworld.com/s/article/9101738/Scientists_build_robot_that_can_replicate_itself

[Gauis2000] gauis (sic). May 1, 2000. "Things to do in Ciscoland when you're dead". *Phrack.* Volume 0xa, Issue 0x38. http://www.phrack.org/phrack/56/p56-0x0a

[Geer2003] Geer, Dan, Rebecca Bace, Peter Gutmann, Perry Metzger, Charles P. Pfleeger, John S. Quarterman, and Bruce Schneier. 2003. *Cyber Insecurity: The Cost of Monopoly.* Computer and Communications Industry Association (CCIA).
http://www.ccianet.org/CCIA/files/ccLibraryFiles/Filename/000000000061/cyberinsecurity.pdf or
http://cryptome.org/cyberinsecurity.htm

[GNU2002] GNU. 2002. *Using and Porting the GNU Compiler Collection (GCC)* (version 3.0.4). http://gcc.gnu.org/onlinedocs/gcc-3.0.4/gcc.html.
176

[Schwartau1994] Schwartau, Winn. 1994. *Information Warfare: Chaos on the Electronic Superhighway*. New York: Thunder's Mouth Press. ISBN 1-56025-080-1.

[SDIO1993] Strategic Defense Initiative Organization (SDIO). July 2, 1993. "Appendix A: Trust Principles". A revised appendix of *Trusted Software Methodology Volume 1: Trusted Software program Demonstration, Assessment and Refinement*. SDI-S-SD-91-000007, June 17, 1992. Washington, DC: SDIO. Prepared by GE Aerospace, Strategic Systems Department, Blue Bell, PA. CDRL A075-101B.

[Shankland2001] Shankland, Stephen. January 11, 2001. "Borland InterBase backdoor detected". ZDNet News. http://news.zdnet.com/2100-9595_22-527115.html

[Singh2002] Singh, Prabhat K., and Arun Lakhotia. February 2002. Analysis and Detection of Computer Viruses and Worms: An Annotated Bibliography. *ACM SIGPLAN Notices*. Volume 37, Issue 2. pp. 29 – 35.

[Spencer1998] Henry Spencer. November 23, 1998. "Re: LWN - The Trojan Horse (Bruce Perens)". *Robust Open Source mailing list* (open-source at csl.sri.com) established by Peter G. Neumann.

[Spencer2005] Henry Spencer, private communication.

[Spinellis2003] Spinellis, Diomidis. June 2003. "Reflections on Trusting Trust Revisited," *Communications of the ACM*. Volume 46, Number 6.
http://www.dmst.aueb.gr/dds/pubs/jrnl/2003-CACM-Reflections2/html/reflections2.pdf

[Stoll1979] Stoll, Robert R. 1979. *Set Theory and Logic*. Mineola, NY: Dover Publications, Inc. (This is the Dover edition, first published in 1979, that is a corrected republication of the work originally published in 1963 by W.H. Freeman and Company.) ISBN 0-486-63829-4.

[Stringer-Calvert1998] David William John Stringer-Calvert. March 1998. "Mechanical Verification of Compiler Correctness" (PhD thesis). University of York, Department of Computer Science. http://www.csl.sri.com/users/dave_sc/papers/thesis.ps.gz

[Thompson1984] Thompson, Ken. April 1984. "Reflections on Trusting Trust". *Communications of the ACM*. Volume 27, Number 8. pp. 761-763. http://www.acm.org/classics/sep95

[Thornburg2000] Thornburg, Jonathan. April 18, 2000. "?Backdoor in Microsoft web server?". Newsgroup sci.crypt. http://groups-beta.google.com/group/sci.crypt/msg/9305502fd7d4ee6f.

[Ulsch2000] Ulsch, MacDonnell. July 2000. "Security Strategies for E-Companies (EC Does it series)". *Information Security Magazine*.
http://infosecuritymag.techtarget.com/articles/july00/columns2_ec_doesit.shtml

[vonHagen2006] von Hagen, William. *The Definitive Guide to GCC*, Second Edition. 2006. New York: Springer-Verlag. ISBN 978-1-59059-585-5.
182

# Curriculum Vitae

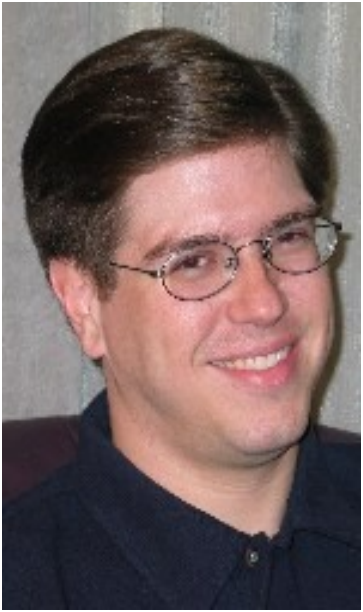

David A. Wheeler was born May 1965 in the United States of America and is an American citizen. He completed his B.S. in Electronics Engineering (with distinction) at George Mason University (GMU) in 1987 (awarded January 1988). He received his M.S. in Computer Science and a certificate for Software Engineering at GMU in 1994, when he also received a Computer Science graduate honor roll award. In 2000 he received a certificate in Information Systems Security from GMU. In 2009 he completed his requirements for a PhD in Information Technology from GMU.

From 1982 on he worked as a computer consultant, solving a variety of problems. He also spent time as the maintainer of the U.S.' first commercial multi-user role-playing game. In 1988 he joined the Institute for Defense Analyses (IDA), where he continues to solve challenging problems. His numerous awards include the Ada Programming Contest Award, membership in the Eta Kappa Nu Honor Society, and the George Washington University Engineering Award; he is also an Eagle Scout. His books include *Software Inspection: An Industry Best Practice* (IEEE Computer Society Press), *Ada 95: The Lovelace Tutorial* (Springer-Verlag), and *Secure Programming for Linux and Unix HOWTO* (self-published). His numerous articles include his developerWorks column "Secure Programmer", the article *Why Open Source Software / Free Software? Look at the Numbers!,* and "Countering Trusting Trust through Diverse Double-Compiling (DDC)" in *Proceedings of the Twenty-First Annual Computer Security Applications Conference* (ACSAC 2005). He has long worked on tasks related to large or high-risk systems, and in particular specializes in developing secure software, Free-libre/open source software (FLOSS), and open standards.

For more information, including contact information, see David A. Wheeler's personal website at <http://www.dwheeler.com>.